%% file: chgf1.tex
\documentclass[reprint,aps,twoside,prd,showkeys,superscriptaddress]{revtex4-2}
\usepackage[T1]{fontenc}
\usepackage{amsmath}
\usepackage{amssymb}
\usepackage{txfonts}
\usepackage{bm}
\usepackage{tensor}
\usepackage{verbatim}
\usepackage[makeroom]{cancel}
\newcommand{\HCd}{\mathcal{H}}
\newcommand{\LCd}{\mathcal{L}}
\newcommand{\FCd}{\tilde{\mathcal{F}}}
\renewcommand{\d}{\,\mathrm{d}}
\newcommand{\onehalf}{{\textstyle\frac{1}{2}}}
\newcommand{\onethird}{{\textstyle\frac{1}{3}}}

\newcommand{\quarter}{{\textstyle\frac{1}{4}}}
\newcommand{\pfrac}[2]{\frac{\partial{#1}}{\partial{#2}}}
\newcommand{\ppfrac}[3]{\frac{\partial^{2}{#1}}{\partial{#2}\partial{#3}}}
\newcommand{\pppfrac}[4]{\frac{\partial^{3}{#1}}{\partial{#2}\partial{#3}\partial{#4}}}
\newcommand{\afffias}{Frankfurt Institute for Advanced Studies (FIAS), Ruth-Moufang-Strasse~1, 60438 Frankfurt am Main, Germany}
\newcommand{\affguni}{Goethe Universit\"at, Max-von-Laue-Strasse~1, 60438 Frankfurt am Main, Germany}
\newcommand{\affgsi}{GSI Helmholtzzentrum f\"ur Schwerionenforschung GmbH, Planckstrasse~1, 64291 Darmstadt, Germany}
\begin{document}
\title{Canonical transformation path to gauge theories of gravity}
\author{J.~Struckmeier}\email{struckmeier@fias.uni-frankfurt.de}
\affiliation{\afffias}
\affiliation{\affguni}
\affiliation{\affgsi}
\author{J.~Muench}
\affiliation{\afffias}
\affiliation{\affguni}
\author{D.~Vasak}
\author{J.~Kirsch}
\affiliation{\afffias}
\author{M.~Hanauske}
\affiliation{\afffias}
\affiliation{\affguni}
\author{H.~Stoecker}
\affiliation{\afffias}
\affiliation{\affguni}
\affiliation{\affgsi}
\received{25 April 2017}
\published{27 June 2017}
\begin{abstract}
In this paper, the generic part of the gauge theory of gravity is derived,
based on the action principle and on the general principle of relativity.
We apply the canonical transformation framework to formulate geometrodynamics as a gauge theory.
The starting point of our paper is constituted by the general De~Donder-Weyl
Hamiltonian of a system of scalar and vector fields, which is supposed to be
form-invariant under (global) Lorentz transformations.
Following the reasoning of gauge theories, the corresponding locally
form-invariant system is worked out by means of canonical transformations.
The canonical transformation approach ensures by construction
that the form of the action functional is maintained.
We thus encounter amended Hamiltonian systems which are
form-invariant under arbitrary spacetime transformations.
This amended system complies with the general principle of relativity
and describes both, the dynamics of the given physical system's fields and
their coupling to those quantities which describe the dynamics of the spacetime geometry.
In this way, it is unambiguously determined how spin-$0$ and
spin-$1$ fields couple to the dynamics of spacetime.

A term that describes the dynamics of the free gauge fields must finally be added
to the amended Hamiltonian, as common to all gauge theories, to allow for a dynamic spacetime geometry.
The choice of this ``dynamics Hamiltonian'' is outside of the scope of gauge theory as presented in this paper.
It accounts for the remaining indefiniteness of any gauge theory
of gravity and must be chosen ``by hand'' on the basis of physical reasoning.
The final Hamiltonian of the gauge theory of gravity is shown to be at least
quadratic in the conjugate momenta of the gauge fields---this is
beyond the Einstein-Hilbert theory of general relativity.

\bigskip\noindent
DOI:~10.1103/PhysRevD.95.124048 (This version amended by Appendices~C, D, and E)
\end{abstract}
\maketitle
\section{Introduction}
The theory of general relativity, as proposed by A.~Einstein in~1915~\cite{einstein15c}---in
conjunction with the vacuum solution of K.~Schwarzschild~\cite{schwarzschild16}---has
provided a stunningly accurate description of the dynamics of celestial bodies.
This fact becomes even more surprising as Einstein's approach was in fact an
``educated guess'', or---in H.~Weyl's words---``a purely speculative theory''~\cite{weyl21}.
The modern comprehension of general relativity as a gauge theory arises from
requiring a given Lorentz-invariant theory to become diffeomorphism-invariant.
This approach was pioneered by Utiyama~\cite{utiyama56} in 1956.
On the other hand, a rigorous derivation of the theory that describes the interaction
of matter/energy with the spacetime fabric on the basis of the action principle
and the requirement that the description of any system should be form-invariant
under general spacetime transformations has not yet been delivered.

Moreover, Einstein's theory has severe limitations:
\begin{itemize}
\item[(i)] The theory is not scale-invariant
as the pertaining coupling constant is not dimensionless.
\item[(ii)] The underlying coarse-grained
energy-momentum balance equation of the theory appears to be inaccessible to quantization.
A more detailed and quantizable theory would describe the \emph{direct} interaction of
individual elementary particle fields with the gravitational field, the latter described
by the (uncontracted) Riemann tensor---similarly to the form of the Maxwell equation.
\item[(iii)] The observed dynamics of clusters of galaxies and stars in galaxies led to postulates
of the existence of ``dark matter'', in order to fit into the solutions of the Einstein equations.
\end{itemize}
In our previous attempts~\cite{struckmeier13,struckvasak15}, we have advocated a strategy
by which a formalism of extended canonical transformations is constructed in the
realm of covariant Hamiltonian field theory~\cite{dedonder30,weyl35}, which enables a
description of canonical transformations of fields under general mappings of the spacetime geometry.
Any theory derived from an action principle must maintain the general
form of the action principle under transformations of its dynamic quantities.
Consequently, those mappings are most naturally formulated as canonical transformations,
hence as transformations whose rules are derived from generating functions.

Any theory which conforms to the general principle of relativity---hence, which respects the
requirement of form-invariance under general mappings of spacetime geometry---must then
be formulated as a canonical transformation along the well-established procedures
of gauge transformations, dating back to H.~Weyl~\cite{weyl18} and W.~Fock~\cite{fock26}.
The particular transformation rule for the Hamiltonian then ``automatically'' provides the
structure of the ``gauge Hamiltonian'' which renders the original system locally form-invariant.
The form-invariance of the action functional is achieved
by simultaneously defining both, the appropriate transformation rules for the fields,
the conjugate momentum fields, and the transformation rule for the Hamiltonian.
In the present context, the particular gauge Hamiltonian is to be isolated which
renders our given Lorentz-invariant Hamiltonian system of scalar
and vector fields into a diffeomorphism-invariant, hence general coordinate-covariant system.
The gauge Hamiltonian thus describes uniquely the coupling between the dynamic
quantities of the given physical system with those describing the spacetime dynamics.
This conforms to the procedure generally pursued in gauge theories of gravity~\cite{ryder09}
and is in stark contrast with a postulation of a particular Lagrangian---the
latter procedure was first presented by D.~Hilbert~\cite{hilbert15}
with the appropriate Lagrangian that led to the postulated Einstein equation.
For the reader's convenience, an outline of the gauge procedure
in given in Sec.~\ref{sec:gp}.

The canonical transformation approach is presented starting
in Sec.~\ref{sec:ct-theory} with a brief review of the extended
canonical formalism in the realm of covariant Hamiltonian field theory.
The theory of canonical transformations then isolates exactly
those transformations of the dynamic variables which maintain the general
form of the action principle---and hence the general form of the canonical field equations.

The formalism is then applied to a system of scalar and vector fields,
in general curvilinear spacetime in Sec.~\ref{sec:example-spacetime-trans}.
The Hamiltonian is required to be form-invariant under general spacetime transformations.
This enforces the introduction of the affine connections as the appropriate
``gauge fields'' which must obey their particular transformation rules.
The affine gauge coefficients are not necessarily symmetric in their lower index pair.
Hence, the torsion of spacetime is included in this theory explicitly.
Introducing the affine connections as the gauge quantities promotes, in the language
of gauge theories, the global Lorentz-symmetry of the given system into a local,
general relativistic symmetry.
This renders the action integral invariant under arbitrary mappings of the reference frame.

Yet, at this stage the affine connection coefficients constitute \emph{external}
gauge fields~\cite{utiyama56}, with their dynamics being left open.
The connection coefficients are then treated as internal dynamic quantities by letting
their transformation rules follow from a particularly crafted generating function.
The subsequent transformation rule for the Hamiltonian then yields the particular gauge Hamiltonian
that amends the given Lorentz-invariant Hamiltonian to become a generally covariant Hamiltonian.
The gauge Hamiltonian is then inserted back into the initial action integral in Sec.~\ref{sec:act-int}.
Remarkably, the Riemann-Cartan tensor is in no way introduced ``by hand'' as in classical GR theories,
but \emph{naturally emerges} at this point of the canonical gauge formalism
as the particular function of the gauge field and its partial derivatives.
This constitutes the main result of our paper: the obtained generally covariant Hamiltonian
represents the generic Hamiltonian that is common to any particular theory of gravity.
Moreover, in order to encounter a closed set of field equations, we show that the final
Hamiltonian with dynamic space-time must be at least quadratic in the conjugate momenta of the gauge fields.
This contrasts with the Einstein-Hilbert theory of general relativity.

The Hamiltonian describing the dynamics of the ``free'' gauge fields is to be
inserted ``by hand'', as is common to all gauge theories.
This Hamiltonian must be chosen on the basis of physical reasoning.
A particular ``free gravity'' Hamiltonian is presented in Sec.~\ref{sec:sample}.
Concluding remarks are given in Sec.~\ref{sec:conclusions}.
\section{Canonical transformation rules under a dynamic spacetime geometry\label{sec:ct-theory}}
The formalism of canonical transformations, in the realm of classical field theory
under dynamic spacetime, was presented earlier~\cite{struckmeier13,struckvasak15}.
Here it is reformulated and thereby simplified considerably.
The extended canonical formalism of field theory involves the description how dynamic
quantities transform under the transition from one reference frame to another, $x\mapsto X$.
To achieve form-invariance of the action integral,
the transformation of the volume form $\d^{4}x$ must be taken into account.
$|\partial x/\partial X|$ denotes the determinant of the Jacobi matrix of the transformation $x\mapsto X$
\begin{equation}\label{eq:jac-matrix}
\left|\pfrac{x}{X}\right|\equiv\pfrac{\left(x^{0},\ldots,x^{3}\right)}{\left(X^{0},\ldots,X^{3}\right)}.
\end{equation}
The volume form $\d^{4}X$ transforms as a relative scalar of weight $w=-1$
\begin{equation}\label{eq:trans-volumeform}
\d^{4}X=\pfrac{\left(X^{0},\ldots,X^{3}\right)}{\left(x^{0},\ldots,x^{3}\right)}\,\d^{4}x=\left|\pfrac{x}{X}\right|^{-1}\,\d^{4}x.
\end{equation}
A general covariant second rank tensor transforms as
\begin{equation*}
G_{\mu\nu}(X)=g_{\alpha\beta}(x)\pfrac{x^{\alpha}}{X^{\mu}}\pfrac{x^{\beta}}{X^{\nu}},
\end{equation*}
and hence its determinant transforms according to
\begin{equation*}
(\det{G_{\mu\nu}})(X)=\left(\det{g_{\mu\nu}}\right)(x)\,\left|\pfrac{x}{X}\right|^{2}.
\end{equation*}
In the following, $g_{\mu\nu}$ is supposed to denote the
covariant representation of the metric tensor.
Then $g_{\mu\nu}$ has maximum rank and $g\equiv\det g_{\mu\nu}<0$.
The transformation rule for the determinant of the covariant
representation of the metric tensor, $g(x)\mapsto G(X)$, follows as:
\begin{equation}\label{eq:detg-trans}
\sqrt{-G}=\sqrt{-g}\,\left|\pfrac{x}{X}\right|.
\end{equation}
$\sqrt{-g}$ thus represents a relative scalar of weight $w=1$, i.e.\ a scalar density.
The product $\sqrt{-g}\,\d^{4}x$ transforms as a scalar of weight $w=0$, hence,
in conjunction with Eq.~(\ref{eq:detg-trans}), as an absolute scalar:
\begin{equation}\label{eq:trans-volform}
\sqrt{-G}\,\d^{4}X=\sqrt{-g}\,\d^{4}x.
\end{equation}
$\sqrt{-g}\,\d^{4}x$ is thus referred to as the \emph{invariant volume form}.

The variation of the action functional for a dynamical system of a scalar field $\phi$ and a
vector field $a_{\mu}$ transforms in conjunction with their respective conjugate momentum field densities
$\tilde{\pi}^{\nu}=\pi^{\nu}\sqrt{-g}$ and $\tilde{p}^{\,\mu\nu}=p^{\,\mu\nu}\sqrt{-g}$ as
\begin{align}
\delta S_{0}\equiv\delta&\int_{\Omega}\left(\tilde{\pi}^{\,\beta}\pfrac{\phi}{x^{\,\beta}}
+\tilde{p}^{\,\alpha\beta}\pfrac{a_{\alpha}}{x^{\,\beta}}-\tilde{\HCd}_{0}-\pfrac{\FCd_{1}^{\alpha}}{x^{\alpha}}\right)\d^{4}x\nonumber\\
\stackrel{!}{=}\delta&\int_{\Omega}\left(\tilde{\Pi}^{\beta}\pfrac{\Phi}{X^{\beta}}
+\tilde{P}^{\,\alpha\beta}\pfrac{A_{\alpha}}{X^{\beta}}-\tilde{\HCd}^{\prime}_{0}\right)\left|\pfrac{X}{x}\right|\,\d^{4}x,
\label{eq:varprinzip}
\end{align}
where $\tilde{\HCd}_{0}=\HCd_{0}\sqrt{-g}$ denotes the Hamiltonian scalar density pertaining
to the scalar $\HCd_{0}\big(\phi,\tilde{\pi}^{\nu},a_{\mu},\tilde{p}^{\,\mu\nu},g_{\mu\eta}\big)$, while
$\tilde{\Pi}^{\nu}=\Pi^{\nu}\sqrt{-G}$, $\tilde{P}^{\,\mu\nu}=P^{\,\mu\nu}\sqrt{-G}$, and
$\tilde{\HCd}_{0}^{\prime}=\HCd_{0}^{\prime}\sqrt{-G}$ denote the respective transformed quantities.
The terms $\tilde{\pi}^{\,\beta}\partial\phi/\partial x^{\,\beta}$ and
$\tilde{p}^{\,\alpha\beta}\partial a_{\alpha}/\partial x^{\,\beta}$ thus are \emph{Lorentz} scalar densities.
Additional gauge quantities must be introduced for general spacetime transformations,
in order to ensure that the integrands in Eq.~(\ref{eq:varprinzip}) are world scalar
densities, thus maintaining their form under general local spacetime transformations.
This corresponds to replacing the partial derivatives of the fields
in~(\ref{eq:varprinzip}) by covariant derivatives.

In other words, the differences of partial and covariant derivatives define the gauge quantities.
This important result is worked out in Sec.~\ref{sec:example-spacetime-trans}.

The action integral is to be varied, therefore Eq.~(\ref{eq:varprinzip})
implies that the integrands may differ by the ordinary
divergence of a \emph{vector density} $\FCd_{1}^{\mu}$,
whose variation vanishes on the boundary $\partial\Omega$ of the integration region $\Omega$ in spacetime.
The generalized version of Gauss' theorem writes for a relative tensor of weight $w=1$
with $n_\alpha$ the normal vector to the surface $S\equiv\partial\Omega$ that encloses the volume $\Omega$ (see~\cite[Eq.~(E.14), for instance]{carroll13}):
\begin{equation}\label{eq:surface-term}
\delta\int_{\Omega}\pfrac{\FCd_{1}^{\alpha}}{x^{\alpha}}\,\d^4x=
\delta\oint_{\partial\Omega}\FCd_{1}^{\alpha}\,n_\alpha\d^3 S\stackrel{!}{=}0.
\end{equation}
The variation of the action integral~(\ref{eq:varprinzip}) is not modified by adding a term
$\partial\FCd_{1}^{\alpha}/\partial x^{\alpha}$ to the integrand which can be converted into
a surface integral according to Eq.~(\ref{eq:surface-term})---commonly denoted briefly as a \emph{surface term}:
the integrand is only determined up to the divergence of vector densities $\FCd_{1}^{\mu}(\Phi,\phi,A,a,x)$.

On the other hand, we are not restricted to a tensorial character of $\FCd_{1}^{\mu}$.
We may confine ourselves to canonical transformations which merely provide \emph{form-invariance} of the action.
For the actual task of promoting a Lorentz-invariant to a diffeo\-mor\-phism-invariant system
by means of incorporating dynamical gauge fields, we encounter couplings of the given fields to the gauge fields.
We thus modify the action functional by finally deriving an additional \emph{gauge Hamiltonian}
$\tilde{\HCd}_{\mathrm{G}}$ that converts the initial action integrand into a world scalar density.
The crucial point is that these gauge Hamiltonians must (and will) have the same functional form in the original
and in the transformed system in order to yield the required preservation of the canonical field equations,
and hence to implement the postulate of diffeo\-mor\-phism-invariance of the gauged system.

The integrand condition for a canonical transformation writes
\begin{align}
&\,\tilde{\pi}^{\beta}\pfrac{\phi}{x^{\beta}}+\tilde{p}^{\,\alpha\beta}\pfrac{a_{\alpha}}{x^{\beta}}-\tilde{\HCd}
-\left(\tilde{\Pi}^{\beta}\pfrac{\Phi}{X^{\beta}}+\tilde{P}^{\,\alpha\beta}\pfrac{A_{\alpha}}{X^{\beta}}
-\tilde{\HCd}^{\prime}\right)\left|\pfrac{x}{X}\right|^{-1}\nonumber\\
=&\,\pfrac{\FCd_{1}^{\beta}}{\phi}\pfrac{\phi}{x^{\beta}}
+\pfrac{\FCd_{1}^{\alpha}}{\Phi}\pfrac{X^{\beta}}{x^{\alpha}}\pfrac{\Phi}{X^{\beta}}
+\pfrac{\FCd_{1}^{\beta}}{a_{\alpha}}\pfrac{a_{\alpha}}{x^{\beta}}
+\pfrac{\FCd_{1}^{\xi}}{A_{\alpha}}\pfrac{X^{\beta}}{x^{\xi}}\pfrac{A_{\alpha}}{X^{\beta}}\nonumber\\
&\mbox{}+\left.\pfrac{\FCd_{1}^{\alpha}}{x^{\alpha}}\right|_{\text{expl}},
\label{eq:integrand-condition}
\end{align}
with the transformation rule of the volume form from Eq.~(\ref{eq:trans-volumeform})
and $\FCd_{1}^{\mu}$ to be taken at $x$.
The transformation rules are obtained by comparing the coefficients
\begin{align}
\tilde{\pi}^{\mu}(x)&=\hphantom{-}\pfrac{\FCd_{1}^{\mu}}{\phi}\nonumber\\
\tilde{\Pi}^{\mu}(X)&=-\pfrac{\FCd_{1}^{\beta}}{\Phi}\pfrac{X^{\mu}}{x^{\beta}}\left|\pfrac{x}{X}\right|\nonumber\\
\tilde{p}^{\,\nu\mu}(x)&=\hphantom{-}\pfrac{\FCd_{1}^{\mu}}{a_{\nu}}\nonumber\\
\tilde{P}^{\,\nu\mu}(X)&=-\pfrac{\FCd_{1}^{\beta}}{A_{\nu}}\pfrac{X^{\mu}}{x^{\beta}}\left|\pfrac{x}{X}\right|\nonumber\\
\tilde{\HCd}^{\prime}\Big|_{X}&=\left(\tilde{\HCd}\Big|_{x}
+\left.\pfrac{\FCd_{1}^{\alpha}}{x^{\alpha}}\right|_{\text{expl}}\right)\left|\pfrac{x}{X}\right|.
\label{eq:f1-ext}
\end{align}
Obviously, $\FCd_{1}^{\mu}$ can be devised to generate specific transformation rules of the
involved fields and their conjugates---this is the reason that $\FCd_{1}^{\mu}$ is called a \emph{generating function}.
The generating function $\FCd_{1}^{\mu}(\Phi,\phi,A,a,x)$ can be Legen\-dre-transformed into
the equivalent generating function $\FCd_{2}^{\mu}(\tilde{\Pi},\phi,\tilde{P},a,x)$ according to
\begin{equation*}
\FCd_{2}^{\mu}=\FCd_{1}^{\mu}+\left(\Phi\,\tilde{\Pi}^{\beta}
+A_{\alpha}\,\tilde{P}^{\alpha\beta}\right)\pfrac{x^{\mu}}{X^{\beta}}\left|\pfrac{x}{X}\right|^{-1}.
\end{equation*}
The transformation rules for $\FCd_{2}^{\mu}$ to be taken at the spacetime event $x$ are:
\begin{align}
\tilde{\pi}^{\mu}(x)&=\pfrac{\FCd_{2}^{\mu}}{\phi}\nonumber\\
\delta_{\nu}^{\mu}\Phi(X)&=\pfrac{\FCd_{2}^{\alpha}}{\tilde{\Pi}^{\nu}}\pfrac{X^{\mu}}{x^{\alpha}}\left|\pfrac{x}{X}\right|\nonumber\\
\tilde{p}^{\,\nu\mu}(x)&=\pfrac{\FCd_{2}^{\mu}}{a_{\nu}}\nonumber\\
\delta_{\nu}^{\mu}A_{\alpha}(X)&=\pfrac{\FCd_{2}^{\beta}}{\tilde{P}^{\alpha\nu}}\pfrac{X^{\mu}}{x^{\beta}}\left|\pfrac{x}{X}\right|\nonumber\\
\tilde{\HCd}^{\prime}\Big|_{X}&=\left(\tilde{\HCd}\Big|_{x}
+\left.\pfrac{\FCd_{2}^{\alpha}}{x^{\alpha}}\right|_{\text{expl}}\right)\left|\pfrac{x}{X}\right|.
\label{eq:f2-ext}
\end{align}
While the Hamiltonians $\tilde{\HCd}$ and $\tilde{\HCd}^{\prime}$ do not necessarily represent scalar densities,
the total integrands in the action integrals~(\ref{eq:varprinzip}) must be world scalars
in order to keep their form under general spacetime transformations.
This ensures that the canonical field equations emerge as tensor equations.
\section{General spacetime transformation of systems of scalar-,
vector-, and tensor fields\label{sec:example-spacetime-trans}}
A Hamiltonian $\HCd_{0}(\phi,\pi^{\nu},a_{\mu},p^{\mu\nu},g_{\mu\eta})$ is now considered
which describes the dynamics of distinct classical fields, namely a scalar field $\phi$, and a vector field $a_{\mu}$.
The quantity $g_{\mu\eta}$ in the argument list of the Hamiltonian
is interpreted as the covariant representation of the (symmetric) metric tensor.
The contravariant vector field $\pi^{\nu}$ denotes the canonical conjugate of $\phi$
in the context of covariant Hamiltonian field theory~\cite{struckmeier08}.
Hence, $\pi^{\nu}$ is dual quantity of the covariant vector of spacetime
derivatives $\partial\phi/\partial x^{\nu}$ of the scalar field $\phi$.
Likewise, the $(2,0)$ tensor $p^{\,\mu\nu}$ stands for the canonical conjugate
of the covariant field vector $a_{\mu}$, hence, for  the dual quantity of the derivatives $\partial a_{\mu}/\partial x^{\nu}$.
A $(3,0)$ tensor $k^{\mu\eta\nu}$, accordingly, represents the canonical conjugate of the metric tensor
$g_{\mu\eta}$ and hence is the dual quantity of the partial derivatives $\partial g_{\mu\eta}/\partial x^{\nu}$.
The tensor $k^{\mu\eta\nu}$ will be introduced later, in the action functional, in order to describe
the metric $g_{\mu\eta}$ as an internal dynamic quantity of an amended
Hamiltonian system, rather than as the external field variable in $\HCd_{0}$.
The Hamiltonian $\HCd_{0}$ is assumed to be form-invariant under global spacetime transformations,
hence, $\HCd_{0}$ constitutes a Lorentz scalar.
The scalar field, the vector field, and the metric tensor transform
under passive local coordinate transitions, i.e.\ if the transformation of
the spacetime event $x^{\mu}\mapsto X^{\mu}$ is applied, according to
\begin{align}
\Phi(X)&=\phi(x)\nonumber\\
A_{\mu}(X)&=a_{\xi}(x)\pfrac{x^{\xi}}{X^{\mu}}\label{eq:vec-trans}\\
G_{\mu\eta}(X)&=g_{\xi\zeta}(x)\pfrac{x^{\xi}}{X^{\mu}}\pfrac{x^{\zeta}}{X^{\eta}}\nonumber.
\end{align}
These transformations are generated, in the context of the extended canonical
transformation formalism of covariant Hamiltonian field theory~\cite{struckmeier15}, by
\begin{equation}\label{eq:gen-ex5}
\begin{split}
{\left.\FCd_{2}^{\mu}\right|}_x&=\Bigg(\tilde{\Pi}^{\beta}(X)\,\phi(x)
+\tilde{P}^{\alpha\beta}(X)\,a_{\xi}(x)\pfrac{x^{\xi}}{X^{\alpha}}\\
&\qquad+\tilde{K}^{\alpha\lambda\beta}(X)g_{\xi\zeta}(x)\pfrac{x^{\xi}}{X^{\alpha}}
\pfrac{x^{\zeta}}{X^{\lambda}}\Bigg)\pfrac{x^{\mu}}{X^{\beta}}\left|\pfrac{x}{X}\right|^{-1}.
\end{split}
\end{equation}
Here $x^{\mu}$ and $X^{\nu}$ denote the independent variables in the two distinct reference frames.
With Eqs.~(\ref{eq:jac-matrix}) and~(\ref{eq:detg-trans}), the tensor density $\tilde{\Pi}^{\beta}(X)=\Pi^{\beta}(X)\sqrt{-G}$
denotes the canonical conjugate of the transformed scalar fields $\Phi(X)$.

In analogy, the $\tilde{P}^{\alpha\beta}(X)=P^{\alpha\beta}(X)\sqrt{-G}$
stand for the corresponding conjugates of the transformed vector fields, $A_{\alpha}(X)$,
and $\tilde{K}^{\alpha\beta\xi}(X)=K^{\alpha\beta\xi}(X)\sqrt{-G}$ denote
the momenta of the transformed tensor field $G_{\alpha\beta}(X)$.

The particular generating function~(\ref{eq:gen-ex5}) embodies a
contravariant vector of weight $w=1$ and, hence, a vector density.
But this need not in general be the case:
if transformations of nontensorial quantities are defined---such as
the connection coefficients---then the corresponding $\FCd_{2}^{\mu}$ cannot represent a vector density.

The crucial requirement is that the total integrand of the action functional
represents a world scalar density.

Equation~(\ref{eq:vec-trans}) constitutes a global symmetry transformation if its
coefficients $\partial x^{\xi}/\partial X^{\mu}$ do not depend on the spacetime event $x$
and if $\HCd_{0}$ is form-invariant under this transformation.
In contrast, the transformation is referred to as being local
if the coefficients \emph{do} depend on spacetime.
The Hamiltonian $\HCd_{0}$ is then no longer form-invariant under the corresponding canonical transformation rule.
Appropriate dynamic gauge quantities must then be introduced to restore the
form-invariance of a then amended Hamiltonian system---as is usual for all gauge theories.

Explicitly, the new canonical transformation rules, which emerge from
$\FCd_{2}^{\mu}\big(\phi,\tilde{\Pi}^{\nu},a_{\mu},\tilde{P}^{\mu\nu},g_{\mu\eta},\tilde{K}^{\mu\eta\nu}\big)$
of Eq.~(\ref{eq:gen-ex5}), are
\begin{align*}
\tilde{\pi}^{\mu}&=\pfrac{\FCd_{2}^{\mu}}{\phi}
=\tilde{\Pi}^{\beta}\pfrac{x^{\mu}}{X^{\beta}}\left|\pfrac{x}{X}\right|^{-1}\\
\delta_{\nu}^{\mu}\Phi&=\pfrac{\FCd_{2}^{\kappa}}{\tilde{\Pi}^{\nu}}\pfrac{X^{\mu}}{x^{\kappa}}\left|\pfrac{x}{X}\right|
=\delta_{\nu}^{\beta}\pfrac{x^{\kappa}}{X^{\beta}}\pfrac{X^{\mu}}{x^{\kappa}}\phi=\delta_{\nu}^{\mu}\phi\\
\tilde{p}^{\nu\mu}&=\pfrac{\FCd_{2}^{\mu}}{a_{\nu}}
=\tilde{P}^{\alpha\beta}\delta_{\xi}^{\nu}\pfrac{x^{\xi}}{X^{\alpha}}\pfrac{x^{\mu}}{X^{\beta}}\left|\pfrac{x}{X}\right|^{-1}\\
&=\tilde{P}^{\alpha\beta}\pfrac{x^{\nu}}{X^{\alpha}}\pfrac{x^{\mu}}{X^{\beta}}\left|\pfrac{x}{X}\right|^{-1}\\
\delta_{\beta}^{\mu}A_{\alpha}&=\pfrac{\FCd_{2}^{\kappa}}{\tilde{P}^{\alpha\beta}}\pfrac{X^{\mu}}{x^{\kappa}}\left|\pfrac{x}{X}\right|
=a_{\xi}\pfrac{x^{\xi}}{X^{\alpha}}\pfrac{x^{\kappa}}{X^{\beta}}\pfrac{X^{\mu}}{x^{\kappa}}\\
&=\delta_{\beta}^{\mu}a_{\xi}\pfrac{x^{\xi}}{X^{\alpha}}\\
\tilde{k}^{\xi\zeta\mu}&=\pfrac{\FCd_{2}^{\mu}}{g_{\xi\zeta}}
=\tilde{K}^{\alpha\lambda\beta}\pfrac{x^{\xi}}{X^{\alpha}}
\pfrac{x^{\zeta}}{X^{\lambda}}\pfrac{x^{\mu}}{X^{\beta}}\left|\pfrac{x}{X}\right|^{-1}\\
\delta_{\beta}^{\mu}G_{\alpha\lambda}
&=\pfrac{\FCd_{2}^{\kappa}}{\tilde{K}^{\alpha\lambda\beta}}\pfrac{X^{\mu}}{x^{\kappa}}\left|\pfrac{x}{X}\right|
=g_{\xi\zeta}\,\pfrac{x^{\xi}}{X^{\alpha}}\pfrac{x^{\zeta}}{X^{\lambda}}\delta_{\beta}^{\mu}.
\end{align*}
Obviously, the required transformation rules~(\ref{eq:vec-trans}) of the fields are reproduced.
By virtue of Eq.~(\ref{eq:jac-matrix}), the momentum fields obey the rules
required for relative tensors of weight $w=1$.
Hence, for tensor densities
\begin{align}
\tilde{\Pi}^{\mu}(X)&=\tilde{\pi}^{\beta}(x)\,\pfrac{X^{\mu}}{x^{\beta}}\left|\pfrac{x}{X}\right|\nonumber\\
\tilde{P}^{\nu\mu}(X)&=\tilde{p}^{\alpha\beta}(x)\pfrac{X^{\nu}}{x^{\alpha}}\pfrac{X^{\mu}}{x^{\beta}}
\left|\pfrac{x}{X}\right|\label{eq:mom-trans}\\
\tilde{K}^{\xi\zeta\mu}(X)&=\tilde{k}^{\alpha\lambda\beta}(x)\pfrac{X^{\xi}}{x^{\alpha}}
\pfrac{X^{\zeta}}{x^{\lambda}}\pfrac{X^{\mu}}{x^{\beta}}\left|\pfrac{x}{X}\right|.\nonumber
\end{align}
The general transformation rule for the Hamiltonian densities is given by Eq.~(\ref{eq:f2-ext}).
For the actual generating function, Eq.~(\ref{eq:gen-ex5}),
the divergence of the explicit $x$-dependent terms of $\FCd_{2}^{\mu}$ follows as
(see Appendix~\ref{sec:app1} for the vanishing first term)
\begin{align}
\left.\pfrac{\FCd_{2}^{\alpha}}{x^{\alpha}}\right|_{\text{expl}}
&=\left(\tilde{\Pi}^{\beta}\,\phi+\tilde{P}^{\alpha\beta}\,a_{\xi}\pfrac{x^{\xi}}{X^{\alpha}}
+\tilde{K}^{\alpha\lambda\beta}g_{\xi\zeta}\pfrac{x^{\xi}}{X^{\alpha}}\pfrac{x^{\zeta}}{X^{\lambda}}\right)\nonumber\\
&\qquad\times\underbrace{\pfrac{}{x^{\mu}}\left(\pfrac{x^{\mu}}{X^{\beta}}\left|\pfrac{x}{X}\right|^{-1}
\right)}_{\equiv0\,\,(\mathrm{Eq.}~(\ref{eq:H-deri-expr-2}))}\nonumber\\
&\quad\mbox{}+\Bigg[\tilde{P}^{\alpha\beta}\,a_{\xi}\ppfrac{x^{\xi}}{X^{\alpha}}{X^{\beta}}
+\tilde{K}^{\alpha\lambda\beta}g_{\xi\zeta}\,\times\nonumber\\
&\qquad\left(\ppfrac{x^{\xi}}{X^{\alpha}}{X^{\beta}}\pfrac{x^{\zeta}}{X^{\lambda}}
+\ppfrac{x^{\zeta}}{X^{\lambda}}{X^{\beta}}\pfrac{x^{\xi}}{X^{\alpha}}\right)\Bigg]\left|\pfrac{x}{X}\right|^{-1}.\label{eq:pext-st}
\end{align}
The divergence~(\ref{eq:pext-st}) can be expressed completely in terms of the original dynamic quantities
by inserting the transformation rules for the momentum fields from Eq.~(\ref{eq:mom-trans}):
\begin{align}
\left.\pfrac{\FCd_{2}^{\alpha}}{x^{\alpha}}\right|_{\text{expl}}&=\left(
\tilde{p}^{\,\alpha\beta}a_{\xi}+\tilde{k}^{\alpha\lambda\beta}g_{\xi\lambda}
+\tilde{k}^{\lambda\alpha\beta}g_{\lambda\xi}\right)
\ppfrac{x^{\xi}}{X^{\kappa}}{X^{\eta}}\pfrac{X^{\kappa}}{x^{\alpha}}\pfrac{X^{\eta}}{x^{\beta}}.
\label{eq:excess-term}
\end{align}
For arbitrary dynamical fields in Cartesian coordinate description the divergence of $\FCd_{2}^{\alpha}$
vanishes if and only if the second derivatives of the $x^{\xi}(X)$ do all vanish:
\begin{equation}\label{eq:global-trans}
\ppfrac{x^{\xi}}{X^{\kappa}}{X^{\eta}}=0\qquad\Leftrightarrow\qquad
\left.\pfrac{\FCd_{2}^{\alpha}}{x^{\alpha}}\right|_{\text{expl}}=0.
\end{equation}
In this case, the Hamiltonian is kept unchanged.
The transformation then does \emph{not} depend on the spacetime
location and is, therefore, referred to as \emph{global}.
Otherwise, for the case of a \emph{local} transformation,
the Hamiltonians are \emph{not} form-invariant.
This shows already at this level that the postulation of diffeomorphism invariance of our system---being derived from an action
principle---is related to the second derivative of $x(X)$ and hence---as will be shown in the following---to a \emph{curvature of spacetime}.

To recover the form-invariance of the Hamiltonian, a ``gauge Hamiltonian'' $\tilde{\HCd}_{\mathrm{G}}$ must be defined which
matches the $x(X)$ dependencies of Eq.~(\ref{eq:excess-term}).
This finally yields the particular amended Hamiltonian which is supposed to be form-invariant under local transformations:
\begin{align}
\tilde{\HCd}_{1}&=\tilde{\HCd}_{0}+\tilde{\HCd}_{\mathrm{G}}\label{eq:gauge-ham2}\\
\tilde{\HCd}_{\mathrm{G}}&=\left(\tilde{p}^{\alpha\beta}a_{\xi}+\tilde{k}^{\alpha\lambda\beta}g_{\xi\lambda}
+\tilde{k}^{\lambda\alpha\beta}g_{\lambda\xi}\right)\gamma\indices{^{\xi}_{\alpha\beta}}.\nonumber
\end{align}
Herein, the $\gamma\indices{^{\xi}_{\alpha\beta}}$ formally denote the gauge quantities,
whose physical meaning will be clarified below setting up their transformation rule.
Of course, the ``gauge Hamiltonian'' $\tilde{\HCd}_{\mathrm{G}}^{\prime}$ of the transformed system must have the same
form in order to work out the locally form-invariant amended Hamiltonian $\tilde{\HCd}_{1}$
\begin{align}
\tilde{\HCd}_{1}^{\prime}&=\tilde{\HCd}_{0}^{\prime}+\tilde{\HCd}_{\mathrm{G}}^{\prime}\label{eq:gauge-ham1}\\
\tilde{\HCd}_{\mathrm{G}}^{\prime}&=\left(\tilde{P}^{\alpha\beta}A_{\xi}
+\tilde{K}^{\alpha\lambda\beta}G_{\xi\lambda}
+\tilde{K}^{\lambda\alpha\beta}G_{\lambda\xi}\right)\Gamma\indices{^{\xi}_{\alpha\beta}}.\nonumber
\end{align}
The transformation rule of the gauge quantities $\gamma\indices{^{\xi}_{\alpha\beta}}$
and $\Gamma\indices{^{\xi}_{\alpha\beta}}$ is determined by expressing the transformed
gauge Hamiltonian~(\ref{eq:gauge-ham1}) in terms of the original fields according to
the canonical transformation rules~(\ref{eq:vec-trans}) and~(\ref{eq:mom-trans}):
\begin{equation}
\tilde{\HCd}_{\mathrm{G}}^{\prime}=\left(\tilde{p}^{\alpha\beta}a_{\xi}
+\tilde{k}^{\alpha\lambda\beta}g_{\xi\lambda}+\tilde{k}^{\lambda\alpha\beta}g_{\lambda\xi}\right)
\pfrac{x^{\xi}}{X^{\eta}}\pfrac{X^{\kappa}}{x^{\alpha}}
\pfrac{X^{\tau}}{x^{\beta}}\Gamma\indices{^{\eta}_{\kappa\tau}}\left|\pfrac{x}{X}\right|.
\label{eq:gauge-ham1a}
\end{equation}
The demanded correlation of the formally introduced gauge quantities
$\gamma\indices{^{\xi}_{\alpha\beta}}$ and $\Gamma\indices{^{\xi}_{\alpha\beta}}$
is obtained by inserting the transformed gauge Hamiltonian in the representation of Eq.~(\ref{eq:gauge-ham1a})
and the original gauge Hamiltonian~(\ref{eq:gauge-ham2}) with~(\ref{eq:excess-term}) into Eq.~(\ref{eq:f2-ext})
\begin{align*}
&\quad\left(\tilde{p}^{\alpha\beta}a_{\xi}+\tilde{k}^{\alpha\lambda\beta}g_{\xi\lambda}
+\tilde{k}^{\lambda\alpha\beta}g_{\lambda\xi}\right)\pfrac{x^{\xi}}{X^{\eta}}\pfrac{X^{\kappa}}{x^{\alpha}}
\pfrac{X^{\tau}}{x^{\beta}}\Gamma\indices{^{\eta}_{\kappa\tau}}\\
&=\left(\tilde{p}^{\alpha\beta}a_{\xi}+\tilde{k}^{\alpha\lambda\beta}g_{\xi\lambda}
+\tilde{k}^{\lambda\alpha\beta}g_{\lambda\xi}\right)\left(\gamma\indices{^{\xi}_{\alpha\beta}}
+\ppfrac{x^{\xi}}{X^{\kappa}}{X^{\eta}}\pfrac{X^{\kappa}}{x^{\alpha}}\pfrac{X^{\eta}}{x^{\beta}}\right).
\end{align*}
The coefficients are compared to yield the condition
\begin{equation*}
\pfrac{x^{\xi}}{X^{\eta}}\pfrac{X^{\kappa}}{x^{\alpha}}\pfrac{X^{\tau}}{x^{\beta}}
\Gamma\indices{^{\eta}_{\kappa\tau}}=\gamma\indices{^{\xi}_{\alpha\beta}}
+\ppfrac{x^{\xi}}{X^{\kappa}}{X^{\eta}}\pfrac{X^{\kappa}}{x^{\alpha}}\pfrac{X^{\eta}}{x^{\beta}}.
\end{equation*}
The transformation rule for the ``gauge fields'' follows, after solving for
the $\Gamma\indices{^{\kappa}_{\alpha\beta}}$, as
\begin{equation}\label{eq:trans-conn-coeff}
\Gamma\indices{^{\kappa}_{\alpha\beta}}(X)
=\gamma\indices{^{\xi}_{\eta\tau}}(x)\pfrac{x^{\eta}}{X^{\alpha}}
\pfrac{x^{\tau}}{X^{\beta}}\pfrac{X^{\kappa}}{x^{\xi}}
+\ppfrac{x^{\xi}}{X^{\alpha}}{X^{\beta}}\pfrac{X^{\kappa}}{x^{\xi}}.
\end{equation}
This transformation rule corresponds to the transformation rule of the affine connection coefficients.
In the following, we identify the gauge fields $\gamma\indices{^{\xi}_{\alpha\beta}}$---formally
introduced in Eq.~(\ref{eq:gauge-ham2})---with the affine connection coefficients.
In this aspect, we follow the approach of Palatini~\cite{palatini19},
who first treated the metric and the connection coefficients as separate dynamic quantities,
which entails an additional equation of motion providing their mutual correlation.

A Hamiltonian system, $\tilde{\HCd}_{0}=\HCd_{0}(\phi,\pi^{\nu},a_{\mu},p^{\mu\nu},g_{\mu\eta})\sqrt{-g}$,
which is supposed to be invariant under Lorentz transformations as the
global symmetry group, is rendered form-invariant under the
local diffeomorphism group if and only if it is amended according to Eq.~(\ref{eq:gauge-ham2}),
provided that the gauge quantities $\gamma\indices{^{\xi}_{\alpha\beta}}$
transform according to Eq.~(\ref{eq:trans-conn-coeff}):
\begin{align}
\pfrac{\FCd_{2}^{\alpha}}{x^{\alpha}}\bigg|_{\text{expl}}
&=\left(\tilde{P}^{\alpha\beta}A_{\xi}+\tilde{K}^{\alpha\lambda\beta}G_{\xi\lambda}
+\tilde{K}^{\lambda\alpha\beta}G_{\lambda\xi}\right)\Gamma\indices{^{\xi}_{\alpha\beta}}\left|\pfrac{X}{x}\right|\nonumber\\
&\quad\mbox{}-\left(\tilde{p}^{\alpha\beta}a_{\xi}+\tilde{k}^{\alpha\lambda\beta}g_{\xi\lambda}
+\tilde{k}^{\lambda\alpha\beta}g_{\lambda\xi}\right)\gamma\indices{^{\xi}_{\alpha\beta}}\nonumber\\
&=\left(\tilde{\HCd}^{\prime}_0+\tilde{\HCd}^{\prime}_{\mathrm{G}}\right)\left|\pfrac{X}{x}\right|-\left(\tilde{\HCd}_0+\tilde{\HCd}_{\mathrm{G}}\right)\nonumber\\
&=\tilde{\HCd}^{\prime}_1\left|\pfrac{X}{x}\right|-\tilde{\HCd}_1.
\label{eq:trans-hg1}
\end{align}
The requirement of form-invariance of the given Hamiltonian
under local spacetime transformations thus induces a coupling term
of the vector- and the tensor fields and their conjugates
via the gauge coefficients $\gamma\indices{^{\xi}_{\alpha\beta}}$ in $\tilde{\HCd}_{\mathrm{G}}$.
The $\gamma\indices{^{\xi}_{\alpha\beta}}$---interpreted as \emph{connection coefficients}---act
in a way to convert the partial derivatives in the action
functional~(\ref{eq:varprinzip}) into covariant derivatives.
Furthermore, the metric $g_{\mu\nu}$ is promoted by the gauge procedure from an external
quantity in $\tilde{\HCd}_{0}$ to an internal spacetime-dependent quantity, whose
coupling to the vector and tensor fields is described by $\tilde{\HCd}_{1}$:
\begin{align}
S_{2}&=\int_{\Omega}\bigg(\tilde{\pi}^{\beta}\pfrac{\phi}{x^{\beta}}
+\tilde{p}^{\,\alpha\beta}\pfrac{a_{\alpha}}{x^{\beta}}
+\tilde{k}^{\,\alpha\lambda\beta}\pfrac{g_{\alpha\lambda}}{x^{\beta}}-\tilde{\HCd}_{1}\bigg)\d^{4}x\nonumber\\
&=\int_{\Omega}\bigg(\tilde{\pi}^{\beta}\phi_{;\beta}
+\tilde{p}^{\,\alpha\beta}a_{\alpha;\beta}
+\tilde{k}^{\,\alpha\lambda\beta}g_{\alpha\lambda;\beta}-\tilde{\HCd}_{0}\bigg)\d^{4}x.
\label{eq:action-integral-first_step}
\end{align}
Generally speaking, the form of the coupling term in gauge theories is uniquely determined by the
particular global symmetry property of the system, which is rendered local.
The $\gamma\indices{^{\xi}_{\alpha\beta}}$ in Eq.~(\ref{eq:trans-conn-coeff})
need not be symmetric in the lower indices $\alpha$ and $\beta$~\cite{hehl76}.
Yet, restricting the theory to only the symmetric part of $\gamma\indices{^{\xi}_{\alpha\beta}}$---which
is equivalent to postulating a vanishing torsion of spacetime---greatly simplifies the field equations for the spacetime dynamics.

We have now successfully eliminated the arbitrary coefficients $\partial x^\xi/\partial X^\mu$ and their derivatives
by replacing them with the physical gauge fields $\gamma\indices{^{\xi}_{\alpha\beta}}$, i.e., the connection coefficients.
Yet, these coefficients constitute \emph{external} gauge fields whose dynamics
are not determined by the amended Hamiltonian $\tilde{\HCd}_{1}$.
In order to include the dynamics of the gauge fields into
the description of the dynamical system,
the generating function~(\ref{eq:gen-ex5}) must be amended accordingly to
also define the transformation rule~(\ref{eq:trans-conn-coeff}).
The set of canonical transformation rules then also yields the rules
for the conjugates of the gauge fields and the rule for
a second amended Hamiltonian $\tilde{\HCd}_{2}$.
In other words, the gauge fields are now treated as \emph{internal} fields
such that the system described by $\tilde{\HCd}_{2}$ is closed.
The ensuing set of canonical equations will turn out to establish a closed set
of coupled field equations, hence, no further gauge quantities need to be introduced.
This ``miracle,'' as an important and welcome surprise, will be the topic of the next section.
\section{Include the dynamics of the gauge fields\label{sec:example-spacetime-trans-curv}}
The extended generating function $\FCd_{2}^{\mu}$ from Eq.~(\ref{eq:gen-ex5})
will now be amended to define the transformation law~(\ref{eq:trans-conn-coeff}),
i.e., the canonical transformation which maps reference frame~$x$ to frame~$X$
\begin{align}
&\bar{\FCd}_{2}^{\mu}(x)=\FCd_{2}^{\mu}(x)
+\tilde{Q}\indices{_{\eta}^{\alpha\xi\beta}}(X)\pfrac{x^{\mu}}{X^{\beta}}\left|\pfrac{x}{X}\right|^{-1}
\label{eq:gen-conn-coeff}\\
&\quad\times\left(\gamma\indices{^{k}_{ij}}(x)\pfrac{X^{\eta}}{x^{k}}\pfrac{x^{i}}{X^{\alpha}}
\pfrac{x^{j}}{X^{\xi}}+\pfrac{X^{\eta}}{x^{k}}\ppfrac{x^{k}}{X^{\alpha}}{X^{\xi}}\right).\nonumber
\end{align}
The quantities $\tilde{Q}\indices{_{\eta}^{\alpha\xi\nu}}(X)=Q\indices{_{\eta}^{\alpha\xi\nu}}(X)\sqrt{-G}$
denote in this definition of an extended generating function of type $\FCd_{2}^{\mu}(x)$,
formally the canonical conjugates of the $\Gamma\indices{^{\eta}_{\alpha\xi}}(X)$ of the
transformed system and, hence, the dual quantities to the $X^{\nu}$-derivatives of the
$\Gamma\indices{^{\eta}_{\alpha\xi}}(X)$.
As the $\gamma\indices{^{\eta}_{\alpha\xi}}(x)$ stand for the gauge
coefficients of the original system, the quantities
$\tilde{q}\indices{_{\eta}^{\alpha\xi\nu}}(x)\equiv q\indices{_{\eta}^{\alpha\xi\nu}}(x)\sqrt{-g}$
denote, accordingly, the dual quantities of the $x^{\nu}$-derivatives of the gauge
coefficients $\gamma\indices{^{\eta}_{\alpha\xi}}(x)$ of the original system.
No prediction with respect to the physical meaning of $q\indices{_{\eta}^{\alpha\xi\nu}}$
and $Q\indices{_{\eta}^{\alpha\xi\nu}}$ is made, at this point.
Rather, their physical meaning will be determined in Sec.~\ref{sec:can-field-equations} by setting
up the canonical field equations of the final, locally form-invariant Hamiltonian.

The amended generating function~(\ref{eq:gen-conn-coeff}) entails the
following additional transformation rules
\begin{align*}
\delta_{\nu}^{\mu}\Gamma\indices{^{\eta}_{\alpha\xi}}
&=\pfrac{\bar{\FCd}_{2}^{\kappa}(x)}{\tilde{Q}\indices{_{\eta}^{\alpha\xi\nu}}}
\pfrac{X^{\mu}}{x^{\kappa}}\left|\pfrac{x}{X}\right|\\
&=\delta_{\nu}^{\lambda}\pfrac{x^{\kappa}}{X^{\lambda}}\pfrac{X^{\mu}}{x^{\kappa}}
\left(\gamma\indices{^{k}_{ij}}\pfrac{X^{\eta}}{x^{k}}\pfrac{x^{i}}{X^{\alpha}}\pfrac{x^{j}}{X^{\xi}}
+\pfrac{X^{\eta}}{x^{k}}\ppfrac{x^{k}}{X^{\alpha}}{X^{\xi}}\right)\\
&=\delta_{\nu}^{\mu}\left(\gamma\indices{^{k}_{ij}}
\pfrac{X^{\eta}}{x^{k}}\pfrac{x^{i}}{X^{\alpha}}\pfrac{x^{j}}{X^{\xi}}
+\pfrac{X^{\eta}}{x^{k}}\ppfrac{x^{k}}{X^{\alpha}}{X^{\xi}}\right)
\end{align*}
and
\begin{equation*}
\tilde{q}\indices{_{k}^{ij\mu}}=\pfrac{\bar{\FCd}_{2}^{\mu}}{\gamma\indices{^{k}_{ij}}}
=\tilde{Q}\indices{_{\eta}^{\alpha\xi\lambda}}\pfrac{X^{\eta}}{x^{k}}
\pfrac{x^{i}}{X^{\alpha}}\pfrac{x^{j}}{X^{\xi}}\pfrac{x^{\mu}}{X^{\lambda}}\left|\pfrac{x}{X}\right|^{-1},
\end{equation*}
hence, solved for $\tilde{Q}\indices{_{\eta}^{\alpha\xi\lambda}}$,
\begin{equation}\label{eq:tr-gen-spacetime}
\tilde{Q}\indices{_{\eta}^{\alpha\xi\lambda}}(X)=\tilde{q}\indices{_{n}^{mrs}}(x)\,\pfrac{x^{n}}{X^{\eta}}
\pfrac{X^{\alpha}}{x^{m}}\pfrac{X^{\xi}}{x^{r}}\pfrac{X^{\lambda}}{x^{s}}\left|\pfrac{x}{X}\right|.
\end{equation}
We observe that the \emph{inhomogeneous transformation rule} for the gauge
coefficients from Eq.~(\ref{eq:trans-conn-coeff}) is recovered.
Furthermore, the canonical conjugates of the gauge coefficients, introduced formally
in the generating function~(\ref{eq:gen-conn-coeff}), turn out to transform as tensor densities.

The transformation rule for the Hamiltonians is again obtained by taking
the divergence of $\bar{\FCd}_{2}^{\mu}$ from Eq.~(\ref{eq:gen-conn-coeff})
\begin{align}
&\left.\pfrac{\bar{\FCd}_{2}^{\mu}}{x^{\mu}}\right|_{\text{expl}}
=\left.\pfrac{\FCd_{2}^{\mu}}{x^{\mu}}\right|_{\text{expl}}
+\tilde{Q}\indices{_{\eta}^{\alpha\xi\beta}}\Gamma\indices{^{\eta}_{\alpha\xi}}
\underbrace{\pfrac{}{x^{\mu}}\left(\pfrac{x^{\mu}}{X^{\beta}}\left|\pfrac{x}{X}\right|^{-1}
\right)}_{\equiv0\,\,(\mathrm{Eq.}~(\ref{eq:H-deri-expr-2}))}\nonumber\\
&\mbox{}+\frac{\tilde{Q}\indices{_{\eta}^{\alpha\xi\beta}}}{\left|\pfrac{x}{X}\right|}
\left[\gamma\indices{^{k}_{ij}}\pfrac{}{X^{\beta}}\left(
\pfrac{X^{\eta}}{x^{k}}\pfrac{x^{i}}{X^{\alpha}}\pfrac{x^{j}}{X^{\xi}}\right)
+\pfrac{}{X^{\beta}}\left(\pfrac{X^{\eta}}{x^{k}}\ppfrac{x^{k}}{X^{\alpha}}{X^{\xi}}\right)\right].
\label{eq:ham-conn-coeff}
\end{align}
The first term on the right-hand side of Eq.~(\ref{eq:ham-conn-coeff}) is given by Eq.~(\ref{eq:trans-hg1}).
The transformation rule for the Hamiltonians shall be expressed
completely in terms of the dynamic variables:
All derivatives of the functions $x^{\mu}(X)$ and $X^{\mu}(x)$
in~(\ref{eq:ham-conn-coeff}) are to be expressed in terms of the original
and transformed gauge coefficients $\gamma\indices{^{\eta}_{\alpha\xi}}(x)$ and
$\Gamma\indices{^{\eta}_{\alpha\xi}}(X)$, and their conjugates,
$\tilde{q}\indices{_{\eta}^{\alpha\xi\mu}}$ and $\tilde{Q}\indices{_{\eta}^{\alpha\xi\mu}}$,
by making use of the respective canonical transformation rules~(\ref{eq:trans-conn-coeff}) and~(\ref{eq:tr-gen-spacetime}).

This calculation was worked out earlier~\cite{struckvasak15} and is
rewritten in Appendix~\ref{sec:app1} in a notation adapted to the actual context.
In summary, the coefficients in the transformation rule~(\ref{eq:ham-conn-coeff})
can completely and symmetrically be expressed in terms of the physical
fields of the original and of the transformed system as
\begin{align}
\left.\pfrac{\bar{\FCd}_{2}^{\mu}}{x^{\mu}}\right|_{\text{expl}}\!\!\!\!\!&
=\left(\tilde{P}^{\alpha\beta}A_{\xi}+\tilde{K}^{\alpha\lambda\beta}G_{\xi\lambda}
+\tilde{K}^{\lambda\alpha\beta}G_{\lambda\xi}\right)\Gamma\indices{^{\xi}_{\alpha\beta}}\left|\pfrac{X}{x}\right|\nonumber\\
&-\left(\tilde{p}^{\alpha\beta}a_{\xi}+\tilde{k}^{\alpha\lambda\beta}g_{\xi\lambda}
+\tilde{k}^{\lambda\alpha\beta}g_{\lambda\xi}\right)\gamma\indices{^{\xi}_{\alpha\beta}}\nonumber\\
&+\onehalf\tilde{Q}\indices{_{\eta}^{\alpha\xi\beta}}\left(
\pfrac{\Gamma\indices{^{\eta}_{\alpha\xi}}}{X^{\beta}}+\pfrac{\Gamma\indices{^{\eta}_{\alpha\beta}}}{X^{\xi}}
+\Gamma\indices{^{k}_{\alpha\beta}}\Gamma\indices{^{\eta}_{k\xi}}
-\Gamma\indices{^{k}_{\alpha\xi}}\Gamma\indices{^{\eta}_{k\beta}}\right)\left|\pfrac{X}{x}\right|\nonumber\\
&-\onehalf\tilde{q}\indices{_{\eta}^{\alpha\xi\beta}}\left(
\pfrac{\gamma\indices{^{\eta}_{\alpha\xi}}}{x^{\beta}}+\pfrac{\gamma\indices{^{\eta}_{\alpha\beta}}}{x^{\xi}}
+\gamma\indices{^{k}_{\alpha\beta}}\gamma\indices{^{\eta}_{k\xi}}
-\gamma\indices{^{k}_{\alpha\xi}}\gamma\indices{^{\eta}_{k\beta}}\right)\nonumber\\
&=\tilde{\HCd}_{2}^{\prime}\left|\pfrac{X}{x}\right|-\tilde{\HCd}_{2}.\label{eq:ham-conn-coeff1}
\end{align}
The terms on the right-hand side can be regarded as amendments to the given system
Hamiltonians, $\tilde{\HCd}_{0}=\HCd_{0}\sqrt{-g}$ and its transformed counterpart
$\tilde{\HCd}_{0}^{\prime}=\HCd_{0}^{\prime}\sqrt{-G}$, which promote the given globally
form-invariant Hamiltonians $\tilde{\HCd}_{0}$ and $\tilde{\HCd}_{0}^{\prime}$ to
\emph{locally} form-invariant Hamiltonians $\tilde{\HCd}_{2}$ and $\tilde{\HCd}_{2}^{\prime}$.
Accordingly, the amended Hamiltonians are form-invariant under the extended
canonical transformation generated by Eq.~(\ref{eq:gen-conn-coeff}).

Amending the Hamiltonian $\tilde{\HCd}_{1}$ from Eq.~(\ref{eq:gauge-ham2}) further on
according to~(\ref{eq:ham-conn-coeff1}) yields a second amended Hamiltonian $\tilde{\HCd}_{2}$
\begin{align}
\tilde{\HCd}_{2}&=\tilde{\HCd}_{0}+\tilde{\HCd}_{\mathrm{G}}\label{eq:H2}\\
\tilde{\HCd}_{\mathrm{G}}&=\left(\tilde{p}^{\alpha\beta}a_{\xi}+\tilde{k}^{\alpha\lambda\beta}g_{\xi\lambda}
+\tilde{k}^{\lambda\alpha\beta}g_{\lambda\xi}\right)\gamma\indices{^{\xi}_{\alpha\beta}}\label{eq:gauge-ham3}\\
&+\onehalf\tilde{q}\indices{_{\eta}^{\alpha\xi\beta}}\left(
\pfrac{\gamma\indices{^{\eta}_{\alpha\xi}}}{x^{\beta}}+\pfrac{\gamma\indices{^{\eta}_{\alpha\beta}}}{x^{\xi}}
+\gamma\indices{^{\eta}_{\tau\xi}}\gamma\indices{^{\tau}_{\alpha\beta}}
-\gamma\indices{^{\eta}_{\tau\beta}}\gamma\indices{^{\tau}_{\alpha\xi}}\right)\nonumber.
\end{align}
In the first stage of Sec.~\ref{sec:example-spacetime-trans},
the metric was rendered an internal system variable.
The gauge coefficients $\gamma\indices{^{\eta}_{\alpha\xi}}$
had to be introduced to restore form-invariance of the system.
In contrast, no further gauge fields had to be introduced in the actual second stage, where the gauge
coefficients $\gamma\indices{^{\eta}_{\alpha\xi}}$ are promoted to internal dynamic quantities.
Rather, the gauge fields $\gamma\indices{^{\xi}_{\alpha\beta}}$ now interact with themselves,
which induces the terms quadratic in $\gamma$ to finally yield the locally form-invariant Hamiltonian $\tilde{\HCd}_{2}$.

Observe that the gauge terms occurring in Eq.~(\ref{eq:H2})---hence the terms that must be added to the
given Lorentz-invariant system Hamiltonian $\tilde{\HCd}_{0}$---have exactly the same structure as those
of the SU$(N)$ (Yang-Mills) gauge theory, in the canonical formalism~\cite{StrRei12}
\begin{align*}
\HCd_{2}&=\HCd_{0}(\pi,\phi)+\mathrm{i}g_{\mathrm{_{YM}}}\left(\overline{\pi}_{K}^{\,\alpha}\,a_{KJ\alpha}\,\phi_{J}
-\overline{\phi}_{J}\,\overline{a}_{JK\alpha}\,\pi_{K}^{\alpha}\right)\nonumber\\
&\quad\mbox{}+\onehalf p_{JK}^{\alpha\beta}\left[\pfrac{a_{KJ\alpha}}{x^{\beta}}
\!+\!\pfrac{a_{KJ\beta}}{x^{\alpha}}\!+\!\mathrm{i}g_{\mathrm{_{YM}}}\left(a_{KI\beta}\,a_{IJ\alpha}-a_{KI\alpha}\,a_{IJ\beta}\right)\right].
\end{align*}
The set of fermionic fields $\phi_{J}$ for the Yang-Mills case corresponds to the vector field $a_{\xi}$ of Eq.~(\ref{eq:H2}),
whereas the hermitian $N\times N$ matrix of bosonic Yang-Mills $4$-vector gauge fields $a_{KJ\mu}$
now reappear anew as the connection coefficients $\gamma\indices{^{\eta}_{\alpha\xi}}$.
In contrast the gauge fields $a_{KJ\mu}$ of the SU$(N)$ theory, the connection coefficients are no tensors.
Yet, this is a \emph{necessary} property here: the non-tensorial quantities in~(\ref{eq:gauge-ham3}) will be shown in the next section
to exactly complement the non-tensorial terms in the initial action functional to render the final integrand a \emph{world scalar density}.
\section{Insert the amended Hamiltonian $\tilde{\HCd}_{2}$ into the action integral\label{sec:act-int}}
The derivative of the nontensorial dynamic quantity $\gamma\indices{^{\eta}_{\alpha\xi}}$
now additionally appears in the amended action integral (for the notation see Sec.~\ref{sec:gp})
\begin{align}
S_{4}&=\!\!\int_{\Omega}\!\bigg(\tilde{\pi}^{\beta}\pfrac{\phi}{x^{\beta}}
\!+\!\tilde{p}^{\,\alpha\beta}\pfrac{a_{\alpha}}{x^{\beta}}
\!+\!\tilde{k}^{\,\alpha\lambda\beta}\pfrac{g_{\alpha\lambda}}{x^{\beta}}
\!+\!\tilde{q}\indices{_{\eta}^{\alpha\xi\beta}}
\pfrac{\gamma\indices{^{\eta}_{\alpha\xi}}}{x^{\beta}}\!-\!\tilde{\HCd}_{2}\bigg)\d^{4}x,
\label{eq:action-integral}
\end{align}
as the affine connection $\gamma\indices{^{\eta}_{\alpha\xi}}$, in conjunction with their conjugates,
$\tilde{q}\indices{_{\eta}^{\alpha\xi\beta}}$, are now internal dynamic variables of the system.

Now it is obvious why the second amended $\tilde{\HCd}_{2}$ cannot represent a world scalar density by itself:
the terms in $\tilde{\HCd}_{2}$ must complement the non-tensorial terms in~(\ref{eq:action-integral}),
such that the total integrand is rendered a world scalar density.
Expressing $\HCd_{2}$ in Eq.~(\ref{eq:action-integral}) according to Eqs.~(\ref{eq:H2}) and~(\ref{eq:gauge-ham3})
in terms of $\tilde{\HCd}_{0}$ and $\tilde{\HCd}_{\mathrm{G}}$ yields:
\begin{align}
S_{4}&=\int_{\Omega}\Bigg[\tilde{\pi}^{\beta}\pfrac{\phi}{x^{\beta}}
+\tilde{p}^{\,\alpha\beta}\pfrac{a_{\alpha}}{x^{\beta}}
+\tilde{k}^{\,\alpha\lambda\beta}\pfrac{g_{\alpha\lambda}}{x^{\beta}}-\tilde{\HCd}_{0}\nonumber\\
&\qquad\,\,-\left(\tilde{p}^{\alpha\beta}a_{\xi}+\tilde{k}^{\alpha\lambda\beta}g_{\xi\lambda}
+\tilde{k}^{\lambda\alpha\beta}g_{\lambda\xi}\right)\gamma\indices{^{\xi}_{\alpha\beta}}\label{eq:action-integral2}\\
&-\onehalf\tilde{q}\indices{_{\eta}^{\alpha\xi\beta}}\Bigg(
\pfrac{\gamma\indices{^{\eta}_{\alpha\beta}}}{x^{\xi}}-\pfrac{\gamma\indices{^{\eta}_{\alpha\xi}}}{x^{\beta}}
+\gamma\indices{^{\eta}_{\tau\xi}}\gamma\indices{^{\tau}_{\alpha\beta}}
-\gamma\indices{^{\eta}_{\tau\beta}}\gamma\indices{^{\tau}_{\alpha\xi}}\Bigg)\Bigg]\d^{4}x.\nonumber
\end{align}
The terms linear in $\gamma\indices{^{\xi}_{\alpha\beta}}$ convert the the partial derivatives of $a_\alpha$
and $g_{\alpha\lambda}$ into covariant derivatives.
Remarkably, the terms proportional to $\tilde{q}\indices{_{\eta}^{\alpha\xi\beta}}$ sum up
to the Riemann-Cartan curvature tensor $R\indices{^{\eta}_{\alpha\xi\beta}}$:
\begin{equation}\label{eq:riemann-tensor}
R\indices{^{\eta}_{\alpha\xi\beta}}=\pfrac{\gamma\indices{^{\eta}_{\alpha\beta}}}{x^{\xi}}
-\pfrac{\gamma\indices{^{\eta}_{\alpha\xi}}}{x^{\beta}}
+\gamma\indices{^{\eta}_{\tau\xi}}\gamma\indices{^{\tau}_{\alpha\beta}}
-\gamma\indices{^{\eta}_{\tau\beta}}\gamma\indices{^{\tau}_{\alpha\xi}}.
\end{equation}
Hence, the Riemann-Cartan tensor is \emph{not} postulated at this point, but naturally emerges from the gauge procedure:
the gauge fields $\gamma\indices{^{\eta}_{\alpha\beta}}$---introduced in Eqs.~(\ref{eq:gauge-ham2})
and~(\ref{eq:gauge-ham1})---come about in the action $S_4$ exactly in the arrangement of Eq.~(\ref{eq:riemann-tensor}).
It is the total contraction of $R\indices{^{\eta}_{\alpha\xi\beta}}$ with the tensor density
$\tilde{q}\indices{_{\eta}^{\alpha\xi\beta}}$ which actually yields a world scalar density.
As the Riemann-Cartan tensor~(\ref{eq:riemann-tensor}) is skew-symmetric in its last index pair, $\xi$ and $\beta$,
only the skew-symmetric part in $\xi$ and $\beta$ of the---as yet undetermined---conjugate field
$\tilde{q}\indices{_{\eta}^{\alpha\xi\beta}}$ contributes to the action integral.
Therefore, $\tilde{q}\indices{_{\eta}^{\alpha\xi\beta}}$ can be assumed
to be skew-symmetric in its last index pair $\xi$ and $\beta$ as well
\begin{equation}\label{eq:q-skew}
\tilde{q}\indices{_{\eta}^{\alpha\xi\beta}}=-\tilde{q}\indices{_{\eta}^{\alpha\beta\xi}}.
\end{equation}
The action~(\ref{eq:action-integral2}) now directly reveals how the gauge procedure works.
To set up a generally form-invariant action integral on the basis of a Lorentz-invariant one,
the partial derivatives of the fields $a_{\mu}$ and $g_{\mu\nu}$ are amended by the
linear terms in the connection coefficients $\gamma$ to yield covariant derivatives.
In contrast, the partial derivative of the connection coefficient $\gamma$---which
cannot be converted into a covariant derivative due to its nontensorial character---is
amended by the quadratic terms in $\gamma$ to yield the Riemann-Cartan tensor.
In both cases, the gauge procedure provides tensor quantities in place of partial derivatives.
This result is derived here via the extended canonical transformation approach of field theories.

With semicolons denoting covariant derivatives and the Riemann-Cartan tensor~(\ref{eq:riemann-tensor})
constituted by the terms in the third line of Eq.~(\ref{eq:action-integral2}),
this action takes on the concise form:
\begin{align}
S_{4}&=\!\!\int_{\Omega}\!\Big(\tilde{\pi}^{\,\beta}\phi_{;\beta}
\!+\!\tilde{p}^{\,\alpha\beta}\,a_{\alpha;\beta}\!+\!\tilde{k}^{\,\alpha\lambda\beta}\,g_{\alpha\lambda;\beta}
\!-\!\onehalf\tilde{q}\indices{_{\eta}^{\alpha\xi\beta}}R\indices{^{\eta}_{\alpha\xi\beta}}\!-\!\tilde{\HCd}_{0}\Big)\d^{4}x.
\label{eq:action-integral4}
\end{align}
The given system Hamiltonian $\tilde{\HCd}_{0}$ is a scalar density by presupposition,
hence, the entire integrand consists of contracted tensor quantities.
It thus makes up a world scalar density, as required for a generally relativistic form-invariant action integral.
The action~(\ref{eq:action-integral4}) is \emph{not} postulated, but
emerges from the gauge principle, which here means to amend a given (globally) Lorentz-invariant
system Hamiltonian $\tilde{\HCd}_{0}$ in a way to render it diffeomorphism-invariant.
In other words, the curvature of spacetime, described by the Riemann-Cartan tensor $R\indices{^{\eta}_{\alpha\xi\beta}}$,
\emph{emerges} rather than being imposed by fiat.
\section{Add the ``free-field'' Hamiltonian\label{sec:free-field}}
As usual, the gauge formalism fixes the coupling of the gauge fields with
the fields described by the given system Hamiltonian $\HCd_{0}$ but does \emph{not}
provide the dynamics of the ``free'' gauge fields, i.e.\ their dynamics in the absence of any coupling.
If the respective gauge fields are considered dynamic (propagating) quantities, the obtained generally
form-invariant Hamiltonian $\tilde{\HCd}_{0}+\tilde{\HCd}_{\mathrm{G}}$ must be further
amended in order for the canonical equations to yield nonstatic solutions for the gauge fields.
Hence, the above form-invariant Hamiltonians~(\ref{eq:ham-conn-coeff1})
must be further amended by ``free field'' Hamiltonians $\tilde{\HCd}^{\prime}_{\mathrm{Gr}}(G,\tilde{K},\tilde{Q})$
and $\tilde{\HCd}_{\mathrm{Gr}}(g,\tilde{k},\tilde{q})$ that obey the transformation rule
\begin{equation}\label{eq:H-dyn}
\tilde{\HCd}^{\prime}_{\mathrm{Gr}}(G,\tilde{K},\tilde{Q})=\tilde{\HCd}_{\mathrm{Gr}}(g,\tilde{k},\tilde{q})\left|\pfrac{x}{X}\right|
\end{equation}
in order for the final extended Hamiltonians to describe the dynamics of the gauge fields and
to maintain the general invariance of the action integral~(\ref{eq:action-integral4}).

In conjunction with the precondition~(\ref{eq:global-trans}) of a globally, i.e., Lorentz-invariant Hamiltonian
$\tilde{\HCd}_{0}$, the final extended Hamiltonian $\tilde{\HCd}_{3}$, which is form-invariant under the
corresponding local transformation generated by~(\ref{eq:gen-conn-coeff}), now reads
\begin{align}
\tilde{\HCd}_{3}\big(\phi,\tilde{\pi},a,\tilde{p},g,\tilde{k},\gamma,\tilde{q}\big)&=
\tilde{\HCd}_{0}\big(\phi,\tilde{\pi},a,\tilde{p},g\big)\label{eq:eham-conn-coeff-inv}\\
&\quad\mbox{}+\tilde{\HCd}_{\mathrm{G}}\big(a,\tilde{p},g,\tilde{k},\gamma,\tilde{q}\big)
+\tilde{\HCd}_{\mathrm{Gr}}\big(g,\tilde{k},\tilde{q}\big)\nonumber
\end{align}
with $\tilde{\HCd}_{\mathrm{G}}$ given by Eq.~(\ref{eq:gauge-ham3}).
The correlation of the derivatives of the fields with their momenta will be identified in
Sec.~\ref{sec:can-field-equations} by setting up the respective canonical field equations.
\section{Summary of the gauge procedure\label{sec:gp}}
The starting point is a classical Hamiltonian system of a real or complex scalar field $\phi$ and a vector field $a_{\nu}$,
with the metric $g_{\mu\nu}$ assumed to be of Minkowski type, $g_{\mu\nu}\equiv\eta_{\mu\nu}=\mathrm{const}$.
The particular Hamiltonian $\HCd_{0}$ of this system needs not to be specified except for its property to be form-invariant
under global coordinate transformations $x\mapsto X$ with $\partial X^{\,\alpha}/\partial x^{\,\beta}=\mathrm{const.}$
\begin{equation}\label{eq:S_0}
S_{0}=\int_{\Omega}\left(\pi^{\,\beta}\pfrac{\phi}{x^{\,\beta}}
+p^{\,\alpha\beta}\pfrac{a_{\alpha}}{x^{\,\beta}}-\HCd_{0}(\pi,\phi,p,a,g\equiv\eta)\right)\d^{4}x.
\end{equation}
The globally form-invariant action $S_{0}$ is to be amended to yield a locally form-invariant
action that describes in addition to the scalar and vector fields the dynamics of the gauge quantities.
In the first step, the metric in Eq.~(\ref{eq:S_0}) is redefined as an arbitrary
spacetime-dependent function, $g_{\mu\nu}=g_{\mu\nu}(x)$.
Consequently, the invariant volume form is now given by $\sqrt{-g}\d^{4}x$.
In the following, the factor $\sqrt{-g}$ is absorbed into defining the canonical momenta as
tensor \emph{densities} rather than absolute tensors, hence $\tilde{\pi}^{\beta}=\pi^{\beta}\sqrt{-g}$
and $\tilde{p}^{\alpha\beta}=p^{\alpha\beta}\sqrt{-g}$.
Correspondingly, the scalar $\HCd_{0}$ in Eq.~(\ref{eq:S_0}) is converted into a \emph{scalar density}
$\tilde{\HCd}_{0}=\HCd_{0}\sqrt{-g}$.
Moreover, the conjugate momentum $\tilde{k}^{\,\mu\nu\lambda}$ as the dual
of the derivative of the metric must be introduced into the action functional
\begin{equation}\label{eq:S_1}
S_{1}\!=\!\int_{\Omega}\left(\tilde{\pi}^{\,\beta}\pfrac{\phi}{x^{\,\beta}}
\!+\!\tilde{p}^{\,\alpha\beta}\pfrac{a_{\alpha}}{x^{\,\beta}}\!+\!\tilde{k}^{\,\mu\nu\beta}\pfrac{g_{\mu\nu}}{x^{\,\beta}}
\!-\!\tilde{\HCd}_{0}(\tilde{\pi},\phi,\tilde{p},a,g)\right)\!\d^{4}x.
\end{equation}
The metric $g_{\mu\nu}$ is now formally treated as as an internal quantity in the action functional~(\ref{eq:S_1}) of the system.
As the partial derivatives of tensors do not transform as tensors, the integrand of~(\ref{eq:S_1}) is now no longer a scalar.
The scalar property of the integrand is restored by adding the gauge Hamiltonian
$\tilde{\HCd}_{\mathrm{G}}$ to the integrand
\begin{align}
S_{2}=\int_{\Omega}&\left(\tilde{\pi}^{\,\beta}\pfrac{\phi}{x^{\,\beta}}
+\tilde{p}^{\,\alpha\beta}\pfrac{a_{\alpha}}{x^{\,\beta}}+\tilde{k}^{\,\mu\nu\beta}\pfrac{g_{\mu\nu}}{x^{\,\beta}}
-\tilde{\HCd}_{0}(\tilde{\pi},\phi,\tilde{p},a,g)\right.\nonumber\\
&\left.\quad\mbox{}-\tilde{\HCd}_{\mathrm{G}}(\tilde{p},a,\tilde{k},g,\gamma)\vphantom{\pfrac{\phi}{x^{\,\beta}}}\right)\d^{4}x
\label{eq:S_2}
\end{align}
with
\begin{equation*}
\tilde{\HCd}_{\mathrm{G}}=\left(\tilde{p}^{\,\alpha\beta}a_{\xi}+\tilde{k}^{\,\alpha\lambda\beta}g_{\xi\lambda}
+\tilde{k}^{\,\lambda\alpha\beta}g_{\lambda\xi}\right)\gamma\indices{^{\xi}_{\alpha\beta}}.
\end{equation*}
This amounts to introducing the connection coefficients $\gamma\indices{^{\,\xi}_{\lambda\alpha}}$
as external gauge quantities and then promoting the \emph{partial} derivatives in~(\ref{eq:S_2})
to covariant derivatives.
The action~(\ref{eq:S_2}) is thus equivalently expressed as
\begin{equation}\label{eq:S_2a}
S_{2}=\!\int_{\Omega}\left(\tilde{\pi}^{\,\beta}\phi_{;\,\beta}
+\tilde{p}^{\,\alpha\beta}a_{\alpha;\,\beta}+\tilde{k}^{\,\mu\nu\beta}g_{\mu\nu;\,\beta}
-\tilde{\HCd}_{0}(\tilde{\pi},\phi,\tilde{p},a,g)\right)\!\d^{4}x.
\end{equation}
In the next step, the so far external gauge fields $\gamma\indices{^{\,\xi}_{\alpha\beta}}$
are treated as internal fields, which means that the description of their dynamics is to be
included in a further amended action functional.
This requires to define the canonical momentum $\tilde{q}\indices{_{\xi}^{\lambda\alpha\beta}}$
as the dual of the partial $x^{\,\beta}$-derivative of the gauge field $\gamma\indices{^{\,\xi}_{\lambda\alpha}}$
and to add the contraction of both terms to the action
\begin{align}
S_{3}=\int_{\Omega}&\left(\tilde{\pi}^{\,\beta}\phi_{;\,\beta}
+\tilde{p}^{\,\alpha\beta}a_{\alpha;\,\beta}+\tilde{k}^{\,\mu\nu\beta}g_{\mu\nu;\,\beta}
+\tilde{q}\indices{_{\xi}^{\lambda\alpha\beta}}\pfrac{\gamma\indices{^{\,\xi}_{\lambda\alpha}}}{x^{\,\beta}}\right.\nonumber\\
&\left.\quad\mbox{}-\tilde{\HCd}_{0}(\tilde{\pi},\phi,\tilde{p},a,g)
\vphantom{\pfrac{\gamma\indices{^{\,\xi}_{\lambda\alpha}}}{x^{\,\beta}}}\right)\d^{4}x.
\label{eq:S_3}
\end{align}
As the connection $\gamma\indices{^{\,\xi}_{\lambda\alpha}}$ is no tensor, neither is its partial derivative.
Again, the invariance property of the integrand~(\ref{eq:S_3}) is restored by supplementing
the appropriate term of the gauge formalism---hence the term quadratic in $\gamma$---to the integrand of Eq.~(\ref{eq:S_3})
\begin{align}
S_{4}=\int_{\Omega}&\left(\tilde{\pi}^{\,\beta}\phi_{;\,\beta}
+\tilde{p}^{\,\alpha\beta}a_{\alpha;\,\beta}+\tilde{k}^{\,\mu\nu\beta}g_{\mu\nu;\,\beta}
+\tilde{q}\indices{_{\xi}^{\lambda\alpha\beta}}\pfrac{\gamma\indices{^{\,\xi}_{\lambda \alpha}}}{x^{\,\beta}}\right.\nonumber\\
&\left.\quad\mbox{}-\tilde{q}\indices{_{\xi}^{\lambda\alpha\beta}}\gamma\indices{^{\tau}_{\lambda\beta}}\gamma\indices{^{\,\xi}_{\tau\alpha}}
-\tilde{\HCd}_{0}(\tilde{\pi},\phi,\tilde{p},a,g)\vphantom{\pfrac{\gamma\indices{^{\,\xi}_{\lambda\alpha}}}{x^{\,\beta}}}\right)\d^{4}x.
\label{eq:S_4}
\end{align}
By virtue of the skew-symmetry of $\tilde{q}\indices{_{\xi}^{\lambda\alpha\beta}}$ in its last index
pair---which follows from the gauge formalism---the terms
proportional to $\tilde{q}$ can be merged to yield the Riemann curvature tensor~(\ref{eq:riemann-tensor})
\begin{align}
S_{4}=\int_{\Omega}&\left(\tilde{\pi}^{\,\beta}\phi_{;\,\beta}
+\tilde{p}^{\,\alpha\beta}a_{\alpha;\,\beta}+\tilde{k}^{\,\mu\nu\beta}g_{\mu\nu;\,\beta}
-\onehalf\tilde{q}\indices{_{\xi}^{\lambda\alpha\beta}}R\indices{^{\,\xi}_{\lambda\alpha\beta}}\right.\nonumber\\
&\left.\quad\mbox{}-\tilde{\HCd}_{0}(\tilde{\pi},\phi,\tilde{p},a,g)\right)\d^{4}x.
\label{eq:S_4a}
\end{align}
As a result of the gauge procedure, the action is obtained completely in terms
of tensor quantities and thus represents a world scalar density.
We observe that the gauge term for the partial derivative of the connection coefficients
in Eq.~(\ref{eq:S_4}) is given by the term quadratic in $\gamma$.
This is a consequence of the fact that the coefficients in the transformation rule
for the Hamiltonian from Eq.~(\ref{eq:ham-conn-coeff2}) can completely be
expressed in term of the connection coefficients.
Therefore, no new gauge fields were needed---which is the reason why
the procedure to promote the globally covariant coordinate description to a
diffeomorphism-covariant one truncates here and does \emph{not} produce an infinite hierarchy
of gauge fields with their pertaining transformation conditions.

In the last step, a Hamiltonian $\tilde{\HCd}_{\mathrm{Gr}}$ of the ``free'' momenta $\tilde{k}$ and $\tilde{q}$ must
be introduced ``by hand'' in order for the corresponding fields $g$ and $\gamma$ to be dynamic
\begin{align}
S_{5}=\int_{\Omega}&\left(\tilde{\pi}^{\,\beta}\phi_{;\,\beta}
+\tilde{p}^{\,\alpha\beta}\,a_{\alpha;\,\beta}+\tilde{k}^{\,\alpha\lambda\beta}\,g_{\alpha\lambda;\,\beta}
-\onehalf\tilde{q}\indices{_{\xi}^{\lambda\alpha\beta}}R\indices{^{\,\xi}_{\lambda\alpha\beta}}\right.\nonumber\\
&\left.\quad\mbox{}-\tilde{\HCd}_{0}(\tilde{\pi},\phi,\tilde{p},a,g)-\tilde{\HCd}_{\mathrm{Gr}}\big(\tilde{k},g,\tilde{q}\big)\right)\d^{4}x.
\label{eq:S_final}
\end{align}
As a consequence, different choices for $\tilde{\HCd}_{\mathrm{Gr}}$ define different gravitational theories.
One option for defining $\tilde{\HCd}_{\mathrm{Gr}}$ will be discussed in Sec.~\ref{sec:sample}.
The canonical field equations, summarized in Sec.~\ref{sec:summary-feq},
follow from the action principle $\delta S_{5}\stackrel{!}{=}0$.
\section{Canonical~field~equations:\newline arbitrary $\tilde{\HCd}_{\mathrm{Gr}}$\label{sec:can-field-equations}}
\subsection{Field equations for $\phi$ and $\tilde{\pi}^{\nu}$}
As the locally form-invariant extended Hamiltonian~(\ref{eq:eham-conn-coeff-inv}) does not
contain additional terms involving $\phi$ and $\tilde{\pi}^{\nu}$, the dynamics of these fields
is determined by the globally form-invariant Hamiltonian $\HCd$ only.
The respective field equations are
\begin{align*}
\pfrac{\phi}{x^{\mu}}&=\hphantom{-}\pfrac{\tilde{\HCd}_{3}}{\tilde{\pi}^{\mu}}=\hphantom{-}\pfrac{\tilde{\HCd}_{0}}{\tilde{\pi}^{\mu}}\\
\pfrac{\tilde{\pi}^{\alpha}}{x^{\alpha}}&=-\pfrac{\tilde{\HCd}_{3}}{\phi}=-\pfrac{\tilde{\HCd}_{0}}{\phi}.
\end{align*}
For a vector density $\tilde{\pi}^{\mu}=\pi^{\mu}\sqrt{-g}$, the ordinary divergence
$\partial\tilde{\pi}^{\alpha}/\partial x^{\alpha}$ of  can be expressed in terms of the covariant divergence as
\begin{equation*}
\tilde{\pi}\indices{^{\alpha}_{;\,\alpha}}=\pfrac{\tilde{\pi}^{\alpha}}{x^{\alpha}}+\tilde{\pi}^{\,\beta}
\gamma\indices{^{\alpha}_{\beta\alpha}}-\tilde{\pi}^{\alpha}\gamma\indices{^{\beta}_{\beta\alpha}}
=\pfrac{\tilde{\pi}^{\alpha}}{x^{\alpha}}+2\tilde{\pi}^{\,\beta}s\indices{^{\,\alpha}_{\beta\alpha}},
\end{equation*}
wherein $s\indices{^{\alpha}_{\beta\alpha}}$ denotes the contraction of the torsion tensor
\begin{equation}\label{eq:def-torsion}
s\indices{^{\,\xi}_{\beta\alpha}}=\onehalf\left(\gamma\indices{^{\,\xi}_{\beta\alpha}}-\gamma\indices{^{\,\xi}_{\alpha\beta}}\right).
\end{equation}
Thus, for connection coefficients that are symmetric in their lower index pair,
the torsion tensor vanishes identically.
The second field equation follows as the tensor equation
\begin{equation*}
\tilde{\pi}\indices{^{\alpha}_{;\,\alpha}}=-\pfrac{\tilde{\HCd}_{0}}{\phi}+2\tilde{\pi}^{\,\beta}s\indices{^{\,\alpha}_{\beta\alpha}}.
\end{equation*}
Both field equations thus emerge as tensor equations.
\subsection{Field equations for $a_{\mu}$ and $\tilde{p}^{\,\mu\nu}$\label{sec:feq-a-p}}
Due to the coupling term $\tilde{p}^{\alpha\beta}\,a_{\eta}\,\gamma\indices{^{\eta}_{\alpha\beta}}$
in the extended Hamiltonian~(\ref{eq:eham-conn-coeff-inv})
the field equations for $a_{\mu}$ and $\tilde{p}^{\,\mu\nu}$ acquire an additional term.
The respective field equations are
\begin{equation}\begin{split}
\pfrac{a_{\nu}}{x^{\mu}}&=\hphantom{-}\pfrac{\tilde{\HCd}_{3}}{\tilde{p}^{\nu\mu}}
=\hphantom{-}\pfrac{\tilde{\HCd}_{0}}{\tilde{p}^{\nu\mu}}+a_{\xi}\,\gamma\indices{^{\xi}_{\nu\mu}}\\
\pfrac{\tilde{p}^{\,\nu\beta}}{x^{\beta}}&=-\pfrac{\tilde{\HCd}_{3}}{a_{\nu}}
=-\pfrac{\tilde{\HCd}_{0}}{a_{\nu}}-\tilde{p}^{\alpha\beta}\gamma\indices{^{\nu}_{\alpha\beta}}.
\end{split}\label{eq:feq-vecfields0}
\end{equation}
The partial derivatives of the fields and the terms proportional to the
affine connections $\gamma\indices{^{\nu}_{\alpha\beta}}$ can be
combined to yield covariant derivatives
\begin{align*}
a_{\nu;\,\mu}&=\pfrac{a_{\nu}}{x^{\mu}}-a_{\eta}\gamma\indices{^{\eta}_{\nu\mu}}\\
\tilde{p}\indices{^{\nu\beta}_{;\,\beta}}&=\pfrac{\tilde{p}^{\,\nu\beta}}{x^{\beta}}
+\tilde{p}^{\,\alpha\beta}\gamma\indices{^{\nu}_{\alpha\beta}}+\tilde{p}^{\,\nu\alpha}
\gamma\indices{^{\,\beta}_{\alpha\beta}}-\tilde{p}^{\,\nu\beta}\gamma\indices{^{\,\alpha}_{\alpha\beta}},
\end{align*}
which yields the tensor equations
\begin{equation}\label{eq:feq-vecfields}
a_{\nu;\,\mu}=\pfrac{\tilde{\HCd}_{0}}{\tilde{p}^{\nu\mu}},\qquad
\tilde{p}\indices{^{\nu\beta}_{;\,\beta}}=-\pfrac{\tilde{\HCd}_{0}}{a_{\nu}}
+2\tilde{p}^{\,\nu\beta}s\indices{^{\,\alpha}_{\beta\alpha}}.
\end{equation}
The coupling term $\tilde{p}^{\alpha\beta}a_{\xi}\,\gamma\indices{^{\xi}_{\alpha\beta}}$
in the extended Hamiltonian $\tilde{\HCd}_{3}$ thus converts the nontensor equations
for $a_{\mu}$ and $\tilde{p}^{\,\mu\nu}$ which emerge from the system's Hamiltonian $\tilde{\HCd}_{0}$
into tensor equations which hold in any reference frame.
\subsection{Field equations for $g_{\alpha\beta}$ and $\tilde{k}\indices{^{\alpha\beta\mu}}$\label{sec:feq:g-k}}
The canonical equation for the metric $g_{\alpha\beta}$ is
\begin{equation}\label{eq:nonmetricity0}
\pfrac{g_{\alpha\lambda}}{x^{\beta}}=\pfrac{\tilde{\HCd}_{3}}{\tilde{k}\indices{^{\alpha\lambda\beta}}}
=g_{\kappa\lambda}\gamma\indices{^{\kappa}_{\alpha\beta}}+g_{\alpha\kappa}\gamma\indices{^{\kappa}_{\lambda\beta}}
+\pfrac{\tilde{\HCd}_{\mathrm{Gr}}}{\tilde{k}\indices{^{\alpha\lambda\beta}}},
\end{equation}
hence
\begin{equation}\label{eq:nonmetricity}
g_{\alpha\lambda;\,\beta}=\pfrac{g_{\alpha\lambda}}{x^{\beta}}
-g_{\lambda\kappa}\gamma\indices{^{\kappa}_{\alpha\beta}}
-g_{\alpha\kappa}\gamma\indices{^{\kappa}_{\lambda\beta}}
=\pfrac{\tilde{\HCd}_{\mathrm{Gr}}}{\tilde{k}\indices{^{\alpha\lambda\beta}}}.
\end{equation}
The field equation thus means that the covariant derivative of the
metric is the dual of the \emph{nonmetricity tensor}---which describes
the length change of a vector under parallel transport.

The canonical equation for the conjugate of the metric follows as
\begin{equation}\label{eq:k-div}
\pfrac{\tilde{k}^{\xi\lambda\beta}}{x^{\beta}}=-\pfrac{\tilde{\HCd}_{3}}{g_{\xi\lambda}}
=-\tilde{k}^{\alpha\lambda\beta}\gamma\indices{^{\xi}_{\alpha\beta}}
-\tilde{k}^{\xi\alpha\beta}\gamma\indices{^{\lambda}_{\alpha\beta}}-\pfrac{\tilde{\HCd}_{0}}{g_{\xi\lambda}}
-\pfrac{\tilde{\HCd}_{\mathrm{Gr}}}{g_{\xi\lambda}},
\end{equation}
hence
\begin{equation}\label{eq:em-tensor-den}
\tilde{k}\indices{^{\,\xi\lambda\beta}_{;\,\beta}}=-\pfrac{\tilde{\HCd}_{0}}{g_{\xi\lambda}}
-\pfrac{\tilde{\HCd}_{\mathrm{Gr}}}{g_{\xi\lambda}}+2\tilde{k}^{\,\xi\lambda\beta}s\indices{^{\,\alpha}_{\beta\alpha}},
\qquad\tilde{k}\indices{^{\xi\lambda\beta}}=\tilde{k}\indices{^{\lambda\xi\beta}}.
\end{equation}
The coupling terms $\tilde{k}^{\alpha\lambda\beta}g_{\xi\lambda}\gamma\indices{^{\xi}_{\alpha\beta}}$
and $\tilde{k}^{\lambda\alpha\beta}g_{\lambda\xi}\gamma\indices{^{\xi}_{\alpha\beta}}$ in the
gauge-invariant extended Hamiltonian $\tilde{\HCd}_{3}$ from Eq.~(\ref{eq:eham-conn-coeff-inv})
thus convert the nontensor equations for $g_{\alpha\lambda}$ and $\tilde{k}^{\xi\lambda\beta}$ into tensor equations.
As $g_{\xi\lambda}$ is symmetric, $\tilde{k}^{\,\xi\lambda\beta}$ is induced to be symmetric in its first index pair, $\xi,\lambda$.

The given Lorentz-invariant system Hamiltonian $\tilde{\HCd}_{0}$ describes the dynamics in a static spacetime background.
For this reason, $\tilde{\HCd}_{0}$ is supposed not to depend on the conjugate of the metric, $\tilde{k}^{\,\xi\alpha\beta}$.
The derivative of the Hamiltonian density $\tilde{\HCd}_{0}$ with respect to
the metric $g_{\xi\lambda}$ then represents the symmetric energy-momentum tensor density
$\tilde{T}^{\lambda\xi}$ of $\tilde{\HCd}_{0}$
\begin{equation}\label{eq:em-tensor}
\tilde{T}\indices{^{\lambda\xi}}=2\pfrac{\tilde{\HCd}_{0}}{g_{\xi\lambda}}.
\end{equation}
Thus, $\tilde{T}\indices{^{\lambda\xi}}$ does not describe the energy-momentum
contributed by a gravitational field and a dynamic spacetime.
\subsection{Field equations for $\gamma\indices{^{\eta}_{\alpha\beta}}$ and $\tilde{q}\indices{_{\eta}^{\alpha\beta\nu}}$\label{sec:feq-gamma-q}}
The canonical equation that provides the correlation of the $x^{\beta}$-derivative of the $\gamma\indices{^{\eta}_{\alpha\xi}}$
with their duals, $\tilde{q}\indices{_{\eta}^{\alpha\xi\beta}}$, follows as
\begin{align}
\onehalf\left(\pfrac{\gamma\indices{^{\eta}_{\alpha\xi}}}{x^{\beta}}-\pfrac{\gamma\indices{^{\eta}_{\alpha\beta}}}{x^{\xi}}\right)
&=\pfrac{\tilde{\HCd}_{3}}{\tilde{q}\indices{_{\eta}^{\alpha\xi\beta}}}
=\pfrac{\tilde{\HCd}_{\mathrm{Gr}}}{\tilde{q}\indices{_{\eta}^{\alpha\xi\beta}}}
+\pfrac{\tilde{\HCd}_{\mathrm{G}}}{\tilde{q}\indices{_{\eta}^{\alpha\xi\beta}}}\nonumber\\
&=\pfrac{\tilde{\HCd}_{\mathrm{Gr}}}{\tilde{q}\indices{_{\eta}^{\alpha\xi\beta}}}
+\onehalf\left(\gamma\indices{^{\eta}_{\tau\xi}}\gamma\indices{^{\tau}_{\alpha\beta}}
-\gamma\indices{^{\eta}_{\tau\beta}}\gamma\indices{^{\tau}_{\alpha\xi}}\right),
\label{eq:gamma-deri}
\end{align}
since after inserting the gauge Hamiltonian~(\ref{eq:gauge-ham3}) into the action~(\ref{eq:action-integral2}),
only the skew-symmetric part of the kinetic term enters into the canonical equation in question.
Solved for $\partial\tilde{\HCd}_{\mathrm{Gr}}/\partial\tilde{q}\indices{_{\eta}^{\alpha\xi\beta}}$, one finds
\begin{align}
2\pfrac{\tilde{\HCd}_{\mathrm{Gr}}}{\tilde{q}\indices{_{\eta}^{\alpha\xi\beta}}}
&=\pfrac{\gamma\indices{^{\eta}_{\alpha\xi}}}{x^{\beta}}
-\pfrac{\gamma\indices{^{\eta}_{\alpha\beta}}}{x^{\xi}}
+\gamma\indices{^{\eta}_{\tau\beta}}\gamma\indices{^{\tau}_{\alpha\xi}}
-\gamma\indices{^{\eta}_{\tau\xi}}\gamma\indices{^{\tau}_{\alpha\beta}}\nonumber\\
&=-R\indices{^{\,\eta}_{\alpha\xi\beta}}.
\label{eq:riemann-tensor2}
\end{align}
On the right-hand side, the connection coefficients $\gamma\indices{^{\,\eta}_{\alpha\beta}}$
and their derivatives sum up to the combination which represents the Riemann curvature tensor~(\ref{eq:riemann-tensor}).
The field equation~(\ref{eq:riemann-tensor2}) thus states that the Riemann tensor vanishes identically
everywhere---and thus any curvature of spacetime---if there is no ``free-field'' Hamiltonian
$\tilde{\HCd}_{\mathrm{Gr}}$.
Therefore, it must then be added ``by hand'' to the \emph{derived} gauge Hamiltonian
$\tilde{\HCd}_{\mathrm{G}}$ in order to allow for a consistent spacetime dynamics.

The divergence of $\tilde{q}\indices{_{\xi}^{\alpha\beta\lambda}}$ is given by the derivative of the gauge Hamiltonian
$\tilde{\HCd}_{\mathrm{G}}$ from Eq.~(\ref{eq:gauge-ham3}) with respect to the
$\gamma\indices{^{\xi}_{\alpha\beta}}$
\begin{equation*}
\pfrac{\tilde{q}\indices{_{\xi}^{\alpha\beta\lambda}}}{x^{\lambda}}
=-\pfrac{\tilde{\HCd}_{3}}{\gamma\indices{^{\xi}_{\alpha\beta}}}
=-\pfrac{\tilde{\HCd}_{\mathrm{G}}}{\gamma\indices{^{\xi}_{\alpha\beta}}}.
\end{equation*}
This equation does not depend on the particular choice of $\tilde{\HCd}_{\mathrm{Gr}}$
as the latter is supposed to not depend on the gauge fields $\gamma\indices{^{\xi}_{\alpha\beta}}$.
With the gauge Hamiltonian from Eq.~(\ref{eq:gauge-ham3}), we find
\begin{align}
\pfrac{\tilde{q}\indices{_{\xi}^{\alpha\beta\lambda}}}{x^{\lambda}}
&=-\tilde{p}^{\,\alpha\beta}a_{\xi}-2\tilde{k}^{\,\lambda\alpha\beta}g_{\lambda\xi}
+\tilde{q}\indices{_{\eta}^{\alpha\beta\lambda}}\gamma\indices{^{\eta}_{\xi\lambda}}
+\tilde{q}\indices{_{\xi}^{\eta\lambda\beta}}\gamma\indices{^{\alpha}_{\eta\lambda}}.
\label{eq:feq-conn-coeff}
\end{align}
In order to express Eq.~(\ref{eq:feq-conn-coeff}) manifestly as a tensor equation, we write the
covariant divergence of the tensor density $\tilde{q}\indices{_{\xi}^{\alpha\beta\lambda}}$
\begin{align*}
\tilde{q}\indices{_{\xi}^{\alpha\beta\lambda}_{;\,\lambda}}
=\pfrac{\tilde{q}\indices{_{\xi}^{\alpha\beta\lambda}}}{x^{\lambda}}
&-\tilde{q}\indices{_{\eta}^{\alpha\beta\lambda}}\gamma\indices{^{\eta}_{\xi\lambda}}
+\tilde{q}\indices{_{\xi}^{\eta\beta\lambda}}\gamma\indices{^{\alpha}_{\eta\lambda}}\\
&\mbox{}+\tilde{q}\indices{_{\xi}^{\alpha\eta\lambda}}\gamma\indices{^{\beta}_{\eta\lambda}}
+\tilde{q}\indices{_{\xi}^{\alpha\beta\eta}}\gamma\indices{^{\lambda}_{\eta\lambda}}
-\tilde{q}\indices{_{\xi}^{\alpha\beta\lambda}}\gamma\indices{^{\eta}_{\eta\lambda}}.
\end{align*}
As $\tilde{q}\indices{_{\xi}^{\alpha\eta\lambda}}$ is skew-symmetric in $\eta$ and $\lambda$,
the first term in the second line can be expressed as well in terms of the torsion tensor
\begin{equation*}
\tilde{q}\indices{_{\xi}^{\alpha\beta\lambda}_{;\,\lambda}}
\!=\!\pfrac{\tilde{q}\indices{_{\xi}^{\alpha\beta\lambda}}}{x^{\lambda}}
-\tilde{q}\indices{_{\eta}^{\alpha\beta\lambda}}\gamma\indices{^{\eta}_{\xi\lambda}}
-\tilde{q}\indices{_{\xi}^{\eta\lambda\beta}}\gamma\indices{^{\alpha}_{\eta\lambda}}
+\tilde{q}\indices{_{\xi}^{\alpha\eta\lambda}}s\indices{^{\beta}_{\eta\lambda}}
+2\tilde{q}\indices{_{\xi}^{\alpha\beta\eta}}s\indices{^{\lambda}_{\eta\lambda}}
\end{equation*}
The field equation~(\ref{eq:feq-conn-coeff}) thus actually represents the tensor density equation
\begin{equation}\label{eq:feq-conn-coeff3}
\tilde{q}\indices{_{\xi}^{\alpha\beta\lambda}_{;\lambda}}
+\tilde{p}^{\,\alpha\beta}a_{\xi}+2\tilde{k}^{\,\lambda\alpha\beta}g_{\lambda\xi}
-\tilde{q}\indices{_{\xi}^{\alpha\eta\lambda}}s\indices{^{\,\beta}_{\eta\lambda}}
+2\tilde{q}\indices{_{\xi}^{\alpha\eta\beta}}s\indices{^{\lambda}_{\eta\lambda}}=0.
\end{equation}
We thus found that \emph{all} field equations emerging from $\HCd_{3}$ are
tensor equations, hence, their forms are the same in any reference frame.
\subsection{Summary of the coupled set of field equations\label{sec:summary-feq}}
With the abbreviations~(\ref{eq:riemann-tensor}) and~(\ref{eq:def-torsion})
for particular combinations of the gauge fields $\gamma\indices{^{\,\eta}_{\xi\lambda}}$
and their partial derivatives, the complete set of coupled field equations is summarized as
\begin{align}
\phi_{;\,\mu}&=\pfrac{\tilde{\HCd}_{0}}{\tilde{\pi}^{\mu}},&
\tilde{\pi}\indices{^{\,\beta}_{;\,\beta}}&=-\pfrac{\tilde{\HCd}_{0}}{\phi}
+2\tilde{\pi}^{\,\beta}s\indices{^{\,\alpha}_{\beta\alpha}}\nonumber\\
a_{\nu;\,\mu}&=\pfrac{\tilde{\HCd}_{0}}{\tilde{p}^{\nu\mu}},&
\tilde{p}\indices{^{\nu\beta}_{;\,\beta}}&=-\pfrac{\tilde{\HCd}_{0}}{a_{\nu}}
+2\tilde{p}^{\,\nu\beta}s\indices{^{\,\alpha}_{\beta\alpha}}\nonumber\\
g_{\xi\lambda;\,\mu}&=\pfrac{\tilde{\HCd}_{\mathrm{Gr}}}{\tilde{k}\indices{^{\xi\lambda\mu}}},&
\tilde{k}\indices{^{\,\xi\lambda\beta}_{;\,\beta}}&=-\pfrac{\left(\tilde{\HCd}_{0}
+\tilde{\HCd}_{\mathrm{Gr}}\right)}{g_{\xi\lambda}}+2\tilde{k}^{\,\xi\lambda\beta}s\indices{^{\,\alpha}_{\beta\alpha}}\nonumber\\
-\frac{R\indices{^{\eta}_{\xi\lambda\mu}}}{2}&=\pfrac{\tilde{\HCd}_{\mathrm{Gr}}}{\tilde{q}\indices{_{\eta}^{\xi\lambda\mu}}},&
\tilde{q}\indices{_{\eta}^{\xi\lambda\beta}_{;\,\beta}}&=
-\tilde{p}^{\,\xi\lambda}a_{\eta}-2\tilde{k}^{\,\beta\xi\lambda}g_{\beta\eta}\nonumber\\
&&&\quad\mbox{}+\tilde{q}\indices{_{\eta}^{\xi\beta\alpha}}s\indices{^{\,\lambda}_{\beta\alpha}}
+2\tilde{q}\indices{_{\eta}^{\xi\lambda\beta}}s\indices{^{\alpha}_{\beta\alpha}}
\label{eq:feqs-Hdyn-1}
\end{align}
The canonical equations with respect to the momenta in the left column of Eqs.~(\ref{eq:feqs-Hdyn-1}) show that for a
dynamic spacetime the \emph{covariant} derivatives of the fields actually represent the canonical conjugates of the momenta.
The surprising fact in the last line of Eqs.~(\ref{eq:feqs-Hdyn-1}) is that the
na\"{\i}vely expected covariant derivatives of the gauge fields, i.e., the covariant
derivatives of the connection coefficients $\gamma\indices{^{\eta}_{\xi\lambda}}$---which
do not exist---are replaced by the Riemann tensor as a result of the canonical gauge procedure---and
thus again establish a tensor equation, as is required for a generally covariant theory.

Only with $\tilde{\HCd}_{\mathrm{Gr}}$ given, the entire set of eight canonical field equations
for the fields $\phi,a_{\nu},g_{\xi\lambda}$, and $\gamma\indices{^{\eta}_{\xi\lambda}}$
and their respective conjugates $\tilde{\pi}^{\mu},\tilde{p}^{\nu\mu},\tilde{k}^{\xi\lambda\mu}$,
and $\tilde{q}\indices{_{\eta}^{\xi\lambda\mu}}$ is closed and can then
be integrated to yield the combined dynamics of fields and spacetime geometry.
As $\tilde{\HCd}_{\mathrm{Gr}}$ does \emph{not} emerge from the gauge formalism,
it must be chosen on the basis of physical reasoning.
The field equations~(\ref{eq:feqs-Hdyn-1}) thus depend on this choice.
A particular choice of $\tilde{\HCd}_{\mathrm{Gr}}$ will be discussed in Sec.~\ref{sec:sample}.
\subsection{Consistency relation\label{sec:consistency}}
Similar to U$(1)$ and SU$(N)$ gauge theories, the set of field equations brings about a \emph{consistency condition},
namely an on-shell integrability and energy-momentum balance relation.
Differentiating the canonical equation~(\ref{eq:feq-conn-coeff}) with respect to $x^{\,\beta}$, the left-hand side
vanishes due to the skew-symmetry of $\tilde{q}\indices{_{\xi}^{\alpha\beta\lambda}}$
in its last index pair, as stated in Eq.~(\ref{eq:q-skew}).
Accordingly, the right-hand side of~(\ref{eq:feq-conn-coeff}) yields the condition
\begin{equation}\label{eq:consistency0}
\pfrac{}{x^{\,\beta}}\left(
\tilde{p}^{\,\alpha\beta}a_{\xi}+2\tilde{k}^{\,\lambda\alpha\beta}g_{\lambda\xi}
+\tilde{q}\indices{_{\eta}^{\alpha\lambda\beta}}\gamma\indices{^{\eta}_{\xi\lambda}}
-\tilde{q}\indices{_{\xi}^{\eta\lambda\beta}}\gamma\indices{^{\alpha}_{\eta\lambda}}\right)=0.
\end{equation}
The partial derivative representations of the canonical field equations
(\ref{eq:feq-vecfields0}), (\ref{eq:nonmetricity0}), (\ref{eq:k-div}), (\ref{eq:gamma-deri}),
and (\ref{eq:feq-conn-coeff}) from Secs.~\ref{sec:feq-a-p},
\ref{sec:feq:g-k}, and~\ref{sec:feq-gamma-q} can now be inserted to yield the consistency condition (see Appendix~\ref{sec:app2})
\begin{align}
&\quad\;2\tilde{k}^{\lambda\alpha\beta}\pfrac{\tilde{\HCd}_{\mathrm{Gr}}}{\tilde{k}^{\lambda\xi\beta}}
-2g_{\xi\beta}\pfrac{\tilde{\HCd}_{\mathrm{Gr}}}{g_{\alpha\beta}}
+\tilde{q}\indices{_{\tau}^{\alpha\beta\lambda}}\pfrac{\tilde{\HCd}_{\mathrm{Gr}}}{\tilde{q}\indices{_{\tau}^{\xi\beta\lambda}}}
-\tilde{q}\indices{_{\xi}^{\tau\beta\lambda}}\pfrac{\tilde{\HCd}_{\mathrm{Gr}}}{\tilde{q}\indices{_{\alpha}^{\tau\beta\lambda}}}\nonumber\\
&=a_{\xi}\pfrac{\tilde{\HCd}_{0}}{a_{\alpha}}-\tilde{p}^{\,\alpha\beta}\pfrac{\tilde{\HCd}_{0}}{\tilde{p}^{\,\xi\beta}}
+2g_{\xi\beta}\pfrac{\tilde{\HCd}_{0}}{g_{\alpha\beta}}\label{eq:consistency}.
\end{align}
This is a second rank tensor equation involving both the Hamiltonian $\tilde{\HCd}_{\mathrm{Gr}}$ describing the
dynamics of the free gravitational field and the Hamiltonian $\tilde{\HCd}_{0}$ of the matter field.
It follows from the closed set of field equations~(\ref{eq:feqs-Hdyn-1}), hence does not constitute
an independent equation, as outlined in Appendix~\ref{sec:cov-div}.
In conjunction with Eq.~(\ref{eq:riemann-tensor2}), it relates the source terms
of the right-hand side to the Riemann tensor terms on the left-hand side.
Note that Eq.~(\ref{eq:consistency}) holds as well for the cases of nonvanishing
torsion ($s\indices{^{\,\beta}_{\eta\lambda}}\neq0$) and nonmetricity ($g_{\xi\beta\,;\,\mu}\neq0$).
Equation~(\ref{eq:consistency}) represents a generic Einstein equation that holds for
any given system of scalar and vector fields described by $\tilde{\HCd}_{0}$ and the particular
model for the dynamics of the free gravitational fields, as described by $\tilde{\HCd}_{\mathrm{Gr}}$.

At this point, it would be interesting for the reader to find out whether
our theory makes observational predictions similar to general relativity.
As general relativity predicts successfully observations made in the
solar system (weak field), is the gauge theory given by Eqs.~(\ref{eq:feqs-Hdyn-1})
and~(\ref{eq:consistency}) capable of giving something like general relativity
plus small corrections?
We will address these issues briefly in the following section.
\subsection{Sample $\tilde{\HCd}_{\mathrm{Gr}}$\label{sec:sample}}
As an example, we postulate $\tilde{\HCd}_{\mathrm{Gr}}$ as a linear combination of
quadratic and linear terms in $\tilde{q}$
\begin{equation}\label{eq:H-dyn-post}
\tilde{\HCd}_{\mathrm{Gr}}=\frac{1}{4g_{1}}\tilde{q}\indices{_{\eta}^{\alpha\xi\beta}}
\tilde{q}\indices{_{\alpha}^{\eta\tau\lambda}}\,g_{\xi\tau}g_{\beta\lambda}\frac{1}{\sqrt{-g}}
-g_{2}\,\tilde{q}\indices{_{\eta}^{\alpha\eta\beta}}\,g_{\alpha\beta}.
\end{equation}
In contrast to the dimensionless coupling constant $g_{1}$, the coupling constant
$g_{2}$ has the natural dimension $\mathrm{Length}^{-2}$.
Note that the sample Hamiltonian~(\ref{eq:H-dyn-post}) does not depend on $\tilde{k}^{\,\xi\lambda\mu}$
and thus directly induces the metric compatibility condition $g_{\xi\lambda\,;\,\mu}=0$ according to Eqs.~(\ref{eq:feqs-Hdyn-1}).
In a subsequent paper, we will discuss the more general case of a $\tilde{\HCd}_{\mathrm{Gr}}$
which also depends quadratically on $\tilde{k}^{\,\xi\lambda\mu}$.

The correlation of the canonical momentum $q$ to the Riemann tensor then follows from Eq.~(\ref{eq:riemann-tensor2}) as
\begin{equation}\label{q-R-corr}
q_{\eta\alpha\xi\beta}=g_{1}\left(R_{\eta\alpha\xi\beta}-R_{\eta\alpha\xi\beta}\big|_{\max}\right),
\end{equation}
with
\begin{equation*}
R_{\eta\alpha\xi\beta}\big|_{\max}=g_{2}\left(g_{\eta\xi}\,g_{\alpha\beta}-g_{\eta\beta}\,g_{\alpha\xi}\right)
\end{equation*}
the Riemann tensor for a \emph{maximally symmetric} $4$-dimensional manifold with constant Ricci curvature $R=12g_{2}$.
The derivatives of $\tilde{\HCd}_{\mathrm{Gr}}$ with respect to $\tilde{q}$ in Eq.~(\ref{eq:consistency}) then cancel
for a symmetric Ricci tensor, hence for vanishing torsion $s\indices{^{\lambda}_{\xi\beta}}$, as
\begin{align*}
&\quad\,q\indices{_{\tau}^{\alpha\beta\lambda}}\pfrac{\tilde{\HCd}_{\mathrm{Gr}}}{\tilde{q}\indices{_{\tau}^{\xi\beta\lambda}}}
-q\indices{_{\xi}^{\tau\beta\lambda}}\pfrac{\tilde{\HCd}_{\mathrm{Gr}}}{\tilde{q}\indices{_{\alpha}^{\tau\beta\lambda}}}
=g_{1}g_{2}\,g^{\alpha\beta}\left(R_{\xi\beta}-R_{\beta\xi}\right)\\
&=2g_{1}g_{2}\,g^{\alpha\beta}\left(s\indices{^{\tau}_{\tau\xi;\beta}}-s\indices{^{\tau}_{\tau\beta;\xi}}
+s\indices{^{\tau}_{\xi\beta;\tau}}-2s\indices{^{\tau}_{\lambda\tau}}s\indices{^{\lambda}_{\xi\beta}}\right).
\end{align*}
The derivative of $\tilde{\HCd}_{\mathrm{Gr}}$ with respect to $g_{\alpha\beta}$ follows as
\begin{align*}
2g_{\xi\beta}\pfrac{\tilde{\HCd}_{\mathrm{Gr}}}{g_{\alpha\beta}}&=\frac{1}{g_{1}}\left(
q\indices{_{\beta\eta\lambda\xi}}\,q\indices{^{\eta\beta\lambda\alpha}}
-\quarter\delta_{\xi}^{\alpha}\,q\indices{_{\beta\eta\lambda\tau}}q\indices{^{\,\eta\beta\lambda\tau}}\right)\sqrt{-g}\\
&\quad\mbox{}-g_{2}\left(q\indices{_{\eta}^{\alpha\eta}_{\xi}}+q\indices{_{\eta\xi}^{\eta\alpha}}\right)\sqrt{-g}.
\end{align*}
Substituting the $q$-terms according to Eq.~(\ref{q-R-corr}) and
writing the derivative of $\tilde{\HCd}_{0}$ with respect to the metric $g_{\alpha\beta}$
as the given system's \emph{metric} energy-momentum tensor according to Eq.~(\ref{eq:em-tensor})
\begin{equation*}
2g_{\xi\beta}\pfrac{\tilde{\HCd}_{0}}{g_{\beta\alpha}}=T\indices{_{\xi}^{\alpha}}\sqrt{-g},
\end{equation*}
the consistency relation~(\ref{eq:consistency}) for the Hamiltonian~(\ref{eq:H-dyn-post}) emerges as
\begin{align}
g_{1}&\left(R\indices{^{\eta\beta\lambda\alpha}}R\indices{_{\eta\beta\lambda\xi}}
-\quarter\delta_{\xi}^{\alpha}R\indices{^{\eta\beta\lambda\tau}}R\indices{_{\eta\beta\lambda\tau}}\right)
+\frac{1}{8\pi G}\left(R\indices{^{\alpha}_{\xi}}-\onehalf\delta_{\xi}^{\alpha}R+\Lambda\delta_{\xi}^{\alpha}\right)\nonumber\\
&=\pfrac{\HCd_{0}}{a_{\alpha}}a_{\xi}-p^{\,\alpha\beta}\pfrac{\HCd_{0}}{p^{\,\xi\beta}}+T\indices{_{\xi}^{\alpha}}=\theta\indices{_{\xi}^{\alpha}}
\label{eq:final}
\end{align}
and thus represents a generalized Einstein equation.
The scalar (spin-$0$) field contributes to the source merely via its \emph{metric} energy-momentum tensor terms,
whereas the vector (spin-$1$) field contributes in addition with the first two terms on the
right-hand side of Eq.~(\ref{eq:final}), which sum up to the \emph{canonical} energy-momentum tensor $\theta\indices{_{\xi}^{\alpha}}$.
So, for systems with only scalar fields, the right-hand side of
Eq.~(\ref{eq:final}) reduces to that of the Einstein equation.

Without the term proportional to the coupling constant $g_{1}$,
the left-hand side of Eq.~(\ref{eq:final}) reduces to the Einstein tensor.
The coupling constants $g_{1}$ and $g_{2}$ contained in~(\ref{eq:H-dyn-post}) can be expressed
in terms of the cosmological constant $\Lambda$ and the gravitational constant $G$ as~\footnote{
As a consequence of the particular definition of $\HCd_{\mathrm{Gr}}$ from Eq.~(\ref{eq:H-dyn-post}),
the coupling constant in front of the quadratic Riemann tensor terms in Eq.~(\ref{eq:final}) is fixed.
In our forthcoming paper, we investigate a generalization of $\HCd_{\mathrm{Gr}}$, which introduces
a third additional coupling constant $g_{3}$ and hence allows more flexibility in the relations between $g_{1}$, $G$, and $\Lambda$.}
\begin{equation}\label{eq:coupling-constants}
g_{1}=-\frac{3}{16\pi\,G\Lambda},\qquad g_{2}=\onethird\Lambda.
\end{equation}
Solutions of the field equation for particular systems $\HCd_{0}$, namely for Klein-Gordon, Maxwell,
and Proca systems, will be discussed in detail in our subsequent paper~\cite{struckmeier17b}.

Whether or not our theory can provide new insights with respect to the dark matter issue remains to be clarified.

Remarkably, the metrics obtained from Eq.~(\ref{eq:final}) for the
exterior regions of nonrotating black holes or rotating black holes
coincide with those solving the Einstein equation with cosmological constant.
In other words, the vacuum field equation
\begin{equation}\label{eq:r-squared}
R\indices{_{\eta\beta\lambda\xi}}\,R\indices{^{\eta\beta\lambda\alpha}}-
\quarter\delta_{\xi}^{\alpha}\,R\indices{_{\eta\beta\lambda\tau}}\,
R\indices{^{\eta\beta\lambda\tau}}=0
\end{equation}
is likewise satisfied not only by the Schwarzschild metric~\cite{stephenson58}, but also
by the more general Schwarzschild-De~Sitter and the Kerr-De~Sitter metrics.
Thus, for a vanishing right-hand side of Eq.~(\ref{eq:final}), both parts of the left-hand side,
the quadratic part~(\ref{eq:r-squared}) and the ``Einstein part'' are satisfied by the same metrics.
As a consequence, the classical tests of general relativity, namely, the bending of light,
the perihelion shift, and the Newtonian limit are equally passed by
the solutions of the field equation~(\ref{eq:r-squared}).
However, the metrics obtained from Eq.~(\ref{eq:final}) for cases where matter/fields are present will
be shown to be different from those emerging from the Einstein equation.
This changes, for instance, the prediction of measurable observables of neutron stars.
\section{Conclusions\label{sec:conclusions}}
By means of the framework of canonical transformations,
we have demonstrated that any (globally) Lorentz-invariant Lagrangian/Hamiltonian system
can be converted into an amended Lagrangian/Hamiltonian which is form-invariant
under a general local transformation of the reference frame, following the
well-established lines of reasoning of gauge theories.
Remarkably, the Riemann-Cartan tensor is not worked in by hand, but \emph{naturally} crops up from the gauge formalism
as the specific combination of the gauge fields, i.e., the connection coefficients, and their partial derivatives.
In particular, our approach includes a nonvanishing torsion of spacetime
and is not restricted to the usual assumption of metric compatibility.

Thus, the description of the spacetime dynamics emerges from basic principles,
namely the action principle and the requirement of the form-invariance of the action integral
under general spacetime transformations---which ensures the general principle of relativity to be satisfied.
The ensuing coupling of spacetime dynamics with matter fields
involve the affine connection coefficients, which thus act as gauge quantities.
The derivation was worked out in the Hamiltonian framework making use of the canonical
transformation formalism---which by construction ensures the action principle to be maintained in its form.
The integrand~(\ref{eq:action-integral4}) of the final action integral was shown to represent a world scalar density
and thereby meets the requirement to be form-invariant under general spacetime transformations.

For the closed description of the spacetime dynamics, a Hamiltonian $\tilde{\HCd}_{\mathrm{Gr}}$
which describes the dynamics of the ``free'' gauge fields must be postulated.
This is a common feature of all gauge theories and reflects here the residual
indeterminacy of any gauge theory of gravity.
In this sense, we have derived the generic part of the description of geometrodynamics
which is common to all specific theories described by a Hamiltonian $\tilde{\HCd}_{0}$
that are based on a particular $\tilde{\HCd}_{\mathrm{Gr}}$.

Moreover, we found that spin-$1$ fields may contribute with additional source terms
to the equations of motion---depending on possible symmetries of the conjugate momenta.
This does not occur for spin-$0$ fields.
The canonical formulation of the gauge theory of gravity requires
in our description  a term quadratic in the canonical momenta $\tilde{q}$ of the gauge
fields $\gamma$ in order for the set of field equations to be closed.
This contrasts with the Einstein approach, which is restricted---in its Hamiltonian
formulation---to a linear momentum term.
\begin{acknowledgments}
The authors are deeply indebted to the ``Walter Greiner-Gesell\-schaft zur F\"orderung
der physikalischen Grundlagenforschung e.V.'' in Frankfurt for its support.
They thank E.~Guendelman (Ben Gurion University, Israel), A.~Redelbach (University W\"urzburg, Germany),
and A.~Koenigstein (University Frankfurt, Germany) for valuable discussions.
\end{acknowledgments}
\appendix\allowdisplaybreaks
\section{EXPLICIT CALCULATION OF THE TRANSFORMATION RULE~(\ref{eq:ham-conn-coeff1})\label{sec:app1}}
First, we show that the second term on the right-hand side of Eq.~(\ref{eq:ham-conn-coeff}) vanishes identically.
According to the chain rule, we have
\begin{align}
\pfrac{}{x^{\alpha}}&\left(\pfrac{x^{\alpha}}{X^{\beta}}\left|\pfrac{x}{X}\right|^{-1}\right)\nonumber\\
&=\left|\pfrac{x}{X}\right|^{-1}\left(\ppfrac{x^{\alpha}}{X^{\beta}}{X^{\xi}}\pfrac{X^{\xi}}{x^{\alpha}}
-\left|\pfrac{x}{X}\right|^{-1}\pfrac{\left|\pfrac{x}{X}\right|}{X^{\beta}}\right).
\label{eq:H-e-modi}
\end{align}
By virtue of the general identity for the derivative of the determinant of a matrix with respect to a matrix element
\begin{displaymath}
\pfrac{\left|\pfrac{x}{X}\right|}{\left(\pfrac{x^{\alpha}}{X^{\xi}}\right)}=\pfrac{X^{\xi}}{x^{\alpha}}\left|\pfrac{x}{X}\right|,
\end{displaymath}
the $X^{\beta}$-derivative of $\ln\left|\partial{x}/\partial{X}\right|$ in Eq.~(\ref{eq:H-e-modi}) is converted into
\begin{align}
\left|\pfrac{x}{X}\right|^{-1}\pfrac{\left|\pfrac{x}{X}\right|}{X^{\beta}}=\left|\pfrac{x}{X}\right|^{-1}
\pfrac{\left|\pfrac{x}{X}\right|}{\left(\pfrac{x^{\alpha}}{X^{\xi}}\right)}\pfrac{\left(\pfrac{x^{\alpha}}{X^{\xi}}\right)}{X^{\beta}}
=\pfrac{X^{\xi}}{x^{\alpha}}\ppfrac{x^{\alpha}}{X^{\xi}}{X^{\beta}}.\nonumber\\
\label{eq:H-deri-expr}
\end{align}
Inserting Eq.~(\ref{eq:H-deri-expr}) into~(\ref{eq:H-e-modi}) then yields
\begin{equation}\label{eq:H-deri-expr-2}
\pfrac{}{x^{\alpha}}\left(\pfrac{x^{\alpha}}{X^{\beta}}\left|\pfrac{x}{X}\right|^{-1}\right)\equiv0.
\end{equation}
In order to express the third term on the right-hand side of Eq.~(\ref{eq:ham-conn-coeff}),
the partial derivatives are first of all written in expanded form
\begin{widetext}
\begin{align}
\left.\pfrac{\bar{\FCd}_{2}^{\mu}}{x^{\mu}}\right|_{\text{expl}}
=\left.\pfrac{\FCd_{2}^{\mu}}{x^{\mu}}\right|_{\text{expl}}
+\tilde{Q}\indices{_{\eta}^{\alpha\xi\beta}}\Bigg[&\gamma\indices{^{k}_{ij}}\left(
\ppfrac{X^{\eta}}{x^{k}}{x^{n}}\pfrac{x^{n}}{X^{\beta}}\pfrac{x^{i}}{X^{\alpha}}\pfrac{x^{j}}{X^{\xi}}
+\ppfrac{x^{i}}{X^{\alpha}}{X^{\beta}}\pfrac{X^{\eta}}{x^{k}}\pfrac{x^{j}}{X^{\xi}}
+\ppfrac{x^{j}}{X^{\xi}}{X^{\beta}}\pfrac{X^{\eta}}{x^{k}}\pfrac{x^{i}}{X^{\alpha}}\right)\nonumber\\
&+\ppfrac{X^{\eta}}{x^{k}}{x^{n}}\ppfrac{x^{k}}{X^{\alpha}}{X^{\xi}}
\pfrac{x^{n}}{X^{\beta}}+\pppfrac{x^{k}}{X^{\alpha}}{X^{\xi}}{X^{\beta}}\pfrac{X^{\eta}}{x^{k}}\Bigg]\left|\pfrac{x}{X}\right|^{-1}.
\label{eq:ham-conn-coeff2}
\end{align}
This expression is now split into a skew-symmetric and a symmetric part
of $\tilde{Q}\indices{_{\eta}^{\alpha\xi\beta}}$ in the indices $\xi$ and $\beta$ according to
\begin{displaymath}
\tilde{Q}\indices{_{\eta}^{\alpha\xi\beta}}
=\onehalf\left(\tilde{Q}\indices{_{\eta}^{\alpha\xi\beta}}-\tilde{Q}\indices{_{\eta}^{\alpha\beta\xi}}\right)
+\onehalf\left(\tilde{Q}\indices{_{\eta}^{\alpha\xi\beta}}+\tilde{Q}\indices{_{\eta}^{\alpha\beta\xi}}\right)
=\tilde{Q}\indices{_{\eta}^{\alpha[\xi\beta]}}+\tilde{Q}\indices{_{\eta}^{\alpha(\xi\beta)}}.
\end{displaymath}
For the skew-symmetric part, $\tilde{Q}\indices{_{\eta}^{\alpha[\xi\beta]}}$,
the two terms in~(\ref{eq:ham-conn-coeff2}) which are symmetric in $\xi$ and $\beta$ vanish, leaving
\begin{align*}
&\quad\,\tilde{Q}\indices{_{\eta}^{\alpha[\xi\beta]}}\left[\gamma\indices{^{k}_{ij}}\left(
\ppfrac{X^{\eta}}{x^{k}}{x^{n}}\pfrac{x^{n}}{X^{\beta}}\pfrac{x^{i}}{X^{\alpha}}\pfrac{x^{j}}{X^{\xi}}
+\ppfrac{x^{i}}{X^{\alpha}}{X^{\beta}}\pfrac{X^{\eta}}{x^{k}}\pfrac{x^{j}}{X^{\xi}}\right)
+\ppfrac{X^{\eta}}{x^{k}}{x^{n}}\pfrac{x^{n}}{X^{\beta}}\ppfrac{x^{k}}{X^{\alpha}}{X^{\xi}}\right]\\
&=\tilde{Q}\indices{_{\eta}^{\alpha[\xi\beta]}}\left[\ppfrac{X^{\eta}}{x^{k}}{x^{n}}\pfrac{x^{n}}{X^{\beta}}\left(
\gamma\indices{^{k}_{ij}}\pfrac{x^{i}}{X^{\alpha}}\pfrac{x^{j}}{X^{\xi}}+\ppfrac{x^{k}}{X^{\alpha}}{X^{\xi}}\right)
+\gamma\indices{^{k}_{ij}}\ppfrac{x^{i}}{X^{\alpha}}{X^{\beta}}\pfrac{X^{\eta}}{x^{k}}\pfrac{x^{j}}{X^{\xi}}\right]\\
&=\tilde{Q}\indices{_{\eta}^{\alpha[\xi\beta]}}\left[\Gamma\indices{^{j}_{\alpha\xi}}
\ppfrac{X^{\eta}}{x^{k}}{x^{n}}\pfrac{x^{k}}{X^{j}}\pfrac{x^{n}}{X^{\beta}}+
\gamma\indices{^{k}_{ij}}\ppfrac{x^{i}}{X^{\alpha}}{X^{\beta}}\pfrac{X^{\eta}}{x^{k}}\pfrac{x^{j}}{X^{\xi}}\right]\\
&=\tilde{Q}\indices{_{\eta}^{\alpha[\xi\beta]}}\left[\Gamma\indices{^{j}_{\alpha\xi}}\left(
\gamma\indices{^{i}_{kn}}\pfrac{X^{\eta}}{x^{i}}\pfrac{x^{k}}{X^{j}}\pfrac{x^{n}}{X^{\beta}}
-\Gamma\indices{^{\eta}_{j\beta}}\right)+\gamma\indices{^{k}_{ij}}\left(\Gamma\indices{^{a}_{\alpha\beta}}
\pfrac{x^{i}}{X^{a}}-\gamma\indices{^{i}_{ab}}\pfrac{x^{a}}{X^{\alpha}}\pfrac{x^{b}}{X^{\beta}}
\right)\pfrac{X^{\eta}}{x^{k}}\pfrac{x^{j}}{X^{\xi}}\right]\\
&=\tilde{Q}\indices{_{\eta}^{\alpha[\xi\beta]}}\left(
-\Gamma\indices{^{i}_{\alpha\xi}}\Gamma\indices{^{\eta}_{i\beta}}
-\gamma\indices{^{i}_{ab}}\gamma\indices{^{k}_{ij}}\pfrac{x^{a}}{X^{\alpha}}\pfrac{x^{b}}{X^{\beta}}
\pfrac{X^{\eta}}{x^{k}}\pfrac{x^{j}}{X^{\xi}}+\Gamma\indices{^{j}_{\alpha\xi}}\gamma\indices{^{i}_{kn}}
\pfrac{X^{\eta}}{x^{i}}\pfrac{x^{k}}{X^{j}}\pfrac{x^{n}}{X^{\beta}}
+\Gamma\indices{^{j}_{\alpha\beta}}\gamma\indices{^{i}_{kn}}
\pfrac{X^{\eta}}{x^{i}}\pfrac{x^{k}}{X^{j}}\pfrac{x^{n}}{X^{\xi}}\right)\\
&=-\tilde{Q}\indices{_{\eta}^{\alpha[\xi\beta]}}\Gamma\indices{^{i}_{\alpha\xi}}\Gamma\indices{^{\eta}_{i\beta}}
+\gamma\indices{^{i}_{ab}}\gamma\indices{^{k}_{ij}}\tilde{Q}\indices{_{\eta}^{\alpha[\xi\beta]}}
\pfrac{x^{a}}{X^{\alpha}}\pfrac{x^{b}}{X^{\xi}}\pfrac{X^{\eta}}{x^{k}}\pfrac{x^{j}}{X^{\beta}}\\
&=-\tilde{Q}\indices{_{\eta}^{\alpha[\xi\beta]}}\Gamma\indices{^{i}_{\alpha\xi}}\Gamma\indices{^{\eta}_{i\beta}}
+\tilde{q}\indices{_{k}^{a[bj]}}\gamma\indices{^{i}_{ab}}\gamma\indices{^{k}_{ij}}\left|\pfrac{x}{X}\right|\\
&=-\tilde{Q}\indices{_{\eta}^{\alpha[\xi\beta]}}\Gamma\indices{^{k}_{\alpha\xi}}\Gamma\indices{^{\eta}_{k\beta}}
+\tilde{q}\indices{_{\eta}^{\alpha[\xi\beta]}}\gamma\indices{^{k}_{\alpha\xi}}\gamma\indices{^{\eta}_{k\beta}}\left|\pfrac{x}{X}\right|\\
&=-\onehalf\tilde{Q}\indices{_{\eta}^{\alpha\xi\beta}}\left(
\Gamma\indices{^{k}_{\alpha\xi}}\Gamma\indices{^{\eta}_{k\beta}}
-\Gamma\indices{^{k}_{\alpha\beta}}\Gamma\indices{^{\eta}_{k\xi}}\right)
+\onehalf\tilde{q}\indices{_{\eta}^{\alpha\xi\beta}}\left(
\gamma\indices{^{k}_{\alpha\xi}}\gamma\indices{^{\eta}_{k\beta}}
-\gamma\indices{^{k}_{\alpha\beta}}\gamma\indices{^{\eta}_{k\xi}}\right)\left|\pfrac{x}{X}\right|.
\end{align*}
The two mixed terms in $\Gamma$ and $\gamma$ cancel each other due to the skew-symmetry of
$\tilde{Q}\indices{_{\eta}^{\alpha[\xi\beta]}}$ in $\xi$ and $\beta$.

The contribution of~(\ref{eq:ham-conn-coeff}) emerging from the \emph{symmetric}
part $\tilde{Q}\indices{_{\eta}^{\alpha(\xi\beta)}}$ can be expressed in terms
of the derivatives of the connection coefficients, whose transformation rule is
\begin{displaymath}
\pfrac{\Gamma\indices{^{\eta}_{\alpha\xi}}}{X^{\kappa}}\pfrac{X^{\kappa}}{x^{n}}=
\pfrac{\gamma\indices{^{k}_{ij}}}{x^{n}}\pfrac{X^{\eta}}{x^{k}}
\pfrac{x^{i}}{X^{\alpha}}\pfrac{x^{j}}{X^{\xi}}+\gamma\indices{^{k}_{ij}}\pfrac{}{x^{n}}\left(\pfrac{X^{\eta}}{x^{k}}
\pfrac{x^{i}}{X^{\alpha}}\pfrac{x^{j}}{X^{\xi}}\right)+
\pfrac{}{x^{n}}\left(\pfrac{X^{\eta}}{x^{k}}\ppfrac{x^{k}}{X^{\alpha}}{X^{\xi}}\right).
\end{displaymath}
Thus
\begin{align*}
&\quad\,\tilde{Q}\indices{_{\eta}^{\alpha(\xi\beta)}}\pfrac{x^{n}}{X^{\beta}}
\left[\gamma\indices{^{k}_{ij}}\pfrac{}{x^{n}}\left(
\pfrac{X^{\eta}}{x^{k}}\pfrac{x^{i}}{X^{\alpha}}\pfrac{x^{j}}{X^{\xi}}\right)+
\pfrac{}{x^{n}}\left(\pfrac{X^{\eta}}{x^{k}}\ppfrac{x^{k}}{X^{\alpha}}{X^{\xi}}\right)\right]\\
&=\tilde{Q}\indices{_{\eta}^{\alpha(\xi\beta)}}\pfrac{x^{n}}{X^{\beta}}\left(
\pfrac{\Gamma\indices{^{\eta}_{\alpha\xi}}}{X^{\kappa}}\pfrac{X^{\kappa}}{x^{n}}-
\pfrac{\gamma\indices{^{k}_{ij}}}{x^{n}}\pfrac{X^{\eta}}{x^{k}}
\pfrac{x^{i}}{X^{\alpha}}\pfrac{x^{j}}{X^{\xi}}\right)\\
&=\tilde{Q}\indices{_{\eta}^{\alpha(\xi\beta)}}
\pfrac{\Gamma\indices{^{\eta}_{\alpha\xi}}}{X^{\beta}}-
\tilde{q}\indices{_{k}^{i(jn)}}\pfrac{\gamma\indices{^{k}_{ij}}}{x^{n}}\left|\pfrac{x}{X}\right|\\
&=\tilde{Q}\indices{_{\eta}^{\alpha(\xi\beta)}}\pfrac{\Gamma\indices{^{\eta}_{\alpha\xi}}}{X^{\beta}}-
\tilde{q}\indices{_{\eta}^{\alpha(\xi\beta)}}\pfrac{\gamma\indices{^{\eta}_{\alpha\xi}}}{x^{\beta}}\left|\pfrac{x}{X}\right|\\
&=\onehalf\tilde{Q}\indices{_{\eta}^{\alpha\xi\beta}}\left(
\pfrac{\Gamma\indices{^{\eta}_{\alpha\xi}}}{X^{\beta}}+\pfrac{\Gamma\indices{^{\eta}_{\alpha\beta}}}{X^{\xi}}\right)-
\onehalf\tilde{q}\indices{_{\eta}^{\alpha\xi\beta}}\left(
\pfrac{\gamma\indices{^{\eta}_{\alpha\xi}}}{x^{\beta}}+\pfrac{\gamma\indices{^{\eta}_{\alpha\beta}}}{x^{\xi}}\right)\left|\pfrac{x}{X}\right|.
\end{align*}
The total transformation rule~(\ref{eq:ham-conn-coeff2}) expressed in terms
of connection coefficients is then
\begin{align*}
\left.\pfrac{\bar{\FCd}_{2}^{\mu}}{x^{\mu}}\right|_{\text{expl}}=
\left.\pfrac{\FCd_{2}^{\mu}}{x^{\mu}}\right|_{\text{expl}}&+
\frac{1}{2}\tilde{Q}\indices{_{\eta}^{\alpha\xi\beta}}\left(
\pfrac{\Gamma\indices{^{\eta}_{\alpha\xi}}}{X^{\beta}}+\pfrac{\Gamma\indices{^{\eta}_{\alpha\beta}}}{X^{\xi}}-
\Gamma\indices{^{k}_{\alpha\xi}}\Gamma\indices{^{\eta}_{k\beta}}+
\Gamma\indices{^{k}_{\alpha\beta}}\Gamma\indices{^{\eta}_{k\xi}}\right)\left|\pfrac{X}{x}\right|\\
&-\frac{1}{2}\tilde{q}\indices{_{\eta}^{\alpha\xi\beta}}\left(
\pfrac{\gamma\indices{^{\eta}_{\alpha\xi}}}{x^{\beta}}+\pfrac{\gamma\indices{^{\eta}_{\alpha\beta}}}{x^{\xi}}-
\gamma\indices{^{k}_{\alpha\xi}}\gamma\indices{^{\eta}_{k\beta}}+
\gamma\indices{^{k}_{\alpha\beta}}\gamma\indices{^{\eta}_{k\xi}}\right).
\end{align*}
\section{EXPLICIT CALCULATION OF THE CONSISTENCY EQUATION~(\ref{eq:consistency})\label{sec:app2}}
Equation~(\ref{eq:consistency0}) reads, in explicit form:
\begin{equation*}
0=\pfrac{\tilde{p}^{\,\nu\beta}}{x^{\,\beta}}a_{\mu}+\tilde{p}^{\,\nu\beta}\pfrac{a_{\mu}}{x^{\,\beta}}
+2\pfrac{\tilde{k}^{\,\lambda\nu\beta}}{x^{\,\beta}}g_{\lambda\mu}+2\tilde{k}^{\,\lambda\nu\beta}\pfrac{g_{\lambda\mu}}{x^{\,\beta}}+
\pfrac{\tilde{q}\indices{_{\eta}^{\nu\lambda\beta}}}{x^{\,\beta}}\gamma\indices{^{\eta}_{\mu\lambda}}+
\tilde{q}\indices{_{\eta}^{\nu\lambda\beta}}\pfrac{\gamma\indices{^{\eta}_{\mu\lambda}}}{x^{\,\beta}}-
\pfrac{\tilde{q}\indices{_{\mu}^{\eta\lambda\beta}}}{x^{\,\beta}}\gamma\indices{^{\nu}_{\eta\lambda}}-
\tilde{q}\indices{_{\mu}^{\eta\lambda\beta}}\pfrac{\gamma\indices{^{\nu}_{\eta\lambda}}}{x^{\,\beta}}.
\end{equation*}
The partial derivative representations (\ref{eq:feq-vecfields0}), (\ref{eq:nonmetricity0}),
(\ref{eq:k-div}), (\ref{eq:gamma-deri}), and (\ref{eq:feq-conn-coeff}) of the canonical field
equations can now be inserted to replace all derivatives with respect to $x^{\,\beta}$, which yields
\begin{align*}
0&=\left(-\pfrac{\tilde{\HCd}_{0}}{a_{\nu}}-\tilde{p}^{\eta\beta}\gamma\indices{^{\nu}_{\eta\beta}}\right)a_{\mu}+
\tilde{p}^{\,\nu\beta}\left(\pfrac{\tilde{\HCd}_{0}}{\tilde{p}^{\,\mu\beta}}+a_{\eta}\,\gamma\indices{^{\eta}_{\mu\beta}}\right)\\
&\quad\mbox{}-2\left(\tilde{k}^{\eta\nu\beta}\gamma\indices{^{\lambda}_{\eta\beta}}+
\tilde{k}^{\lambda\eta\beta}\gamma\indices{^{\nu}_{\eta\beta}}+\pfrac{\tilde{\HCd}_{0}}{g_{\lambda\nu}}+
\pfrac{\tilde{\HCd}_{\mathrm{Gr}}}{g_{\lambda\nu}}\right)g_{\lambda\mu}+
2\tilde{k}^{\,\lambda\nu\beta}\left(g_{\eta\mu}\gamma\indices{^{\eta}_{\lambda\beta}}+
g_{\lambda\eta}\gamma\indices{^{\eta}_{\mu\beta}}+
\pfrac{\tilde{\HCd}_{\mathrm{Gr}}}{\tilde{k}\indices{^{\lambda\mu\beta}}}\right)\\
&\quad\mbox{}-\left(\tilde{p}^{\,\nu\lambda}a_{\eta}+2\tilde{k}^{\,\beta\nu\lambda}g_{\beta\eta}+
\tilde{q}\indices{_{\beta}^{\nu\tau\lambda}}\gamma\indices{^{\beta}_{\eta\tau}}-
\cancel{\tilde{q}\indices{_{\eta}^{\tau\beta\lambda}}\gamma\indices{^{\nu}_{\tau\beta}}}\,\right)\gamma\indices{^{\eta}_{\mu\lambda}}+
\tilde{q}\indices{_{\eta}^{\nu\lambda\beta}}\left(
\pfrac{\tilde{\HCd}_{\mathrm{Gr}}}{\tilde{q}\indices{_{\eta}^{\mu\lambda\beta}}}+
\gamma\indices{^{\tau}_{\mu\beta}}\gamma\indices{^{\eta}_{\tau\lambda}}\right)\\
&\quad\mbox{}+\left(\tilde{p}^{\,\eta\lambda}a_{\mu}+2\tilde{k}^{\,\tau\eta\lambda}g_{\tau\mu}-
\cancel{\tilde{q}\indices{_{\beta}^{\eta\lambda\tau}}\gamma\indices{^{\beta}_{\mu\tau}}}+
\tilde{q}\indices{_{\mu}^{\tau\lambda\beta}}\gamma\indices{^{\eta}_{\tau\beta}}
\right)\gamma\indices{^{\nu}_{\eta\lambda}}-\tilde{q}\indices{_{\mu}^{\eta\lambda\beta}}\left(
\pfrac{\tilde{\HCd}_{\mathrm{Gr}}}{\tilde{q}\indices{_{\nu}^{\eta\lambda\beta}}}+
\gamma\indices{^{\tau}_{\eta\beta}}\gamma\indices{^{\nu}_{\tau\lambda}}\right).
\end{align*}
All terms depending on the affine connections $\gamma$ cancel, as can be seen after rearranging and relabeling some running indices:
\allowdisplaybreaks[0]
\begin{align*}
0&=-\pfrac{\tilde{\HCd}_{0}}{a_{\nu}}a_{\mu}+\tilde{p}^{\,\nu\beta}\pfrac{\tilde{\HCd}_{0}}{\tilde{p}^{\,\mu\beta}}-
2\pfrac{\tilde{\HCd}_{0}}{g_{\lambda\nu}}g_{\lambda\mu}-
2\pfrac{\tilde{\HCd}_{\mathrm{Gr}}}{g_{\lambda\nu}}g_{\lambda\mu}+
2\tilde{k}^{\,\lambda\nu\beta}\pfrac{\tilde{\HCd}_{\mathrm{Gr}}}{\tilde{k}\indices{^{\lambda\mu\beta}}}+
\tilde{q}\indices{_{\eta}^{\nu\lambda\beta}}
\pfrac{\tilde{\HCd}_{\mathrm{Gr}}}{\tilde{q}\indices{_{\eta}^{\mu\lambda\beta}}}-
\tilde{q}\indices{_{\mu}^{\eta\lambda\beta}}
\pfrac{\tilde{\HCd}_{\mathrm{Gr}}}{\tilde{q}\indices{_{\nu}^{\eta\lambda\beta}}}\\
&\quad\mbox{}-
\tilde{p}^{\,\eta\beta}a_{\mu}\gamma\indices{^{\nu}_{\eta\beta}}+
\tilde{p}^{\,\eta\lambda}a_{\mu}\gamma\indices{^{\nu}_{\eta\lambda}}+
\tilde{p}^{\,\nu\beta}a_{\eta}\gamma\indices{^{\eta}_{\mu\beta}}-
\tilde{p}^{\,\nu\lambda}a_{\eta}\gamma\indices{^{\eta}_{\mu\lambda}}\\
&\quad\mbox{}-
2\tilde{k}^{\,\eta\nu\beta}g_{\lambda\mu}\gamma\indices{^{\lambda}_{\eta\beta}}+
2\tilde{k}^{\,\lambda\nu\beta}g_{\eta\mu}\gamma\indices{^{\eta}_{\lambda\beta}}-
2\tilde{k}^{\,\lambda\eta\beta}g_{\lambda\mu}\gamma\indices{^{\nu}_{\eta\beta}}+
2\tilde{k}^{\,\tau\eta\lambda}g_{\tau\mu}\gamma\indices{^{\nu}_{\eta\lambda}}+
2\tilde{k}^{\,\lambda\nu\beta}g_{\lambda\eta}\gamma\indices{^{\eta}_{\mu\beta}}-
2\tilde{k}^{\,\beta\nu\lambda}g_{\beta\eta}\gamma\indices{^{\eta}_{\mu\lambda}}\\
&\quad\mbox{}-
\tilde{q}\indices{_{\beta}^{\nu\tau\lambda}}\gamma\indices{^{\eta}_{\mu\lambda}}\gamma\indices{^{\beta}_{\eta\tau}}+
\tilde{q}\indices{_{\eta}^{\nu\lambda\beta}}\gamma\indices{^{\tau}_{\mu\beta}}\gamma\indices{^{\eta}_{\tau\lambda}}+
\tilde{q}\indices{_{\eta}^{\tau\beta\lambda}}\gamma\indices{^{\eta}_{\mu\lambda}}\gamma\indices{^{\nu}_{\tau\beta}}-
\tilde{q}\indices{_{\beta}^{\eta\lambda\tau}}\gamma\indices{^{\beta}_{\mu\tau}}\gamma\indices{^{\nu}_{\eta\lambda}}+
\tilde{q}\indices{_{\mu}^{\tau\lambda\beta}}\gamma\indices{^{\eta}_{\tau\beta}}\gamma\indices{^{\nu}_{\eta\lambda}}-
\tilde{q}\indices{_{\mu}^{\eta\lambda\beta}}\gamma\indices{^{\tau}_{\eta\beta}}\gamma\indices{^{\nu}_{\tau\lambda}}.
\end{align*}
The remaining terms constitute the second rank tensor equation~(\ref{eq:consistency})
\begin{equation}\label{eq:consistency2}
\pfrac{\tilde{\HCd}_{0}}{a_{\nu}}a_{\mu}-\tilde{p}^{\,\nu\beta}\pfrac{\tilde{\HCd}_{0}}{\tilde{p}^{\,\mu\beta}}+
2\pfrac{\tilde{\HCd}_{0}}{g_{\lambda\nu}}g_{\lambda\mu}=-2\pfrac{\tilde{\HCd}_{\mathrm{Gr}}}{g_{\lambda\nu}}g_{\lambda\mu}+
2\tilde{k}^{\,\lambda\nu\beta}\pfrac{\tilde{\HCd}_{\mathrm{Gr}}}{\tilde{k}\indices{^{\lambda\mu\beta}}}-
\tilde{q}\indices{_{\mu}^{\eta\lambda\beta}}\pfrac{\tilde{\HCd}_{\mathrm{Gr}}}{\tilde{q}\indices{_{\nu}^{\eta\lambda\beta}}}+
\tilde{q}\indices{_{\eta}^{\nu\lambda\beta}}\pfrac{\tilde{\HCd}_{\mathrm{Gr}}}{\tilde{q}\indices{_{\eta}^{\mu\lambda\beta}}}.
\end{equation}
\section{EQUIVALENT REPRESENTATION AND COVARIANT DIVERGENCE OF THE CONSISTENCY EQUATION\label{sec:cov-div}}
By virtue of the \emph{identities} for scalar density valued functions of tensors and the metric~\cite{struckvasvenn21},
one encounters the following identities from the dependencies of the Hamiltonians $\tilde{\HCd}_{0}$ and $\tilde{\HCd}_{\mathrm{Gr}}$:
\begin{align*}
\pfrac{\tilde{\HCd}_{0}}{\tilde{\pi}^{\mu}}\tilde{\pi}^{\nu}
-\pfrac{\tilde{\HCd}_{0}}{a_{\nu}}a_{\mu}+\pfrac{\tilde{\HCd}_{0}}{\tilde{p}^{\,\mu\beta}}\tilde{p}^{\nu\beta}
+\pfrac{\tilde{\HCd}_{0}}{\tilde{p}^{\,\beta\mu}}\tilde{p}^{\,\beta\nu}
-2\pfrac{\tilde{\HCd}_{0}}{g_{\beta\nu}}g_{\beta\mu}
&\equiv\delta_{\mu}^{\nu}\left(\pfrac{\tilde{\HCd}_{0}}{\tilde{\pi}^{\beta}}\tilde{\pi}^{\beta}
+\pfrac{\tilde{\HCd}_{0}}{\tilde{p}^{\alpha\beta}}\tilde{p}^{\alpha\beta}-\tilde{\HCd}_{0}\right)\nonumber\\
-2\pfrac{\tilde{\HCd}_{\mathrm{Gr}}}{g_{\beta\nu}}g_{\beta\mu}
+2\pfrac{\tilde{\HCd}_{\mathrm{Gr}}}{\tilde{k}^{\beta\mu\alpha}}\tilde{k}^{\beta\nu\alpha}
+\pfrac{\tilde{\HCd}_{\mathrm{Gr}}}{\tilde{k}^{\beta\alpha\mu}}\tilde{k}^{\beta\alpha\nu}
-\pfrac{\tilde{\HCd}_{\mathrm{Gr}}}{\tilde{q}\indices{_{\nu}^{\alpha\beta\lambda}}}\tilde{q}\indices{_{\mu}^{\alpha\beta\lambda}}
+\pfrac{\tilde{\HCd}_{\mathrm{Gr}}}{\tilde{q}\indices{_{\alpha}^{\mu\beta\lambda}}}\tilde{q}\indices{_{\alpha}^{\nu\beta\lambda}}
+2\pfrac{\tilde{\HCd}_{\mathrm{Gr}}}{\tilde{q}\indices{_{\alpha}^{\beta\lambda\mu}}}\tilde{q}\indices{_{\alpha}^{\beta\lambda\nu}}
&\equiv\delta_{\mu}^{\nu}\left(\pfrac{\tilde{\HCd}_{\mathrm{Gr}}}{\tilde{k}^{\beta\alpha\lambda}}\tilde{k}^{\beta\alpha\lambda}
+\pfrac{\tilde{\HCd}_{\mathrm{Gr}}}{\tilde{q}\indices{_{\alpha}^{\beta\lambda\xi}}}\tilde{q}\indices{_{\alpha}^{\beta\lambda\xi}}-\tilde{\HCd}_{\mathrm{Gr}}\right).
\end{align*}
The consistency equation~(\ref{eq:consistency2}) thus has the equivalent representation:
\begin{equation}\label{eq:consistency2b}
\pfrac{\tilde\HCd_{0}}{\tilde\pi^{\mu}}\tilde\pi^{\nu}+\pfrac{\tilde\HCd_{0}}{\tilde p^{\,\beta\mu}}\tilde p^{\,\beta\nu}
-\delta_{\mu}^{\nu}\left(\pfrac{\tilde\HCd_{0}}{\tilde\pi^{\alpha}}\tilde\pi^{\alpha}+\pfrac{\tilde\HCd_{0}}{\tilde p^{\alpha\beta}}\tilde p^{\alpha\beta}-\tilde\HCd_{0}\right)
=-\pfrac{\tilde\HCd_{\mathrm{Gr}}}{\tilde k^{\beta\alpha\mu}}\tilde k^{\beta\alpha\nu}
-2\pfrac{\tilde\HCd_{\mathrm{Gr}}}{\tilde q\indices{_{\alpha}^{\beta\lambda\mu}}}\tilde q\indices{_{\alpha}^{\beta\lambda\nu}}
+\delta_{\mu}^{\nu}\left(\pfrac{\tilde\HCd_{\mathrm{Gr}}}{\tilde k^{\beta\alpha\lambda}}\tilde k^{\beta\alpha\lambda}
+\pfrac{\tilde\HCd_{\mathrm{Gr}}}{\tilde q\indices{_{\alpha}^{\beta\lambda\xi}}}\tilde q\indices{_{\alpha}^{\beta\lambda\xi}}-\tilde\HCd_{\mathrm{Gr}}\right).
\end{equation}
We observe that the left-hand side is exactly the \emph{canonical} energy-momentum tensor $\tilde\theta\indices{_\mu^\nu}$ associated with the Hamiltonian $\tilde\HCd_{0}$
\begin{equation*}
\tilde\theta\indices{_\mu^\nu}:=\pfrac{\tilde\HCd_{0}}{\tilde\pi^{\mu}}\tilde\pi^{\nu}+\pfrac{\tilde\HCd_{0}}{\tilde p^{\,\beta\mu}}\tilde p^{\,\beta\nu}
-\delta_{\mu}^{\nu}\left(\pfrac{\tilde\HCd_{0}}{\tilde\pi^{\alpha}}\tilde\pi^{\alpha}+\pfrac{\tilde\HCd_{0}}{\tilde p^{\alpha\beta}}\tilde p^{\alpha\beta}-\tilde\HCd_{0}\right).
\end{equation*}
If the consistency equation is interpreted as an energy-momentum balance equation, the right-hand side defines the
energy-momentum tensor $\tilde\vartheta\indices{_\mu^\nu}$ that is associated with the Hamiltonian $\tilde\HCd_{\mathrm{Gr}}$:
\begin{equation*}
\tilde\vartheta\indices{_\mu^\nu}:=\pfrac{\tilde\HCd_{\mathrm{Gr}}}{\tilde k^{\beta\alpha\mu}}\tilde k^{\beta\alpha\nu}
+2\pfrac{\tilde\HCd_{\mathrm{Gr}}}{\tilde q\indices{_{\alpha}^{\beta\lambda\mu}}}\tilde q\indices{_{\alpha}^{\beta\lambda\nu}}
-\delta_{\mu}^{\nu}\left(\pfrac{\tilde\HCd_{\mathrm{Gr}}}{\tilde k^{\beta\alpha\lambda}}\tilde k^{\beta\alpha\lambda}
+\pfrac{\tilde\HCd_{\mathrm{Gr}}}{\tilde q\indices{_{\alpha}^{\beta\lambda\xi}}}\tilde q\indices{_{\alpha}^{\beta\lambda\xi}}-\tilde\HCd_{\mathrm{Gr}}\right),
\end{equation*}
so that the consistency equation takes on the concise form:
\begin{equation}\label{eq:ce-concise}
\tilde\theta\indices{_\mu^\nu}+\tilde\vartheta\indices{_\mu^\nu}=0.
\end{equation}
The value of the covariant derivative of~(\ref{eq:ce-concise}), taken along the system path satisfying the canonical equations~(\ref{eq:feqs-Hdyn-1}) should be zero as well.
To verify this, we calculate explicitly
\begin{align*}
\tilde\theta\indices{_\mu^\alpha_;_\alpha}&=\left(\pfrac{\tilde\HCd_{0}}{\tilde\pi^{\mu}}\right)_{;\alpha}\tilde\pi^{\alpha}
+\pfrac{\tilde\HCd_{0}}{\tilde\pi^{\mu}}\tilde\pi\indices{^\alpha_;_\alpha}
+\left(\pfrac{\tilde\HCd_{0}}{\tilde p^{\,\beta\mu}}\right)_{;\alpha}\tilde p^{\,\beta\alpha}+\pfrac{\tilde\HCd_{0}}{\tilde p^{\,\beta\mu}}\tilde p\indices{^{\,\beta\alpha}_;_\alpha}
-\left(\pfrac{\tilde\HCd_{0}}{\tilde\pi^{\alpha}}\right)_{;\mu}\tilde\pi^{\alpha}-\cancel{\pfrac{\tilde\HCd_{0}}{\tilde\pi^{\alpha}}\tilde\pi\indices{^\alpha_;_\mu}}
-\left(\pfrac{\tilde\HCd_{0}}{\tilde p^{\beta\alpha}}\right)_{;\mu}\tilde p^{\beta\alpha}-\bcancel{\pfrac{\tilde\HCd_{0}}{\tilde p^{\alpha\beta}}\tilde p\indices{^\alpha^\beta_;_\mu}}\\
&\quad+\cancel{\pfrac{\tilde\HCd_{0}}{\tilde\pi^{\alpha}}\tilde\pi\indices{^\alpha_;_\mu}}+\pfrac{\tilde\HCd_{0}}{\phi}\phi_{;\mu}
+\bcancel{\pfrac{\tilde\HCd_{0}}{\tilde p^{\alpha\beta}}\tilde p\indices{^{\alpha\beta}_;_\mu}}+\pfrac{\tilde\HCd_{0}}{a_\beta}a_{\beta;\mu}
+\pfrac{\tilde\HCd_{0}}{g_{\beta\alpha}}g_{\beta\alpha;\mu},
\end{align*}
which simplifies along the system path determined by the canonical field equations~(\ref{eq:feqs-Hdyn-1})
\begin{align}
\tilde\theta\indices{_\mu^\alpha_;_\alpha}&=\left(\phi_{;\mu;\alpha}-\phi_{;\alpha;\mu}+2\phi_{;\mu} s\indices{^\xi_\alpha_\xi}\right)\tilde\pi^{\alpha}
+\left(a_{\beta;\mu;\alpha}-a_{\beta;\alpha;\mu}+2a_{\beta;\mu}s\indices{^\xi_\alpha_\xi}\right)\tilde p^{\beta\alpha}
+\pfrac{\tilde\HCd_{0}}{g_{\beta\alpha}}g_{\beta\alpha;\mu}\nonumber\\
&=2\left(\phi_{;\xi}s\indices{^\xi_\alpha_\mu}+\phi_{;\mu}s\indices{^\xi_\alpha_\xi}\right)\tilde\pi^\alpha
+2\left(a_{\beta;\xi}s\indices{^\xi_\alpha_\mu}+a_{\beta;\mu}s\indices{^\xi_\alpha_\xi}\right)\tilde p^{\,\beta\alpha}
+\pfrac{\tilde\HCd_{0}}{g_{\beta\alpha}}g_{\beta\alpha;\mu}-R\indices{^\xi_\beta_\alpha_\mu}\tilde p^{\,\beta\alpha}a_\xi.
\label{eq:emt0deri}
\end{align}
The scalar field interacts with the torsion of spacetime whereas the vector field interacts with both torsion and curvature.
As required, the divergence of the canonical energy-momentum tensor $\tilde\theta\indices{_\mu^\nu}$ vanishes in a flat spacetime.

The covariant divergence of $\tilde\vartheta\indices{_\mu^\nu}$ follows as:
\begin{align*}
\tilde\vartheta\indices{_\mu^\alpha_;_\alpha}&=\left(\pfrac{\tilde\HCd_{\mathrm{Gr}}}{\tilde k^{\beta\alpha\mu}}\right)_{;\nu}\tilde k^{\beta\alpha\nu}
+\pfrac{\tilde\HCd_{\mathrm{Gr}}}{\tilde k^{\beta\alpha\mu}}\tilde k\indices{^{\beta\alpha\nu}_;_\nu}
+2\left(\pfrac{\tilde\HCd_{\mathrm{Gr}}}{\tilde q\indices{_{\alpha}^{\beta\lambda\mu}}}\right)_{;\nu}\tilde q\indices{_{\alpha}^{\beta\lambda\nu}}
+2\pfrac{\tilde\HCd_{\mathrm{Gr}}}{\tilde q\indices{_{\alpha}^{\beta\lambda\mu}}}\tilde q\indices{_{\alpha}^{\beta\lambda\nu}_;_\nu}\\
&\quad-\left(\pfrac{\tilde\HCd_{\mathrm{Gr}}}{\tilde k^{\beta\alpha\nu}}\right)_{;\mu}\tilde k^{\beta\alpha\nu}
-\cancel{\pfrac{\tilde\HCd_{\mathrm{Gr}}}{\tilde k^{\beta\alpha\lambda}}\tilde k\indices{^{\beta\alpha\lambda}_;_\mu}}
-\left(\pfrac{\tilde\HCd_{\mathrm{Gr}}}{\tilde q\indices{_{\alpha}^{\beta\lambda\nu}}}\right)_{;\mu}\tilde q\indices{_{\alpha}^{\beta\lambda\nu}}
-\bcancel{\pfrac{\tilde\HCd_{\mathrm{Gr}}}{\tilde q\indices{_{\alpha}^{\beta\lambda\xi}}}\tilde q\indices{_{\alpha}^{\beta\lambda\xi}_;_\mu}}
+\cancel{\pfrac{\tilde\HCd_{\mathrm{Gr}}}{\tilde k^{\beta\alpha\lambda}}\tilde k\indices{^{\beta\alpha\lambda}_;_\mu}}
+\bcancel{\pfrac{\tilde\HCd_{\mathrm{Gr}}}{\tilde q\indices{_{\alpha}^{\beta\lambda\xi}}}\tilde q\indices{_{\alpha}^{\beta\lambda\xi}_;_\mu}}
+\pfrac{\tilde\HCd_{\mathrm{Gr}}}{g_{\beta\alpha}}g_{\beta\alpha;\mu}.
\end{align*}
Inserting the canonical field equations yields, considering that $k^{\beta\alpha\nu}$ is symmetric in $\beta,\alpha$:
\begin{align}
\tilde\vartheta\indices{_\mu^\alpha_;_\alpha}&=\left(g_{\beta\alpha;\mu;\nu}-g_{\beta\alpha;\nu;\mu}\right)\tilde k^{\beta\alpha\nu}
-\left(\pfrac{\tilde\HCd_0}{g_{\beta\alpha}}+\cancel{\pfrac{\tilde\HCd_{\mathrm{Gr}}}{g_{\beta\alpha}}}-2\tilde k^{\beta\alpha\nu}s\indices{^\xi_\nu_\xi}\right)g_{\beta\alpha;\mu}
+\left(\onehalf R\indices{^\alpha_{\beta\lambda\nu;\mu}}-R\indices{^\alpha_{\beta\lambda\mu;\nu}}\right)\tilde q\indices{_{\alpha}^{\beta\lambda\nu}}
+\cancel{\pfrac{\tilde\HCd_{\mathrm{Gr}}}{g_{\beta\alpha}}g_{\beta\alpha;\mu}}\nonumber\\
&\quad+R\indices{^\alpha_{\beta\lambda\mu}}\left(\tilde p^{\,\beta\lambda}a_\alpha+2\tilde k^{\nu\beta\lambda}g_{\nu\alpha}
-\tilde q\indices{_\alpha^{\beta\xi\nu}}s\indices{^\lambda_\xi_\nu}-2\tilde q\indices{_\alpha^{\beta\lambda\nu}}s\indices{^\xi_\nu_\xi}\right)\nonumber\\
&=\left(g_{\beta\alpha;\mu;\nu}-g_{\beta\alpha;\nu;\mu}+2g_{\beta\alpha;\mu}s\indices{^\xi_\nu_\xi}+2R_{(\beta\alpha)\nu\mu}\right)\tilde k^{\beta\alpha\nu}
-\pfrac{\tilde\HCd_0}{g_{\beta\alpha}}g_{\beta\alpha;\mu}
-\left(\cancel{s\indices{^\xi_\lambda_\nu}R\indices{^\alpha_{\beta\mu\xi}}}+s\indices{^\xi_\nu_\mu}R\indices{^\alpha_{\beta\lambda\xi}}
+s\indices{^\xi_\mu_\lambda}R\indices{^\alpha_{\beta\nu\xi}}\right)\tilde q\indices{_{\alpha}^{\beta\lambda\nu}}\nonumber\\
&\quad+R\indices{^\alpha_{\beta\lambda\mu}}\tilde p^{\,\beta\lambda}a_\alpha
+\cancel{R\indices{^\alpha_{\beta\mu\xi}}\tilde q\indices{_\alpha^{\beta\lambda\nu}}s\indices{^\xi_\lambda_\nu}}
-2R\indices{^\alpha_{\beta\lambda\mu}}\tilde q\indices{_\alpha^{\beta\lambda\nu}}s\indices{^\xi_\nu_\xi}.
\label{eq:emtgrderi0}
\end{align}
The covariant derivative terms of the Riemann-Cartan tensor were converted according to
\begin{align}
&\quad\left(\onehalf R\indices{^\alpha_{\beta\lambda\nu;\mu}}-R\indices{^\alpha_{\beta\lambda\mu;\nu}}\right)\tilde q\indices{_{\alpha}^{\beta\lambda\nu}}\nonumber\\
&=\onehalf\left(R\indices{^\alpha_{\beta\lambda\nu;\mu}}-R\indices{^\alpha_{\beta\lambda\mu;\nu}}+R\indices{^\alpha_{\beta\mu\lambda;\nu}}\right)
\tilde q\indices{_{\alpha}^{\beta\lambda\nu}}\nonumber\\
&=\onehalf\left(R\indices{^\alpha_{\beta\lambda\nu;\mu}}+R\indices{^\alpha_{\beta\nu\mu;\lambda}}+R\indices{^\alpha_{\beta\mu\lambda;\nu}}\right)
\tilde q\indices{_{\alpha}^{\beta\lambda\nu}}\nonumber\\
&=-\left(s\indices{^\xi_\lambda_\nu}R\indices{^\alpha_{\beta\mu\xi}}+s\indices{^\xi_\nu_\mu}R\indices{^\alpha_{\beta\lambda\xi}}
+s\indices{^\xi_\mu_\lambda}R\indices{^\alpha_{\beta\nu\xi}}\right)\tilde q\indices{_{\alpha}^{\beta\lambda\nu}},
\label{eq:cd-RC}
\end{align}
making use of the skew-symmetry of $\tilde q$ and $R$ in their last index pairs and of the Bianchi identity for the covariant derivative
of the Riemann-Cartan tensor in spaces with torsion and non-metricity.

For non-metricity $R_{(\beta\alpha)\nu\mu}$ does not vanish as the Riemann-Cartan tensor is then no longer skew-symmetric in its first index pair.
By direct calculation of the commutator of the second partial derivatives of the metric $g_{\beta\alpha}$ one finds:
\begin{equation}\label{eq:comm-cd-metr}
g_{\beta\alpha;\mu;\nu}-g_{\beta\alpha;\nu;\mu}=-2R_{(\beta\alpha)\nu\mu}+2g_{\beta\alpha;\xi}s\indices{^\xi_\nu_\mu}.
\end{equation}
Equation~(\ref{eq:emtgrderi0}) thus further simplifies since the term proportional to $R_{(\beta\alpha)\nu\mu}$ cancels:
\begin{equation}
\tilde\vartheta\indices{_\mu^\alpha_;_\alpha}=2\left(g_{\beta\lambda;\xi}s\indices{^\xi_\nu_\mu}+g_{\beta\lambda;\mu}s\indices{^\xi_\nu_\xi}\right)\tilde k^{\,\beta\lambda\nu}
-2\left(R\indices{^\alpha_{\beta\nu\xi}}s\indices{^\xi_\mu_\lambda}
+R\indices{^\alpha_{\beta\lambda\mu}}s\indices{^\xi_\nu_\xi}\right)\tilde q\indices{_{\alpha}^{\beta\lambda\nu}}
-\pfrac{\tilde\HCd_{0}}{g_{\beta\alpha}}g_{\beta\alpha;\mu}+R\indices{^\alpha_{\beta\lambda\mu}}\tilde p^{\,\beta\lambda}a_\alpha.
\label{eq:emtgrderi}
\end{equation}
Calculating now the sum of the covariant derivatives of the energy-momentum tensors~(\ref{eq:emt0deri}) and~(\ref{eq:emtgrderi})
yields the desired covariant derivative of the consistency equation:
\begin{align*}
\onehalf\left(\tilde\theta\indices{_\mu^\alpha_;_\alpha}+\tilde\vartheta\indices{_\mu^\alpha_;_\alpha}\right)
&=\left(\phi_{;\xi}s\indices{^\xi_\alpha_\mu}+\phi_{;\mu}s\indices{^\xi_\alpha_\xi}\right)\tilde\pi^\alpha
+\left(a_{\beta;\xi}s\indices{^\xi_\alpha_\mu}+a_{\beta;\mu}s\indices{^\xi_\alpha_\xi}\right)\tilde p^{\,\beta\alpha}
+\left(g_{\lambda\beta;\xi}s\indices{^\xi_\alpha_\mu}+g_{\lambda\beta;\mu}s\indices{^\xi_\alpha_\xi}\right)\tilde k^{\lambda\beta\alpha}
-\left(R\indices{^\tau_{\lambda\beta\xi}}s\indices{^\xi_\alpha_\mu}
+R\indices{^\tau_{\lambda\beta\mu}}s\indices{^\xi_\alpha_\xi}\right)\tilde q\indices{_{\tau}^{\lambda\beta\alpha}}\\
&=\left(\phi_{;\xi}\tilde\pi^\alpha+a_{\beta;\xi}\tilde p^{\,\beta\alpha}+g_{\lambda\beta;\xi}\tilde k^{\lambda\beta\alpha}
-R\indices{^\tau_{\lambda\beta\xi}}\tilde q\indices{_{\tau}^{\lambda\beta\alpha}}\right)s\indices{^\xi_\alpha_\mu}
+\left(\phi_{;\mu}\tilde\pi^\alpha+a_{\beta;\mu}\tilde p^{\,\beta\alpha}+g_{\lambda\beta;\mu}\tilde k^{\lambda\beta\alpha}
-R\indices{^\tau_{\lambda\beta\mu}}\tilde q\indices{_{\tau}^{\lambda\beta\alpha}}\right)s\indices{^\xi_\alpha_\xi}.
\end{align*}
Remarkably, both, the derivative of $\tilde\HCd_0$ with respect to the metric $g_{\beta\alpha}$ as well as
the interaction term $R\indices{^\alpha_{\beta\lambda\mu}}\tilde p^{\,\beta\lambda}a_\alpha$ of vector field and curvature cancel.

Inserting back canonical field equations, the covariant divergence of the consistency equation is expressed
in terms of the derivatives of the Hamiltonians $\tilde\HCd_0$ and $\tilde\HCd_{\mathrm{Gr}}$
\begin{align*}
\onehalf\left(\tilde\theta\indices{_\mu^\alpha_;_\alpha}+\tilde\vartheta\indices{_\mu^\alpha_;_\alpha}\right)
&=\left(\pfrac{\tilde\HCd_{0}}{\tilde\pi^\xi}\tilde\pi^\alpha+\pfrac{\tilde\HCd_0}{\tilde p^{\,\beta\xi}}\tilde p^{\beta\alpha}
+\pfrac{\tilde\HCd_{\mathrm{Gr}}}{\tilde k^{\lambda\beta\xi}}\tilde k^{\lambda\beta\alpha}
+2\pfrac{\tilde\HCd_{\mathrm{Gr}}}{\tilde q\indices{_\tau^{\lambda\beta\xi}}}\tilde q\indices{_\tau^{\lambda\beta\alpha}}\right)s\indices{^\xi_\alpha_\mu}\\
&\,+\left(\pfrac{\tilde\HCd_{0}}{\tilde\pi^\mu}\tilde\pi^\alpha+\pfrac{\tilde\HCd_0}{\tilde p^{\,\beta\mu}}\tilde p^{\beta\alpha}
+\pfrac{\tilde\HCd_{\mathrm{Gr}}}{\tilde k^{\lambda\beta\mu}}\tilde k^{\lambda\beta\alpha}
+2\pfrac{\tilde\HCd_{\mathrm{Gr}}}{\tilde q\indices{_\tau^{\lambda\beta\mu}}}\tilde q\indices{_\tau^{\lambda\beta\alpha}}\right)s\indices{^\xi_\alpha_\xi}.
\end{align*}
Inserting finally the consistency equation~(\ref{eq:consistency2b}) for the terms in parentheses yields due to the skew-symmetry of the torsion tensor
\begin{align*}
\onehalf\left(\tilde\theta\indices{_\mu^\alpha_;_\alpha}+\tilde\vartheta\indices{_\mu^\alpha_;_\alpha}\right)
&=\left(\pfrac{\tilde\HCd_{0}}{\tilde\pi^{\alpha}}\tilde\pi^{\alpha}+\pfrac{\tilde\HCd_{0}}{\tilde p^{\alpha\beta}}\tilde p^{\alpha\beta}-\tilde\HCd_{0}
+\pfrac{\tilde\HCd_{\mathrm{Gr}}}{\tilde k^{\beta\alpha\lambda}}\tilde k^{\beta\alpha\lambda}
+\pfrac{\tilde\HCd_{\mathrm{Gr}}}{\tilde q\indices{_{\alpha}^{\beta\lambda\xi}}}\tilde q\indices{_{\alpha}^{\beta\lambda\xi}}-\tilde\HCd_{\mathrm{Gr}}\right)s\indices{^\xi_\xi_\mu}\\
&\,+\left(\pfrac{\tilde\HCd_{0}}{\tilde\pi^{\alpha}}\tilde\pi^{\alpha}+\pfrac{\tilde\HCd_{0}}{\tilde p^{\alpha\beta}}\tilde p^{\alpha\beta}-\tilde\HCd_{0}
+\pfrac{\tilde\HCd_{\mathrm{Gr}}}{\tilde k^{\beta\alpha\lambda}}\tilde k^{\beta\alpha\lambda}
+\pfrac{\tilde\HCd_{\mathrm{Gr}}}{\tilde q\indices{_{\alpha}^{\beta\lambda\xi}}}\tilde q\indices{_{\alpha}^{\beta\lambda\xi}}-\tilde\HCd_{\mathrm{Gr}}\right)s\indices{^\xi_\mu_\xi}\\
&=0.
\end{align*}
As a consequence, it is obvious that the consistency equation~(\ref{eq:ce-concise}) is covariantly conserved
along the system path given by the solution of the canonical field equations for \emph{arbitrary} torsion and non-metricity
$g_{\lambda\beta;\mu}\neq0$, independently of the specific form of the Hamiltonians $\tilde\HCd_0$ and $\tilde\HCd_{\mathrm{Gr}}$:
\begin{equation}\label{eq:ce-concise-div}
\tilde\theta\indices{_\mu^\alpha}+\tilde\vartheta\indices{_\mu^\alpha}=0\qquad\Rightarrow\qquad
\left(\tilde\theta\indices{_\mu^\alpha}+\tilde\vartheta\indices{_\mu^\alpha}\right)_{;\alpha}=0.
\end{equation}
It follows that the covariant divergences $\tilde\theta\indices{_\mu^\alpha_;_\alpha}$ and $\tilde\vartheta\indices{_\mu^\alpha_;_\alpha}$
of the \emph{individual} energy-momentum tensors along the system path do not vanish in general,
but only their sum---even for the case of metricity and zero torsion if vector fields are present.
\section{LAGRANGIAN DESCRIPTION}
In order to derive the Lagrangian representation of the consistency equation~(\ref{eq:consistency}),
the Lagrangians $\tilde{\LCd}_0$ and $\tilde{\LCd}_{\mathrm{Gr}}$ are determined as the respective Legendre transforms
of $\tilde{\HCd}_0$ and $\tilde{\HCd}_{\mathrm{Gr}}$ from the ``gauged'' action~(\ref{eq:action-integral4})
and the canonical field equations~(\ref{eq:feqs-Hdyn-1}) with respect to the momenta:
\begin{align*}
\tilde{\LCd}_{0}&=\tilde{\pi}^\beta\,\phi_{;\,\beta}+\tilde{p}^{\nu\beta}\,a_{\nu;\,\beta}-\tilde{\HCd}_{0},&
\phi_{;\,\beta}&=\pfrac{\tilde{\HCd}_{0}}{\tilde{\pi}^{\beta}},\qquad\qquad\! a_{\nu;\,\beta}=\pfrac{\tilde{\HCd}_{0}}{\tilde{p}^{\nu\beta}}\\
\tilde{\LCd}_{\mathrm{Gr}}&=\tilde{k}^{\lambda\nu\beta}\,g_{\lambda\nu;\,\beta}
-\onehalf\tilde{q}\indices{_\eta^{\nu\lambda\beta}}\,R\indices{^\eta_{\nu\lambda\beta}}-\tilde{\HCd}_{\mathrm{Gr}},&
g_{\lambda\nu;\,\beta}&=\pfrac{\tilde{\HCd}_{\mathrm{Gr}}}{\tilde{k}\indices{^{\lambda\nu\beta}}},\qquad
-\frac{R\indices{^{\eta}_{\nu\lambda\beta}}}{2}=\pfrac{\tilde{\HCd}_{\mathrm{Gr}}}{\tilde{q}\indices{_{\eta}^{\nu\lambda\beta}}}.
\end{align*}
The derivatives of $\tilde{\HCd}$ and $\tilde{\LCd}$ with respect to the fields differ by the sign
\begin{equation*}
-\pfrac{\tilde{\HCd}_{0}}{a_{\nu}}=\pfrac{\tilde{\LCd}_{0}}{a_{\nu}},\qquad
-\pfrac{\tilde{\HCd}}{g_{\lambda\nu}}g_{\lambda\mu}=\pfrac{\tilde{\LCd}}{g_{\lambda\nu}}g_{\lambda\mu}
=-\pfrac{\tilde{\LCd}}{g^{\lambda\mu}}g^{\lambda\nu}.
\end{equation*}
With the momentum fields as the duals of the corresponding canonical equations,
\begin{equation*}
\tilde{p}^{\nu\beta}=\pfrac{\tilde{\LCd}_{0}}{a_{\nu;\,\beta}},
\qquad\tilde{k}\indices{^{\lambda\nu\beta}}=\pfrac{\tilde{\LCd}_{\mathrm{Gr}}}{g_{\lambda\nu;\,\beta}},\qquad
\tilde{q}\indices{_\eta^{\nu\lambda\beta}}=-2\pfrac{\tilde{\LCd}_{\mathrm{Gr}}}{R\indices{^{\eta}_{\nu\lambda\beta}}},
\end{equation*}
the consistency equation~(\ref{eq:consistency}), resp.~(\ref{eq:consistency2}) acquires the equivalent Lagrangian representation:
\begin{equation}\label{eq:consistency2a}
\pfrac{\tilde{\LCd}_{0}}{a_{\nu}}a_{\mu}+\pfrac{\tilde{\LCd}_{0}}{a_{\nu;\beta}}a_{\mu;\beta}
-2\pfrac{\tilde{\LCd}_{0}}{g^{\lambda\mu}}g^{\lambda\nu}=2\pfrac{\tilde{\LCd}_{\mathrm{Gr}}}{g^{\lambda\mu}}g^{\lambda\nu}
-2\pfrac{\tilde{\LCd}_{\mathrm{Gr}}}{g_{\lambda\nu;\beta}}g_{\lambda\mu;\beta}
-\pfrac{\tilde{\LCd}_{\mathrm{Gr}}}{R\indices{^{\eta}_{\nu\lambda\beta}}}R\indices{^{\eta}_{\mu\lambda\beta}}
+\pfrac{\tilde{\LCd}_{\mathrm{Gr}}}{R\indices{^{\mu}_{\eta\lambda\beta}}}R\indices{^{\nu}_{\eta\lambda\beta}}.
\end{equation}
Note that the \emph{covariant} derivatives of the vector field $a_\nu$ and the metric $g_{\lambda\nu}$
actually constitute the dual quantities of the respective canonical momenta,
$\tilde{p}^{\nu\beta}$ and $\tilde{k}^{\lambda\nu\beta}$.
Remarkably, minus one-half the Riemann-Cartan tensor $R\indices{^\eta_\nu_\lambda_\beta}$ turns up here as the replacement
of the non-existing covariant derivative of the connection field $\gamma\indices{^\eta_\nu_\lambda}$,
with $\tilde{q}\indices{_\eta^{\nu\lambda\beta}}$ the pertaining dual field.

The consistency equation in the form of Eq.~(\ref{eq:consistency2a}) exhibits a clear interpretation of its components.
The two derivatives with respect to the metric represent the classical Einstein approach to vary the Lagrangians of
gravity and matter fields with respect to the metric.
The derivatives of $\tilde{\LCd}_{0}$ with respect to the vector field and its covariant derivative emerge from
the generalization of the theory to also include vector fields in place of the restriction to scalar matter fields.
On the side of the gravitational Lagrangian $\tilde{\LCd}_{\mathrm{Gr}}$, the term proportional to the covariant
derivative of the metric stands for the generalization to ``non-metricity'', hence a not covariantly conserved metric.
Finally, the difference of the terms proportional to the Riemann tensor can be shown on the basis of~(\ref{eq:H-dyn-post}) to vanish
in the case of a torsion-free spacetime, and hence serves to include torsion effects into the generalized description of gravity.

In the Lagrangian representation, the identities for scalar density valued functions of tensors and the metric~\cite{struckvasvenn21},
follow for the Lagrangians $\tilde{\LCd}_{0}$ and $\tilde{\LCd}_{\mathrm{Gr}}$ as:
\begin{align*}
\pfrac{\tilde{\LCd}_{0}}{\left(\pfrac{\phi}{x^{\nu}}\right)}\pfrac{\phi}{x^{\mu}}
+\pfrac{\tilde{\LCd}_{0}}{a_{\nu}}a_{\mu}+\pfrac{\tilde{\LCd}_{0}}{a_{\nu;\beta}}a_{\mu;\beta}
+\pfrac{\tilde{\LCd}_{0}}{a_{\beta;\nu}}a_{\beta;\mu}
-2\pfrac{\tilde{\LCd}_{0}}{g^{\lambda\mu}}g^{\lambda\nu}&\equiv\delta_{\mu}^{\nu}\tilde{\LCd}_{0}\nonumber\\
-2\pfrac{\tilde{\LCd}_{\mathrm{Gr}}}{g^{\lambda\mu}}g^{\lambda\nu}+
2\pfrac{\tilde{\LCd}_{\mathrm{Gr}}}{g_{\lambda\nu;\beta}}g_{\lambda\mu;\beta}
+\pfrac{\tilde{\LCd}_{\mathrm{Gr}}}{g_{\lambda\beta;\nu}}g_{\lambda\beta;\mu}
-\pfrac{\tilde{\LCd}_{\mathrm{Gr}}}{R\indices{^{\mu}_{\eta\lambda\beta}}}R\indices{^{\nu}_{\eta\lambda\beta}}
+\pfrac{\tilde{\LCd}_{\mathrm{Gr}}}{R\indices{^{\eta}_{\nu\lambda\beta}}}R\indices{^{\eta}_{\mu\lambda\beta}}
+2\pfrac{\tilde{\LCd}_{\mathrm{Gr}}}{R\indices{^{\eta}_{\lambda\beta\nu}}}R\indices{^{\eta}_{\lambda\beta\mu}}
&\equiv\delta_{\mu}^{\nu}\tilde{\LCd}_{\mathrm{Gr}}.
\end{align*}
The ``Lagrangian'' consistency equation~(\ref{eq:consistency2a}) thus has the equivalent tensor representation:
\begin{equation}\label{eq:consistency3}
-\left(\pfrac{\tilde\LCd_{0}}{\left(\pfrac{\phi}{x^{\nu}}\right)}\pfrac{\phi}{x^{\mu}}
+\pfrac{\tilde\LCd_{0}}{a_{\beta;\nu}}a_{\beta;\mu}-\delta_{\mu}^{\nu}\tilde\LCd_{0}\right)
=\pfrac{\tilde\LCd_{\mathrm{Gr}}}{g_{\lambda\beta;\nu}}g_{\lambda\beta;\mu}
+2\pfrac{\tilde\LCd_{\mathrm{Gr}}}{R\indices{^{\eta}_{\lambda\beta\nu}}}R\indices{^{\eta}_{\lambda\beta\mu}}
-\delta_{\mu}^{\nu}\tilde\LCd_{\mathrm{Gr}}.
\end{equation}
\end{widetext}
After dividing by $\sqrt{-g}\,$, the left-hand side of Eq.~(\ref{eq:consistency3}) is minus the covariant version
of the \emph{canonical} energy-momentum tensor $\theta\indices{_{\mu}^{\nu}}$ of the Lagrangian system $\LCd_{0}$
of scalar and vector fields.
In a \emph{flat} spacetime this tensor is given by:
\begin{equation*}
\left.\theta\indices{_\mu^\nu}\right|_{\text{flat}}=\pfrac{\LCd_{0,\mathrm{flat}}}{\left(\pfrac{\phi}{x^\nu}\right)}\pfrac{\phi}{x^\mu}
+\pfrac{\LCd_{0,\mathrm{flat}}}{\left(\pfrac{a_\beta}{x^\nu}\right)}\pfrac{a_\beta}{x^\mu}-\delta_\mu^\nu\LCd_{0,\mathrm{flat}}.
\end{equation*}
However, in a \emph{curved} spacetime this is not a tensor!
Remarkably, the canonical gauge formalism yields this expression with
the partial derivatives of the vector field replaced by covariant derivatives.
The consistency equation~(\ref{eq:consistency3}) thus requires to define the
energy-momentum tensor of $\LCd_{0}$ as
\begin{equation*}
\theta\indices{_\mu^\nu}=\pfrac{\LCd_0}{\left(\pfrac{\phi}{x^\nu}\right)}\pfrac{\phi}{x^\mu}
+\pfrac{\LCd_0}{a_{\beta;\nu}}a_{\beta;\mu}-\delta_\mu^\nu\LCd_0,
\end{equation*}
which obviously has tensor property in a curved spacetime.
The covariant derivative of the vector field induces the connection in addition to the metric to act as a coupling quantity.

Provided that right-hand side of Eq.~(\ref{eq:consistency3}) is interpreted as energy-momentum tensor $\vartheta\indices{_\mu^\nu}$
that is associated with the Lagrangian $\LCd_{\mathrm{Gr}}$ of the free gravitational field, hence
\begin{equation*}
\vartheta\indices{_\mu^\nu}=\pfrac{\LCd_{\mathrm{Gr}}}{g_{\lambda\beta;\nu}}g_{\lambda\beta;\mu}+
2\pfrac{\LCd_{\mathrm{Gr}}}{R\indices{^{\eta}_{\lambda\beta\nu}}}R\indices{^{\eta}_{\lambda\beta\mu}}
-\delta_{\mu}^{\nu}\LCd_{\mathrm{Gr}},
\end{equation*}
then the consistency equation represents the energy-momen\-tum balance equation
\begin{equation}\label{eq:consistency4}
\vartheta\indices{_\mu^\nu}+\theta\indices{_{\mu}^{\nu}}=0,
\end{equation}
which conforms for the particular definitions of $\vartheta\indices{_\mu^\nu}$ and $\theta\indices{_{\mu}^{\nu}}$ with the conjecture
of a zero-energy universe~\cite{jordan39,sciama53,feynman62,hawking03} that holds for non-zero torsion as well as non-metricity.
For metric compatibility, hence for $g_{\lambda\beta;\mu}=0$, Eq.~(\ref{eq:consistency4}) reduces to the Einstein-type equation:
\begin{equation}\label{eq:consistency5}
2\pfrac{\LCd_{\mathrm{Gr}}}{R\indices{^{\eta}_{\lambda\beta\nu}}}R\indices{^{\eta}_{\lambda\beta\mu}}
-\delta_{\mu}^{\nu}\LCd_{\mathrm{Gr}}=-\theta\indices{_{\mu}^{\nu}}.
\end{equation}
For the Hilbert Lagrangian $\LCd_{\mathrm{Gr,H}}=R/16\pi G$ as a particular model Lagrangian for the free gravitational field,
one immediately encounters the form of the classical Einstein equation
\begin{equation*}
R\indices{^\nu_\mu}-\onehalf\delta_{\mu}^{\nu}R=-8\pi\,G\,\theta\indices{_{\mu}^{\nu}},
\end{equation*}
which here also holds for non-zero torsion
with a non-symmetric Ricci tensor $R^{\,\mu\nu}$ in the case of a non-symmetric energy-momentum tensor $\theta^{\,\mu\nu}$.

The Lagrangian $\tilde{\LCd}_{\mathrm{Gr}}$ that corresponds to the sample Hamiltonian $\tilde{\HCd}_{\mathrm{Gr}}$,
defined in Eq.~(\ref{eq:H-dyn-post}) is obtained via the Legendre transformation
\begin{align*}
\tilde{\LCd}_{\mathrm{Gr}}&=\tilde{q}\indices{_{\eta}^{\nu\mu\beta}}\pfrac{\tilde{\HCd}_{\mathrm{Gr}}}{\tilde{q}\indices{_{\eta}^{\nu\mu\beta}}}-\tilde{\HCd}_{\mathrm{Gr}}\\
&=\quarter g_{1}\,g_{\mu\tau}g_{\beta\lambda}\,q\indices{_{\eta}^{\nu\mu\beta}}q\indices{_{\nu}^{\eta\tau\lambda}}\sqrt{-g}.
\end{align*}
To obtain the proper Lagrangian, the momenta $q\indices{_{\eta}^{\nu\mu\beta}}$
must be replaced by the Riemann tensor according to Eq.~(\ref{q-R-corr}), which finally yields:
\begin{equation*}
\LCd_{\mathrm{Gr}}=\quarter g_{1}\,R\indices{^\eta_{\nu\mu\beta}}\,R\indices{^\nu_{\eta\tau\lambda}}\,g^{\mu\tau}g^{\beta\lambda}
+g_{1}g_{2}\,R\indices{^\mu_{\nu\mu\beta}}\,g^{\nu\beta}-6g_{1}g_{2}^2.
\end{equation*}
Inserting this Lagrangian into the generic Einstein-type equation~(\ref{eq:consistency5}) yields the pertaining field equation:
\begin{align*}
&g_{1}\left(R\indices{^{\eta\beta\lambda\mu}}R\indices{_{\eta\beta\lambda\nu}}
-\quarter\delta_{\nu}^{\mu}R\indices{^{\eta\beta\lambda\tau}}R\indices{_{\eta\beta\lambda\tau}}\right)\nonumber\\
&-2g_{1}g_{2}\left(R\indices{^{\mu}_{\nu}}-\onehalf\delta_{\nu}^{\mu}R+3g_{2}\,\delta_{\nu}^{\mu}\right)=\theta\indices{_\nu^\mu},
\end{align*}
which coincides with Eq.~(\ref{eq:final}).
\begin{widetext}
\section{COVARIANT DIVERGENCE OF THE ``LAGRANGIAN'' CONSISTENCY EQUATION~(\ref{eq:consistency3})}
In analogy to the Hamiltonian description of Sect.~\ref{sec:cov-div}, we
calculate the covariant derivative of Eq.~(\ref{eq:consistency3}).
We start with the left-hand side:
\begin{align*}
\tilde\theta\indices{_\mu^\nu_;_\nu}&=\left(\pfrac{\tilde\LCd_0}{\phi_{;\nu}}\right)_{;\nu}\phi_{;\mu}+\pfrac{\tilde\LCd_0}{\phi_{;\nu}}\phi_{;\mu;\nu}
+\left(\pfrac{\tilde\LCd_{0}}{a_{\beta;\nu}}\right)_{;\nu}a_{\beta;\mu}+\pfrac{\tilde\LCd_{0}}{a_{\beta;\nu}}a_{\beta;\mu;\nu}
-\pfrac{\tilde\LCd_0}{\phi}\phi_{;\mu}-\pfrac{\tilde\LCd_0}{\phi_{;\nu}}\phi_{;\nu;\mu}
-\pfrac{\tilde\LCd_0}{a_\beta}a_{\beta;\mu}-\pfrac{\tilde\LCd_0}{a_{\beta;\nu}}a_{\beta;\nu;\mu}
-\pfrac{\tilde\LCd_0}{g_{\alpha\beta}}g_{\alpha\beta;\mu}\\
&=\left[\left(\pfrac{\tilde\LCd_0}{\phi_{;\nu}}\right)_{;\nu}-\pfrac{\tilde\LCd_0}{\phi}\right]\phi_{;\mu}
+\left[\left(\pfrac{\tilde\LCd_{0}}{a_{\beta;\nu}}\right)_{;\nu}-\pfrac{\tilde\LCd_0}{a_\beta}\right]a_{\beta;\mu}
+\pfrac{\tilde\LCd_0}{\phi_{;\nu}}\left(\phi_{;\mu;\nu}-\phi_{;\nu;\mu}\right)
+\pfrac{\tilde\LCd_{0}}{a_{\beta;\nu}}\left(a_{\beta;\mu;\nu}-a_{\beta;\nu;\mu}\right)
-\pfrac{\tilde\LCd_0}{g_{\alpha\beta}}g_{\alpha\beta;\mu}.
\end{align*}
The first four canonical equations of the closed set~(\ref{eq:feqs-Hdyn-1}) are equivalent to the Euler-Lagrange equations
\begin{equation*}
\left(\pfrac{\tilde\LCd_0}{\phi_{;\nu}}\right)_{;\nu}-\pfrac{\tilde\LCd_0}{\phi}=2\pfrac{\tilde\LCd_0}{\phi_{;\nu}}s\indices{^\alpha_\nu_\alpha},\qquad
\left(\pfrac{\tilde\LCd_{0}}{a_{\beta;\nu}}\right)_{;\nu}-\pfrac{\tilde\LCd_0}{a_\beta}=2\pfrac{\tilde\LCd_{0}}{a_{\beta;\nu}}s\indices{^\alpha_\nu_\alpha}.
\end{equation*}
With the second covariant derivatives
\begin{equation*}
\phi_{;\mu;\nu}-\phi_{;\nu;\mu}=-2\phi_{;\xi}s\indices{^\xi_\mu_\nu},\qquad
a_{\beta;\mu;\nu}-a_{\beta;\nu;\mu}=a_\xi R\indices{^\xi_{\beta\mu\nu}}-2a_{\beta;\xi}s\indices{^\xi_\mu_\nu},
\end{equation*}
the covariant divergence of $\tilde\theta\indices{_\mu^\nu}$ is given by:
\begin{equation}\label{eq:emt0deri-lag}
\tilde\theta\indices{_\mu^\nu_;_\nu}=2\left(\pfrac{\tilde\LCd_0}{\phi_{;\nu}}\phi_{;\mu}
+\pfrac{\tilde\LCd_{0}}{a_{\beta;\nu}}a_{\beta;\mu}\right)s\indices{^\alpha_\nu_\alpha}
-2\left(\pfrac{\tilde\LCd_0}{\phi_{;\nu}}\phi_{;\alpha}+\pfrac{\tilde\LCd_{0}}{a_{\beta;\nu}}a_{\beta;\alpha}\right)s\indices{^\alpha_\mu_\nu}
+\pfrac{\tilde\LCd_{0}}{a_{\beta;\nu}}a_\alpha R\indices{^\alpha_\beta_\mu_\nu}-\pfrac{\tilde\LCd_0}{g_{\alpha\beta}}g_{\alpha\beta;\mu}.
\end{equation}
\nopagebreak
It is generically nonzero for systems involving vector fields and in the presence of torsion or non-metricity.

Setting up the covariant derivative of the right-hand side of Eq.~(\ref{eq:consistency3}) yields:
\begin{align*}
\tilde\vartheta\indices{_\mu^\nu_;_\nu}&=
\left(\pfrac{\tilde\LCd_{\mathrm{Gr}}}{g_{\lambda\beta;\nu}}\right)_{;\nu}g_{\lambda\beta;\mu}
+\pfrac{\tilde\LCd_{\mathrm{Gr}}}{g_{\lambda\beta;\nu}}g_{\lambda\beta;\mu;\nu}
+2\left(\pfrac{\tilde\LCd_{\mathrm{Gr}}}{R\indices{^{\eta}_{\lambda\beta\nu}}}\right)_{;\nu}R\indices{^{\eta}_{\lambda\beta\mu}}
+2\pfrac{\tilde\LCd_{\mathrm{Gr}}}{R\indices{^{\eta}_{\lambda\beta\nu}}}R\indices{^{\eta}_{\lambda\beta\mu;\nu}}
-\pfrac{\tilde\LCd_{\mathrm{Gr}}}{g_{\lambda\beta}}g_{\lambda\beta;\mu}
-\pfrac{\tilde\LCd_{\mathrm{Gr}}}{g_{\lambda\beta;\nu}}g_{\lambda\beta;\nu;\mu}
-\pfrac{\tilde\LCd_{\mathrm{Gr}}}{R\indices{^{\eta}_{\lambda\beta\nu}}}R\indices{^{\eta}_{\lambda\beta\nu;\mu}}\\
&=\pfrac{\tilde\LCd_{\mathrm{Gr}}}{g_{\lambda\beta;\nu}}\left(g_{\lambda\beta;\mu;\nu}-g_{\lambda\beta;\nu;\mu}\right)
-2\pfrac{\tilde\LCd_{\mathrm{Gr}}}{R\indices{^{\eta}_{\lambda\beta\nu}}}\left(\onehalf R\indices{^{\eta}_{\lambda\beta\nu;\mu}}-R\indices{^{\eta}_{\lambda\beta\mu;\nu}}\right)
+\left[\left(\pfrac{\tilde\LCd_{\mathrm{Gr}}}{g_{\lambda\beta;\nu}}\right)_{;\nu}
-\pfrac{\tilde\LCd_{\mathrm{Gr}}}{g_{\lambda\beta}}\right]g_{\lambda\beta;\mu}
+2\left(\pfrac{\tilde\LCd_{\mathrm{Gr}}}{R\indices{^{\eta}_{\lambda\beta\nu}}}\right)_{;\nu}R\indices{^{\eta}_{\lambda\beta\mu}}.
\end{align*}
The last four canonical equations of the closed set~(\ref{eq:feqs-Hdyn-1}) have the equivalent Euler-Lagrange representations:
\begin{equation}\label{eq:feqs-Ldyn-1}
\left(\pfrac{\tilde\LCd_{\mathrm{Gr}}}{g_{\lambda\beta;\nu}}\right)_{;\nu}-\pfrac{\tilde\LCd_{\mathrm{Gr}}}{g_{\lambda\beta}}=2\pfrac{\tilde\LCd_{\mathrm{Gr}}}{g_{\lambda\beta;\nu}}s\indices{^{\,\alpha}_{\nu\alpha}}+\pfrac{\tilde\LCd_0}{g_{\lambda\beta}},\qquad
2\left(\pfrac{\tilde\LCd_{\mathrm{Gr}}}{R\indices{^{\eta}_{\lambda\beta\nu}}}\right)_{;\nu}=
\pfrac{\tilde\LCd_0}{a_{\lambda;\beta}}a_{\eta}+2\pfrac{\tilde\LCd_{\mathrm{Gr}}}{g_{\lambda\nu;\beta}}g_{\nu\eta}
+2\pfrac{\tilde\LCd_{\mathrm{Gr}}}{R\indices{^{\eta}_{\lambda\nu\alpha}}}s\indices{^{\,\beta}_{\nu\alpha}}
+4\pfrac{\tilde\LCd_{\mathrm{Gr}}}{R\indices{^{\eta}_{\lambda\beta\nu}}}s\indices{^\alpha_{\nu\alpha}}.
\end{equation}
With Eqs.~(\ref{eq:cd-RC}), (\ref{eq:comm-cd-metr}), and~(\ref{eq:feqs-Ldyn-1}), the covariant divergence $\tilde\vartheta\indices{_\mu^\nu_;_\nu}$ simplifies to:
\begin{align}
\tilde\vartheta\indices{_\mu^\nu_;_\nu}&=
2\pfrac{\tilde\LCd_{\mathrm{Gr}}}{g_{\lambda\beta;\nu}}\left(g_{\lambda\beta;\alpha}s\indices{^{\,\alpha}_{\nu\mu}}-\bcancel{R\indices{^\eta_\beta_\nu_\mu}g_{\eta\lambda}}\right)
+2\pfrac{\tilde\LCd_{\mathrm{Gr}}}{R\indices{^\eta_{\lambda\beta\nu}}}\left(\cancel{R\indices{^\eta_{\lambda\mu\xi}}s\indices{^\xi_{\beta\nu}}}
+R\indices{^\eta_{\lambda\beta\xi}}s\indices{^\xi_{\nu\mu}}+R\indices{^\eta_{\lambda\nu\xi}}s\indices{^\xi_{\mu\beta}}\right)\nonumber\\
&\quad+\left(\pfrac{\tilde\LCd_0}{g_{\lambda\beta}}+2\pfrac{\tilde\LCd_{\mathrm{Gr}}}{g_{\lambda\beta;\nu}}s\indices{^{\,\alpha}_{\nu\alpha}}\right)g_{\lambda\beta;\mu}
+\left(\pfrac{\tilde\LCd_0}{a_{\lambda;\beta}}a_{\eta}+\bcancel{2\pfrac{\tilde\LCd_{\mathrm{Gr}}}{g_{\lambda\nu;\beta}}g_{\nu\eta}}
+\cancel{2\pfrac{\tilde\LCd_{\mathrm{Gr}}}{R\indices{^{\eta}_{\lambda\nu\alpha}}}s\indices{^{\,\beta}_{\nu\alpha}}}
+4\pfrac{\tilde\LCd_{\mathrm{Gr}}}{R\indices{^{\eta}_{\lambda\beta\nu}}}s\indices{^\alpha_{\nu\alpha}}\right)R\indices{^{\eta}_{\lambda\beta\mu}}\nonumber\\
&=\pfrac{\tilde\LCd_0}{a_{\lambda;\beta}}a_{\eta}R\indices{^{\eta}_{\lambda\beta\mu}}+\pfrac{\tilde\LCd_0}{g_{\lambda\beta}}g_{\lambda\beta;\mu}
+2\pfrac{\tilde\LCd_{\mathrm{Gr}}}{g_{\lambda\beta;\nu}}\left(g_{\lambda\beta;\alpha}s\indices{^{\,\alpha}_{\nu\mu}}+g_{\lambda\beta;\mu}s\indices{^{\,\alpha}_{\nu\alpha}}\right)
+4\pfrac{\tilde\LCd_{\mathrm{Gr}}}{R\indices{^\eta_{\lambda\beta\nu}}}\left(R\indices{^\eta_{\lambda\beta\mu}}s\indices{^\alpha_{\nu\alpha}}
+R\indices{^\eta_{\lambda\beta\xi}}s\indices{^\xi_{\nu\mu}}\right).\label{eq:emtgrderi-lag}
\end{align}
Again, the divergence $\tilde\vartheta\indices{_\mu^\nu_;_\nu}$ never vanishes in the presence of vector fields, torsion or non-metricity.

Setting up the sum of both divergences~(\ref{eq:emt0deri-lag}) and~(\ref{eq:emtgrderi-lag}), all terms not directly related to the torsion tensor drop out:
\begin{align*}
\tilde\theta\indices{_\mu^\nu_;_\nu}+\tilde\vartheta\indices{_\mu^\nu_;_\nu}&=2\left(\pfrac{\tilde\LCd_0}{\phi_{;\nu}}\phi_{;\mu}
+\pfrac{\tilde\LCd_{0}}{a_{\beta;\nu}}a_{\beta;\mu}\right)s\indices{^\alpha_\nu_\alpha}
-2\left(\pfrac{\tilde\LCd_0}{\phi_{;\nu}}\phi_{;\alpha}+\pfrac{\tilde\LCd_{0}}{a_{\beta;\nu}}a_{\beta;\alpha}\right)s\indices{^\alpha_\mu_\nu}
+\cancel{\pfrac{\tilde\LCd_{0}}{a_{\beta;\nu}}a_\alpha R\indices{^\alpha_\beta_\mu_\nu}}-\bcancel{\pfrac{\tilde\LCd_0}{g_{\alpha\beta}}g_{\alpha\beta;\mu}}\\
&\quad+\cancel{\pfrac{\tilde\LCd_0}{a_{\lambda;\beta}}a_{\eta}R\indices{^{\eta}_{\lambda\beta\mu}}}+\bcancel{\pfrac{\tilde\LCd_0}{g_{\lambda\beta}}g_{\lambda\beta;\mu}}
+2\pfrac{\tilde\LCd_{\mathrm{Gr}}}{g_{\lambda\beta;\nu}}\left(g_{\lambda\beta;\alpha}s\indices{^{\,\alpha}_{\nu\mu}}+g_{\lambda\beta;\mu}s\indices{^{\,\alpha}_{\nu\alpha}}\right)
+4\pfrac{\tilde\LCd_{\mathrm{Gr}}}{R\indices{^\eta_{\lambda\beta\nu}}}\left(R\indices{^\eta_{\lambda\beta\mu}}s\indices{^\alpha_{\nu\alpha}}
+R\indices{^\eta_{\lambda\beta\alpha}}s\indices{^\alpha_{\nu\mu}}\right)\\
&=2\left(\pfrac{\tilde\LCd_0}{\phi_{;\nu}}\phi_{;\mu}+\pfrac{\tilde\LCd_{0}}{a_{\beta;\nu}}a_{\beta;\mu}
+\pfrac{\tilde\LCd_{\mathrm{Gr}}}{g_{\lambda\beta;\nu}}g_{\lambda\beta;\mu}
+2\pfrac{\tilde\LCd_{\mathrm{Gr}}}{R\indices{^\eta_{\lambda\beta\nu}}}R\indices{^\eta_{\lambda\beta\mu}}\right)s\indices{^\alpha_\nu_\alpha}\\
&\quad-2\left(\pfrac{\tilde\LCd_0}{\phi_{;\nu}}\phi_{;\alpha}+\pfrac{\tilde\LCd_{0}}{a_{\beta;\nu}}a_{\beta;\alpha}
+\pfrac{\tilde\LCd_{\mathrm{Gr}}}{g_{\lambda\beta;\nu}}g_{\lambda\beta;\alpha}
+2\pfrac{\tilde\LCd_{\mathrm{Gr}}}{R\indices{^\eta_{\lambda\beta\nu}}}R\indices{^\eta_{\lambda\beta\alpha}}\right)s\indices{^\alpha_\mu_\nu}.
\end{align*}
However, the torsion tensor terms also cancel, as can be seen inserting the Lagrangian representation
of the consistency equation from Eq.~(\ref{eq:consistency3}) for the terms in parentheses:
\begin{align*}
\tilde\theta\indices{_\mu^\nu_;_\nu}+\tilde\vartheta\indices{_\mu^\nu_;_\nu}&=
2\delta_\mu^\nu\left(\tilde\LCd_{0}+\tilde\LCd_{\mathrm{Gr}}\right)s\indices{^\alpha_\nu_\alpha}
-2\delta_\alpha^\nu\left(\tilde\LCd_{0}+\tilde\LCd_{\mathrm{Gr}}\right)s\indices{^\alpha_\mu_\nu}
=0.
\end{align*}
Thus, provided that the consistency equation~(\ref{eq:consistency3}) is satisfied, its covariant divergence vanishes as well
along the system path given by the solution of the Euler-Lagrange field equations,
independently of the specific form of the Lagrangians $\tilde\LCd_0$ and $\tilde\LCd_{\mathrm{Gr}}$:
\begin{equation}\label{eq:ce-concise-div2}
\tilde\theta\indices{_\mu^\alpha}+\tilde\vartheta\indices{_\mu^\alpha}=0\qquad\Rightarrow\qquad
\left(\tilde\theta\indices{_\mu^\alpha}+\tilde\vartheta\indices{_\mu^\alpha}\right)_{;\alpha}=0.
\end{equation}
Remarkably, the \emph{individual} covariant divergences $\tilde\theta\indices{_\mu^\alpha_;_\alpha}$
and $\tilde\vartheta\indices{_\mu^\alpha_;_\alpha}$ do not vanish in general, but only their sum.
This applies even for the case of metricity and zero torsion.
In the presence of a vector field, the term $\pfrac{\tilde\LCd_0}{a_{\lambda;\beta}}a_{\eta}R\indices{^{\eta}_{\lambda\beta\mu}}$
only cancels in the sum~(\ref{eq:ce-concise-div2}).
\end{widetext}
\input{chgf1.bbl}
\end{document}

%% file: chgf1.bbl
%